\documentclass[a4paper,fleqn]{cas-dc}

\usepackage[numbers]{natbib}
\usepackage{algorithmic}
\usepackage{algorithm}
\usepackage{amsmath}
\usepackage{multirow,booktabs}
\usepackage{subcaption}
\usepackage{bm}

\def\tsc#1{\csdef{#1}{\textsc{\lowercase{#1}}\xspace}}
\tsc{WGM}
\tsc{QE}
\tsc{EP}
\tsc{PMS}
\tsc{BEC}
\tsc{DE}

\begin{document}
\let\WriteBookmarks\relax
\def\floatpagepagefraction{1}
\def\textpagefraction{.001}
\shorttitle{Leveraging social media news}
\shortauthors{CV Radhakrishnan et~al.}

\title [mode = title]{Deep Class-guided Hashing for Multi-label Cross-modal Retrieval}                      
\tnotemark[1,2]



\author[1]{Hao Chen} 
\ead{cgh@stu.hunau.edu.cn}
\address[1]{College of Information and Intelligence, Hunan Agricultural University, Changsha, P.R. China, 410125}
\credit{Conceptualization, Methodology, Writing - Original draft preparation}

\author[1]{Lei Zhu}[]
\ead{leizhu@hunau.edu.cn}

\author[1]{Xinghui Zhu}[]
\ead{zhuxh@hunau.edu.cn}












\begin{abstract}
  Deep hashing, due to its low cost and efficient retrieval advantages, is widely valued in cross-modal retrieval. However, existing cross-modal hashing methods either explore the relationships between data points, which inevitably leads to intra-class dispersion, or explore the relationships between data points and categories while ignoring the preservation of inter-class structural relationships, resulting in the generation of suboptimal hash codes. How to maintain both intra-class aggregation and inter-class structural relationships, In response to this issue, this paper proposes a DCGH method. Specifically, we use proxy loss as the mainstay to maintain intra-class aggregation of data, combined with pairwise loss to maintain inter-class structural relationships, and on this basis, further propose a variance constraint to address the semantic bias issue caused by the combination. A large number of comparative experiments on three benchmark datasets show that the DCGH method has comparable or even better performance compared to existing cross-modal retrieval methods. The code for the implementation of our DCGH framework is available at https://github.com/donnotnormal/DCGH.
\end{abstract}

\begin{graphicalabstract}
  \includegraphics[width=1.0\linewidth]{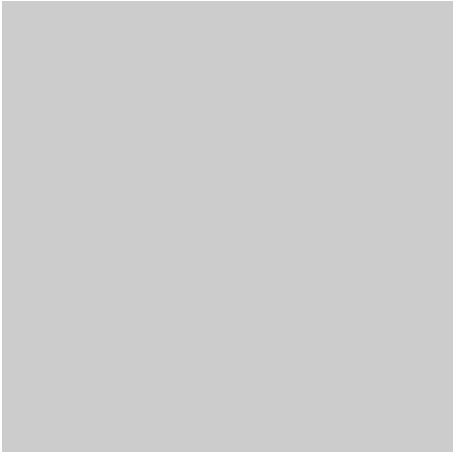}
\end{graphicalabstract}

\begin{highlights}
\item Research highlights item 1
\item Research highlights item 2
\item Research highlights item 3
\end{highlights}

\begin{keywords}
  Deep Hashing;\sep Intra-class aggregation;\sep Inter-class structural relationships;\sep Cross-modal Retrieval;\sep Semantic bias;
\end{keywords}

\maketitle

\section{Introduction} \label{sec:introduction}

In recent years, by employing Approximate Nearest Neighbor (ANN) search technology, cross-modal retrieval has shown great potential in handling large-scale information retrieval tasks. The core challenge of implementing cross-modal retrieval lies in how to effectively bridge the semantic gap between different modalities and accurately capture their semantic connections. Traditional cross-modal retrieval methods usually transform data from different modalities into a shared low-dimensional subspace and calculate the similarity between data within this subspace. However, as the number of data continues to grow, these methods may reduce retrieval efficiency due to increased computational complexity.

To address this challenge, hash-based cross-modal retrieval methods have begun to receive widespread attention. Cross-Modal Hashing methods compress high-dimensional data instances into compact binary codes in the latent space, ensuring that similar data items have similar hash codes. This approach uses efficient OR (XOR) operations to measure the similarity between different modalities, thus providing a fast and effective solution for similarity retrieval of cross-modal data.
Simultaneously investigate feature representation and hash encoding. Compared to shallow hash techniques~\cite{SePH,CMFH,CCA,STMH,LSCM,MLH}, deep hashing methods~\cite{DSPH,EGDH,DN-pH,CMCL,RDPH,SCAHN,MITH,DSeH,LTCH}, leveraging deep learning networks, can extract more discriminative features, thereby generating higher quality hash codes.

In previous research on deep hashing methods, most approaches based on pairwise or triplet losses have considered the relationships between data points, pulling each data point closer to similar ones and pushing it further away from dissimilar ones. This method effectively preserves the inter-class structural relationships. However, the optimization objectives typically focus on the relative distances between sample pairs, lacking global constraints on the overall distribution of intra-class samples, and only optimize locally on sample pairs. Since the model does not learn the global characteristic of keeping intra-class samples clustered together, it often results in intra-class dispersion. This phenomenon, when compounded by the hashing quantization process, can significantly impact the model's retrieval performance.
As a result, hash models that consider the relationships between points and classes have emerged, such as center loss and proxy  loss. Center loss penalizes the distance between each sample and its class center, causing intra-class samples to cluster together and form a compact intra-class distribution. However, in the multi-label retrieval process, the number of categories involved is very large and dynamically changing. In this case, it is difficult to define and maintain a stable center vector for each class, making the use of center loss complex. Proxy  loss generates a proxy vector for each single-label category, learning the hash representation by reducing the distance between samples and their corresponding proxies and increasing the distance between samples and proxies of incorrect categories. The introduction of proxies enables the model to have better generalization capabilities when dealing with unseen categories and transforms the problem of relationships between sample data into a problem of only considering the relationships between data and proxy vectors, thus effectively maintaining intra-class aggregation. However, while they solve the problem of intra-class dispersion, they neglect to maintain inter-class structural relationships.As shown in Figure~\ref{fig:our-method}.

To maintain both intra-class aggregation and inter-class structural relationships, a simple idea is to combine pairwise loss with proxy loss. However, how to effectively combine these two is a matter that warrants consideration. When considering both the relationships between data points and the relationships between data points and class proxies, semantic bias can occur, especially when there is an imbalance in the number of data points associated with certain label categories.
 As illustrated in Figure~\ref{fig:problem}, the pentagrams represent proxies for each single-label class. Class S1 is related to proxy points P1, P2, and P3, while class S2 is related to proxy points P1 and P4. When considering both the relationships between points and proxies, and between points and points, if there are more data points related to proxy P1 than to other proxies, the data points will tend to lean towards proxy P1. However, data points should maintain a consistent distance relationship with each of their relevant proxies; otherwise, it will lead to semantic bias.

Secondly, if the approach is merely a simple combination, the inter-class structural relationships are indeed maintained. However, each data point will still consider its distance relationships with other data points, and merely adding the consideration of the distance relationships with its corresponding class proxies on this basis can still lead to intra-class dispersion, which affects the performance of the hash codes.

In response to the aforementioned issues, we have accordingly provided solutions.We have introduced a variance constraint to ensure that the distances between each data point and its corresponding  proxy  are as consistent as possible, preventing semantic bias. Regarding the issue of intra-class dispersion, considering that in multi-label datasets, the number of irrelevant pairs in the pairwise loss is much smaller than the number of relevant pairs, and that positive sample pairs have a stronger constraint and are more likely to cause intra-class dispersion compared to negative sample pairs, we assign different small weights to the positive and negative sample constraints of the pairwise loss. This allows the aggregation of similar data points to be primarily guided by the proxy loss, thereby addressing the issue of intra-class dispersion.

To sum up, the main contributions of this article are three-folds:

\begin{figure*}
  \centering
  \includegraphics[width=1.0\linewidth]{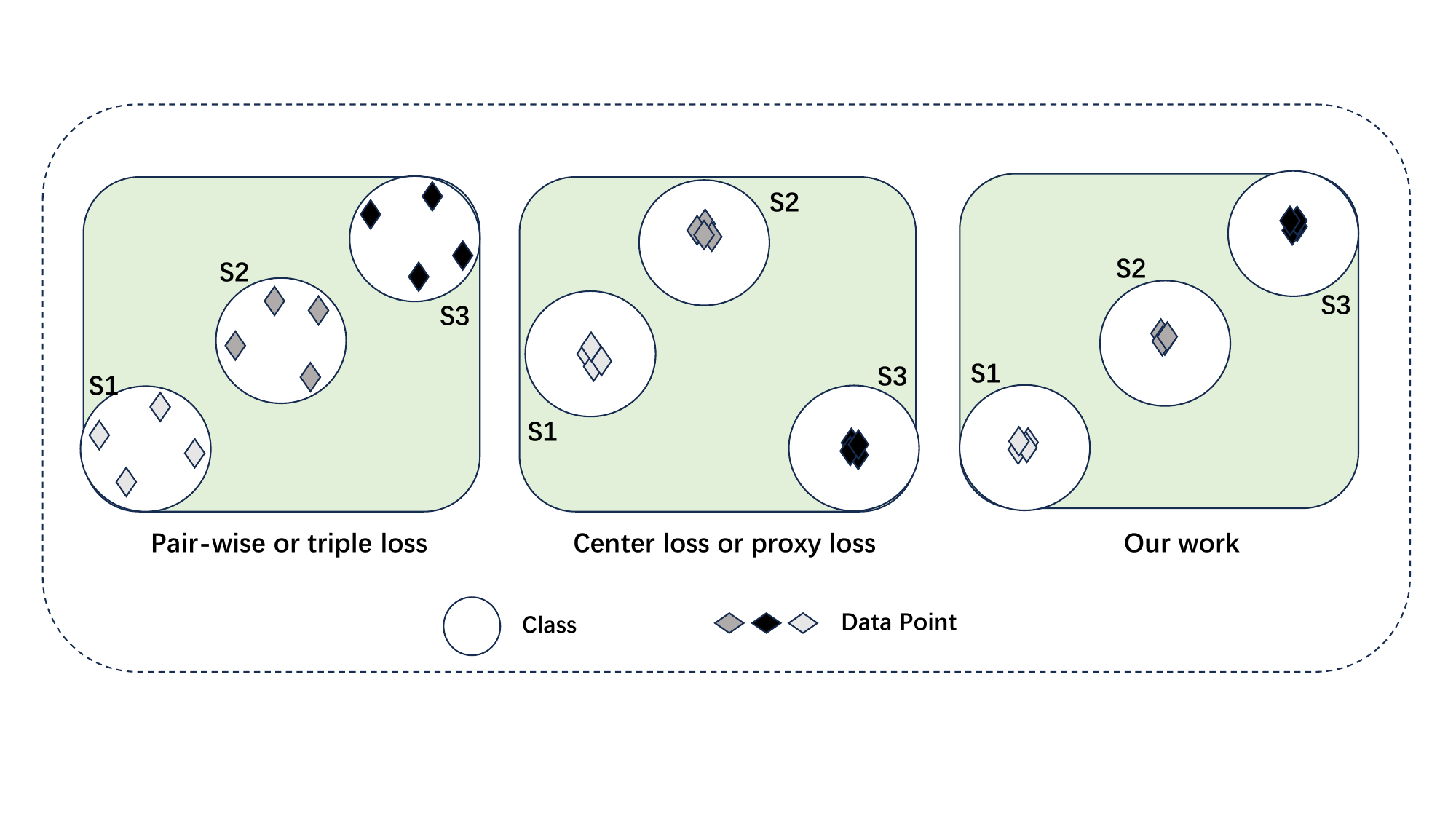}
  \vspace{-15mm}
  \caption{\small Illustration of hash codes learned using different hash losses. Class S1 and S2 have common labels, S2 and S3 have common labels, while S1 and S3 have no common labels. Compared with the pairwise loss, triplet loss, center loss and proxy loss, our approach maintains both intra-class aggregation and inter-class structural relationships.}
  \label{fig:our-method}
\end{figure*}




\begin{itemize}
  \item \textbf{Starting from the perspective of intra-class aggregation and inter-class structural relationship maintenance, this paper proposes a combination of proxy loss and pairwise loss.} 
  \item \textbf{Further considering the issues arising from the combination of proxy loss and pairwise loss, the DCGH method is proposed.} 
  \item \textbf{Extensive experiments on three benchmark datasets demonstrate that the proposed DCGH algorithm outperforms other baseline methods in cross-modal hashing retrieval.}
\end{itemize}

 The remainder of this paper is organized as follows. Section~\ref{sec:related-works} 
introduces representative supervised cross-modal hashing methods and unsupervised cross-modal hashing methods. Section~\ref{sec:methodology} introduces the proposed method in detail. Experimental results and analysis are provided in Section~\ref{sec:experiments}, Section~\ref{sec:conclusion} summarizes the conclusions and future work.
\section{Related Works} \label{sec:related-works}
In this section, we review the related studies from two aspects: supervised cross-modal hashing in~\ref{subsec:supervised} and unsupervised cross-modal hashing in~\ref{subsec:unsupervised}.

\subsection{Supervised Hashing Methods} \label{subsec:supervised}

Supervised cross-modal hashing methods primarily leverage label information to learn hashing functions, achieving satisfactory performance. Existing supervised hashing methods either consider the relative relationships between data points based on pairwise or triplet losses, or learn the hashing functions by considering the relationship between data and public anchor points through center loss or proxy loss. We will briefly introduce previous work from these two perspectives.
\subsubsection{Pairwise loss and triplet loss} \label{Pairwise loss and triplet loss}

Pairwise loss and triplet loss both learn the hashing function by leveraging the relative relationships between data points.DCMH was the first to integrate hash code learning and feature learning into the same framework for each modality, enabling end-to-end learning~\cite{DCMH}.In order to explore the inherent similarity structure between pairwise data points, the GCH method introduces graph convolutional networks to capture structural semantic information.~\cite{GCH}.SSAH incorporates adversarial learning in a self-supervised manner into cross-modal hashing learning, employing two adversarial networks to jointly learn high-dimensional features specific to each modality, and using information obtained from the label network to supervise the learning process~\cite{SSAH}.TDH introduces intra-modal and inter-modal triplets, proposing a triplet-based deep hashing framework to establish semantic relationships between heterogeneous data from different modalities~\cite{TDH}.Additionally, the Deep Adversarial Discrete Hashing framework has developed a weighted cosine triplet loss to explore the similarity relationships between data under multi-label settings~\cite{DADH}.As multi-label cross-modal hashing methods gain significant attention, the Multi-label Enhanced Self-Supervised Deep Cross-Modal Hashing (MESDCH) approach, which employs a multi-label semantic similarity module, is used to maintain pairwise similarity relationships within multi-label data~\cite{MESDCH}.However, optimizing binary hash codes during model training is challenging. In response to this, scholars from home and abroad have proposed a selection mechanism for the Differentiable Cross-Modal Hashing via Multimodal Transformers (DCHMT) method to more effectively optimize the hash codes~\cite{DCHMT}.HHF further explores the incompatible conflict between metric learning and quantization learning in the optimization process, setting different thresholds according to different hash bit lengths and the number of categories to reduce this conflict~\cite{HHF}.To address the issues of solution space compression and loss function oscillation, SCH categorizes sample pairs into fully semantically similar, partially semantically similar, and semantically negative pairs based on their similarity. It imposes different constraints on each category to ensure the entire Hamming space is utilized, and sets upper and lower bounds on the similarity between samples to prevent loss function oscillation~\cite{SCH}.

\subsubsection{Center loss and proxy loss} \label{Center loss and proxy loss}

To address the issue of intra-class dispersion, EGDH employs a label network that encodes non-redundant multi-label annotations into hash codes to form a common classifier, guiding the hash feature learning network to construct features that are easily classified by the classifier and close to their semantic hash code anchors~\cite{EGDH}. SCCGDH utilizes the label information of the data to learn a specific class center for each label, keeping data points close to their corresponding class centers to maintain intra-class aggregation~\cite{SCCGDH}.DCPH first introduced the concept of proxy hash, mapping label or category information to proxy hash codes, which are then used to guide the training of modality-specific hash networks~\cite{DCPH}. Considering that the generation of previous proxies did not take into account the distribution of the data, DAPH integrates data features and labels into a framework to learn category proxies, and then uses the obtained category proxies to guide the training of modality-specific hash networks~\cite{DAPH}.
DSPH employs the Transformer as the feature extraction architecture, and points out that proxy hashing methods have an issue with ambiguous relationships between negative samples, adding a negative sample constraint to the proxy loss to address this issue.~\cite{DSPH}.How to independently preserve each hash bit and learn zero-mean threshold hash functions has always been a very challenging issue in cross-modal hashing ~\cite{gailv}. Moreover, due to the variability of balance conditions across batches, the current binary balance mechanism is not well-suited for batch processing in training methods. To address these issues, DNPH introduces a uniform distribution constraint into the proxy loss to ensure that the distribution of hash codes follows a discrete uniform distribution, and to maintain modality-specific semantic information, it adds a prediction layer for each modality network using cross-entropy loss~\cite{DNPH}.DHaPH further considers the potential semantic hierarchical relationships between multimodal samples, introduces learnable shared hierarchical proxies in hyperbolic space, and proposes a hierarchical perception proxy loss algorithm that effectively mines potential semantic hierarchical information without requiring any hierarchical prior knowledge~\cite{DHaPH}.

\subsection{Unsupervised Hashing Methods} \label{subsec:unsupervised}

Under the condition of lacking supervisory information, deep cross-modal hashing technology has shown excellent performance in cross-modal retrieval tasks. Deep Binary Reconstruction (DBRC)~\cite{DBRC} introduces a deep learning framework designed for binary reconstruction, which aims to model the relationships between different modalities and learn the corresponding hash codes. Unsupervised Deep Cross-Modal Hashing (UDCMH) ~\cite{UDCMH} uses an alternating optimization strategy to solve discrete optimization problems, and dynamically assigns weights to different modalities during the optimization process. Deep Joint Semantic Reconstructing Hashing (DJSRH) ~\cite{DJSRH} reconstructs a structure that captures the joint semantics to learn hash codes, explicitly incorporating the original neighborhood information derived from the multi-modal data. Aggregation-based Graph Convolutional Hashing (AGCH) ~\cite{AGCH} aggregates structural information and applies various similarity measurement methods to construct a comprehensive similarity matrix.Joint-modal Distribution-based Similarity Hashing (JDSH) ~\cite{JDSH} creates a joint-modal similarity matrix and designs a sampling and weighting mechanism to generate more discriminative hash codes.  Correlation-Identity Reconstruction Hashing (CIRH) ~\cite{CIRH} designs a multi-modal collaborative graph to build heterogeneous multi-modal correlations and performs semantic aggregation on graph networks to generate a multi-modal complementary representation.

\begin{table}
	\caption{A summary of frequently-used notations and symbols.}\label{tab:notations}
	\begin{tabular*}{\tblwidth}{@{} LL@{} }
		\toprule
		Notation & Definition \\
		\midrule
		$\bm{D}$ & Training Dataset \\
    $\bm{N}$ & Number of Samples \\
    $\bm{G}$ & Feature Extractor \\
    $\bm{F}$ & Features of Samples \\
		$\bm{B}$ & Hash Code \\
		$K$ & the length of hash codes \\
		$C$ & the number of categories \\
		$\bm{S}$ & Similarity matrix \\
    $\bm{H}$ & Hash Functions \\
    $\bm{x}_{i}$ & i-th Image Sample\\
    $\bm{y}_{i}$ & i-th Text Sample \\
		$\bm{l}_{i}$ & Label Corresponding to i-th Sample \\
    $\bm{P}$ & Learnable Proxies \\
    $\bm{I}$ & Indicator Function \\
    $\operatorname{Var}$ & Variance Function\\
		\bottomrule
	\end{tabular*}
\end{table}

\section{Methodology} \label{sec:methodology}
This section introduces the proposed method. Before delving into the technical details, we present the problem formulation at first. Then, we provide a overview of DCGH, followed by details of each components.

\subsection{Problem Formulation} \label{subsec:problem-formulation}

\textbf{Problem Definition.} 
In this work, we primarily focus on the image-text cross-modal retrieval task. Given an image-text training set $D=\left\{d_{i}\right\}_{i=1}^{N}$ containing N samples, where $d_{i}=\left\{x_{i}, y_{i}, l_{i}\right\}$, $x_{i}$ and $y_{i}$ represent the 
${i}$-th example of the image and text modalities, respectively. $l_{i}=\left[l_{i 1}, l_{i 2}, \ldots, l_{i c}\right]$ is a multi-label annotation for $d_{i}$ , and $C$ represents the number of categories. The goal of deep cross-modal hashing is to learn image and text hashing functions $H^{x}$ and $H^{y}$, which project image data and text data into the Hamming space to generate binary hash codes $B^{x} \in\{-1,1\}^{K}$ and $B^{y} \in\{-1,1\}^{K}$, respectively, ensuring that the Hamming distance between relevant samples is smaller and the Hamming distance between irrelevant samples is larger. Since binary optimization is a typical NP-hard problem, the continuous relaxation strategy ($tanh$)
is employed to learn the binary-like code during training. The frequently used mathematical notations are summarized in Table~\ref{tab:notations} for readability.

\begin{figure*}
  \centering
  \includegraphics[width=1.0\linewidth]{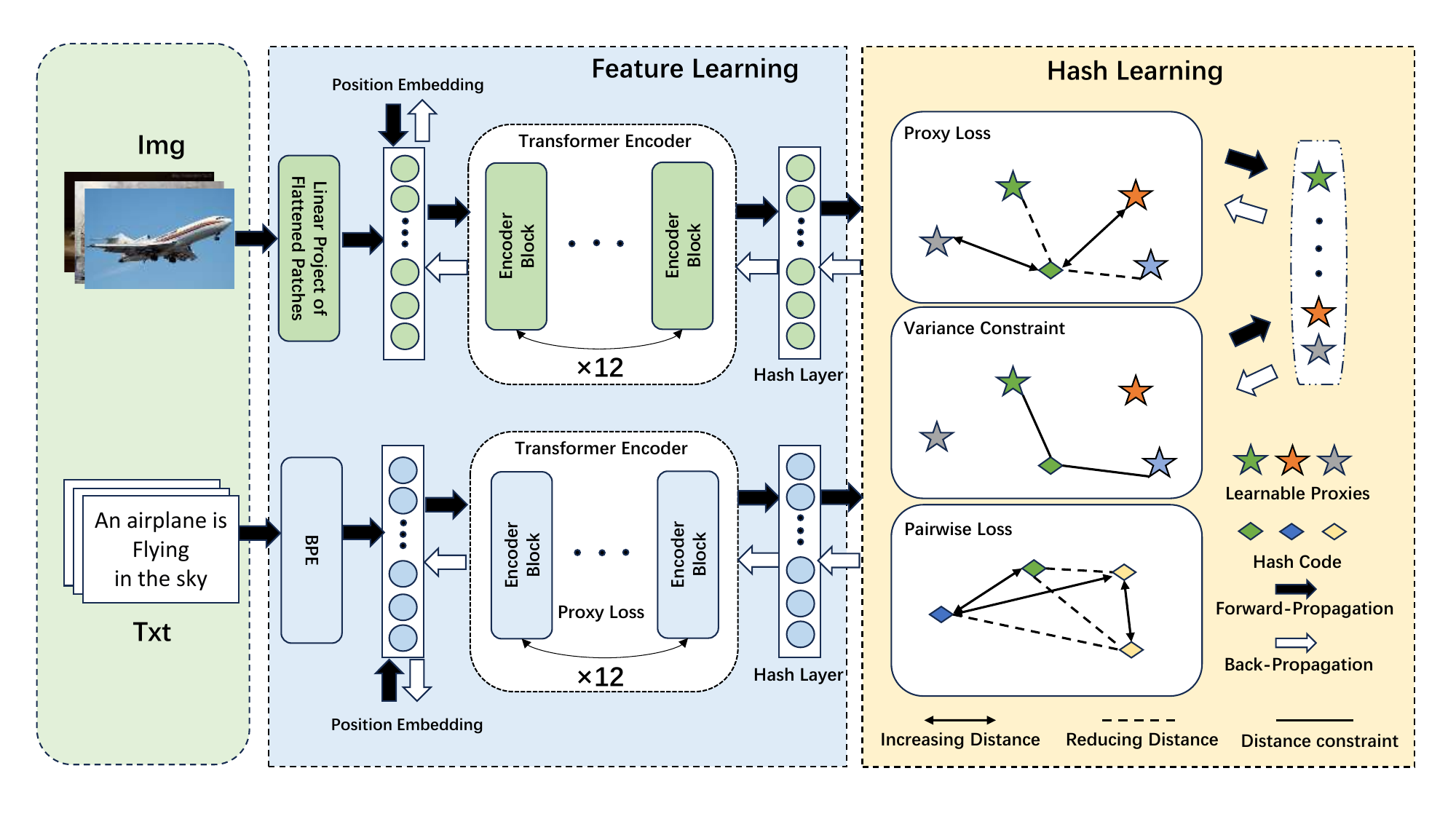}
  \vspace{-2mm}
  \caption{\small  An overview of our proposed DCGH framework, including two parts: (1) Feature Learning: Two feature extractors
with different Transformer Encoders are designed to extract the representative semantic features from image modalities and
text modalities respectively. (2) Hash Learning: Ingeniously combining proxy loss and pairwise loss, while exploring relationships between points and between points and classes, and to prevent semantic bias, variance constraints are introduced to ensure the consistency of the relationship between data and relevant proxy points. }
  \label{fig:framework}
\end{figure*} 

\subsection{Overview of DCGH Framework} \label{subsec:overview}
As illustrated in Fig.~\ref{fig:framework}, the proposed DCGH framework, in general, consists of DCGH modules: Feature Learning, Hash Learning.

\subsection{Feature Learning} \label{subsec:}

In traditional CNN-based image encoders~\cite{RESNET}, convolutional kernels typically only capture local, limited-range features. This means that for global information in images, especially features that span larger spatial distances, CNN may struggle to capture them effectively. This limits the model's ability to understand the global structure and content of images~\cite{Bi_NCMH}. Compared to traditional CNNs, Transformer models~\cite{TRANSFORMER}, through their self-attention mechanism, can capture the relationships between any two positions in an image or text sequence. This global perception capability allows the model to better understand the structure and content of the entire data, enhancing the extraction of semantic information. As research deepens, Transformer encoders are expected to play an increasingly important role in the fields of computer vision and natural language processing in the future.
Inspired by ~\cite{DCHMT} and ~\cite{DSPH}, we introduce feature extraction based on the Transformer encoder into cross-modal retrieval to obtain representative semantic features from both the image and text modalities. Specifically, the image Transformer encoder has the same structure as the VIT encoder ~\cite{VIT}, which is composed of stacked 12 encoder blocks. Each encoder block has the same structure, including Layer Normalization (LN), Multi-Head Self-Attention (MSA), and MLP blocks. The number of MSA is 12. For the image semantic feature $F_{x}=G_{x}\left(X, \theta_{x}\right)$,where $G_{x}$ represents the image semantic encoder, and $theta_{x}$ represents the parameters of the image semantic encoder. The text Transformer encoder consists of 12 encoder blocks, each with 8 MSA. For the text semantic feature $F_{y}=G_{y}\left(Y, \theta_{y}\right)$, where $G_{y}$ represents the text semantic encoder, and $theta_{y}$ is the parameter of the text semantic encoder.

\subsection{Hash Learning} \label{subsec:}

In order to generate hash codes that maintain intra-class aggregation while preserving inter-class structural relationships, a comprehensive objective function composed of proxy loss, pairwise loss, and variance constraint has been designed to optimize the parameters of our proposed DCGH framework. In the following sections, we will detail each component of the DCGH architecture.

\subsubsection{Proxy Loss} \label{Proxy Loss}

Proxy-based methods can achieve satisfactory performance in single-label cross-modal retrieval, but considering multi-label cross-modal retrieval, proxy-based methods have been proven to produce poor performance with limited hash bits because they fail to deeply express multi-label correlations and neglect the preservation of inter-class structural relationships ~\cite{DCPH}. However, since only the relationships between data and proxy points need to be considered, the learned hash codes can well maintain intra-class aggregation. Therefore, we first learn hash codes for intra-class aggregation through proxy loss.

For the proxy loss, a learnable proxy is first generated for each label category. The hash representation is learned by bringing samples closer to their relevant proxies and pulling away irrelevant data-proxy pairs. For $P=\left\{p_{1}, p_{2}, \ldots \ldots p_{C}\right\}$, $P$ is the learnable proxies for each label category, where $p_{i}$ is a $K$-bits vector. When the samples and the proxy are relevant, we could calculate the cosine distance between the binary-like hash codes and relevant proxies by using Eq.( 1).
\begin{equation}
\cos _{+}(h, p)=-\cos \langle h, p\rangle=-\frac{|h \cdot p|}{|h| \cdot|p|}
\end{equation}

For proxies that are not related to the samples, the distance between binary-like hash codes and irrelevant proxies is pushed away by calculating Equation ( 2).
\begin{equation}
\cos _{-}(h, p)  =(\cos \langle h, p\rangle)_{+}  =\max \left(\frac{|h \cdot p|}{|h| \cdot|p|}, 0\right)
\end{equation}

Hence, the  proxy loss of image $\mathcal{L}_{\text {proxy }}^{x}$ could be calculated as shown in Eq.( 3).
\begin{equation}
\begin{aligned}
\mathcal{L}_{\text {proxy }}^{x} & =\frac{\sum_{i=1}^{N} \sum_{j=1}^{C} I\left(l_{i j}=1\right) \cos _{+}\left(h_{i}^{x}, p_{j}\right)}{\sum_{i=1}^{N} \sum_{j=1}^{C} I\left(l_{i j}=1\right)} \\
& +\frac{\sum_{i=1}^{N} \sum_{j=1}^{C} I\left(l_{i j}=0\right) \cos _{-}\left(h_{i}^{x}, p_{j}\right)}{\sum_{i=1}^{N} \sum_{j=1}^{C} I\left(l_{i j}=0\right)}
\end{aligned}
\end{equation}
where I is an indicator function.The denominators represent the number of relevant data-proxy pairs and irrelevant data-proxy pairs, respectively, aiming for normalization. Similarly, the proxy loss of text $\mathcal{L}_{\text {proxy }}^{y}$ can be calculated as shown in Equation (4).
\begin{equation}
\begin{aligned}
\mathcal{L}_{\text {proxy }}^{y} & =\frac{\sum_{i=1}^{N} \sum_{j=1}^{C} I\left(l_{i j}=1\right) \cos _{+}\left(h_{i}^{y}, p_{j}\right)}{\sum_{i=1}^{N} \sum_{j=1}^{C} I\left(l_{i j}=1\right)} \\
& +\frac{\sum_{i=1}^{N} \sum_{j=1}^{C} I\left(l_{i j}=0\right) \cos _{-}\left(h_{i}^{y}, p_{j}\right)}{\sum_{i=1}^{N} \sum_{j=1}^{C} I\left(l_{i j}=0\right)}
\end{aligned}
\end{equation}

The total multi-modal proxy loss $\mathcal{L}_{\text {proxy }}$ is calculated, as shown in Eq.( 5).
\begin{equation}
\begin{aligned}
\mathcal{L}_{\text {proxy }}=\mathcal{L}_{\text {proxy }}^{x}+\mathcal{L}_{\text {proxy }}^{y}
\end{aligned}
\end{equation}

\subsubsection{Pairwise Loss} \label{Pairwise Loss}

In order to maintain inter-class structural relationships, we further explore the relationships between data and data.That is, to bring similar data closer together and to push away dissimilar data.For this, we provide a similarity matrix $S$ which is defined as follows:
\begin{equation}
\begin{aligned}
S_{i j}=\frac{l_{i}}{\left\|l_{i}\right\|_{L_{2}}} \cdot\left(\frac{l_{j}}{\left\|l_{j}\right\| _{L_{2}}}\right)^{T}
\end{aligned}
\end{equation}
where $\|\cdot\|_{L_{2}}$ is the $L_{2}$ norm, $(.)^{T}$ is the transpose of vector(or matrix). The range of $S_{i j}$ is [0, 1]. If $S_{i j}$ > 0, then $x_{i}$(or $y_{i}$) and $x_{j}$(or $y_{j}$) are called an relevant pair. If $S_{i j}$ = 0, then they are considered as  irrelevant pairs. For pulling relevant pairs closer, we can calculate the cosine distance between relevant data pairs using Eq.( 7).
\begin{equation}
\begin{aligned}
\cos _{\text {pos }}\left(h_{i}, h_{j}\right) & =\left(S_{i j}-\cos \left\langle h_{i},h_{j}\right\rangle\right)_{+} \\
& =\max \left(S_{i j}-\frac{\left|h_{i} \cdot h_{j}\right|}{\left|h_{i}\right| \cdot\left|h_{j}\right|}, 0\right)
\end{aligned}
\end{equation}
Inspired by~\cite{SCH}, we relax the constraints on irrelevant data pairs. The distance between irrelevant data pairs is pushed away by calculating Eq.( 8).
\begin{equation}
\begin{aligned}
\cos _{\text {neg }}\left(h_{i}, h_{j}\right) & =(\cos \left\langle h_{i},h_{j}\right\rangle)_{+} \\
& =\max \left(\frac{\left|h_{i} \cdot h_{j}\right|}{\left|h_{i}\right| \cdot\left|h_{j}\right|}, 0\right)
\end{aligned}
\end{equation}
Hence, the relevant loss $\mathcal{L}_{p o s \_p a i r}$ could be calculated by Eq.( 9).
\begin{equation}
\begin{aligned}
\mathcal{L}_{\text {pos pair }}=\frac{\sum_{* \in\{x, y\}} \sum_{i=1, j=1}^{N} I\left(S_{i j}>0\right) \cos _{\text {pos }}\left(h_{i}^{*}, h_{j}^{*}\right)}{\sum_{i=1, j=1}^{N} I\left(S_{i j}>0\right)}
\end{aligned}
\end{equation}
Similarly, irrelevant loss $\mathcal{L}_{n e g \_p a i r}$ is calculated by Eq.( 10).
\begin{equation}
\begin{aligned}
\mathcal{L}_{\text {neg pair }}=\frac{\sum_{* \in\{x, y\}} \sum_{i=1, j=1}^{N} I\left(S_{i j}=0\right) \cos _{\text {neg }}\left(h_{i}^{*}, h_{j}^{*}\right)}{\sum_{i=1, j=1}^{N} I\left(S_{i j}=0\right)}
\end{aligned}
\end{equation}

To address the issue of intra-class dispersion arising from the pairwise loss, we assign a small weight to the pairwise loss, allowing the aggregation of similar data points to be primarily guided by the proxy loss, thereby resolving the issue of intra-class dispersion. Considering that in multi-label datasets, the number of irrelevant pairs in the pairwise loss is significantly smaller than the number of relevant pairs, and that positive sample pairs have a stronger constraint and are more likely to lead to intra-class dispersion compared to negative sample pairs, we assign different small weights to the positive and negative sample constraints of the pairwise loss.Therefore, the overall pairwise loss $\mathcal{L}_{\text {pair }}$ is given by the following formula:
\begin{equation}
\begin{aligned}
\mathcal{L}_{\text {pair }}=\alpha \mathcal{L}_{\text {pos-pair }}+\beta \mathcal{L}_{\text {neg-pair }}
\end{aligned}
\end{equation}
where $\alpha$ and $\beta$ are the small weight hyper-parameters for positive and negative sample constraints, respectively.

\begin{figure}
    \centering
    \includegraphics[width=1.0\linewidth]{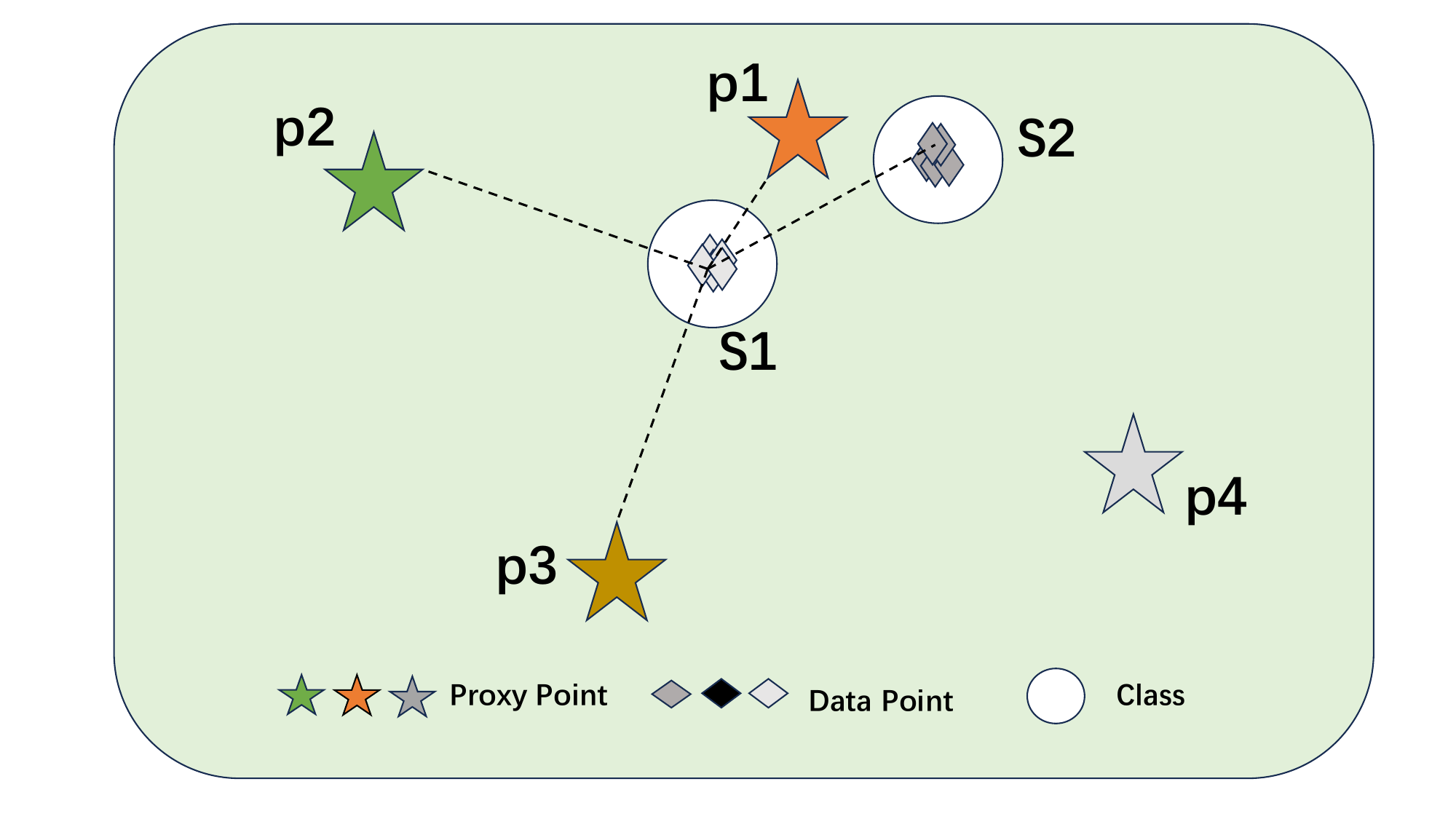}
    \caption{\small Class S1 is related to proxy points P1, P2, and P3, while class S2 is related to proxy points P1 and P4. When considering both point and class proxies, as well as the relationships between points, if there are more data points related to P1 than to other proxies, the data points will be biased towards the P1 proxy. However, the data points should maintain a consistent distance relationship with each of their related proxies.}
    \label{fig:problem}
\end{figure}

\subsubsection{Variance Constraint} \label{Variance Constraint}

When considering both the relationships between points and proxies, and between points and points, if there are more data points related to proxy P1 than to other proxies, the data points will tend to lean towards proxy P1. However, data points should maintain a consistent relationship with each of their relevant proxies. As shown in Fig.~\ref{fig:problem}. Therefore, we use variance constraints to maintain a consistent distance relationship between the data and its relevant proxy points.
For an image data $x_{i}$, $l_{i}$ represents its label, and we use $M_{i}$ to denote the index set of its corresponding relevant proxies, where $M_{i} = \{ j \mid j \in \{1, 2, \ldots, C\} \text{ and } l_{ij} = 1 \}$.Then, the variance constraint for it can be given by the following formula:
\begin{equation}
\operatorname{Var}\left(x_{i}\right) = \frac{1}{U_{i}} \sum_{j \in M_{i}} \left(\cos_{+}\left(h_{i}^{x}, p_{j}\right) - \frac{\sum_{j \in M_{i}} \cos_{+}\left(h_{i}^{x}, p_{j}\right)}{U_{i}}\right)^{2}
\end{equation}
where $U_{i}$ represents the number of elements in the set $M_{i}$. Hence, the overall constraint of the image $\mathcal{L}_{v a r}^{x}$ could be calculated by Eq.( 13).
\begin{equation}
\begin{aligned}
\mathcal{L}_{v a r}^{x} = \frac{1}{N} \sum_{i=1}^{N} \operatorname{Var}\left(x_{i}\right)
\end{aligned}
\end{equation}
Similarly, the overall constraint of the text $\mathcal{L}_{v a r}^{y}$ could be calculated by Eq.( 14).
\begin{equation}
\begin{aligned}
\mathcal{L}_{v a r}^{y} = \frac{1}{N} \sum_{i=1}^{N} \operatorname{Var}\left(y_{i}\right)
\end{aligned}
\end{equation}

The total variance constraint loss $\mathcal{L}_{v a r}$ is calculated, as shown in Eq.( 15).
\begin{equation}
\begin{aligned}
\mathcal{L}_{v a r}=\mathcal{L}_{v a r}^{x}+\mathcal{L}_{v a r}^{y}
\end{aligned}
\end{equation}

\subsection{Objective Function} \label{subsec:Objective Function}

By jointly optimizing Equations (5), (11), and (15), the overall objective function for learning the hash projection is calculated, as follows:
\begin{equation}
\begin{aligned}
\mathcal{L}_{total}=\mathcal{L}_{proxy}+\mathcal{L}_{pair}+\mathcal{L}_{v a r}
\end{aligned}
\end{equation}
By adopting Eq.( 16) to calculate the overall loss, Our DCGH method can effectively learn hash codes that maintain intra-class aggregation while preserving inter-class structural relationships, and prevent semantic bias.

\subsection{Optimization} \label{subsec:optimization}

The training algorithm is a critical component ofour proposed DCGH framework and is presented
in Algorithm 1. The DCGH model is optimized by standard backpropagation algorithms and mini-batch gradient descent methods. The algorithm for generating the hash code is shown in
Algorithm 2. For a query sample, use the well-trained DCGH model to generate binary-like hash codes, and then use the $\operatorname{sign}$ function to generate the final binary hash codes.
\begin{equation}
\begin{aligned}
\operatorname{sign}(x)=\left\{\begin{array}{l}
+1, x>0 \\
-1, x<0
\end{array}\right.
\end{aligned}
\end{equation}

Specifically, for given the query data $x_{i}$($y_{i}$), the compact hash code could be generated by Eq.( 18).
\begin{equation}
\begin{aligned}
\begin{array}{l}
b_{i}^{x}=\operatorname{sign}\left(H^{x}\left(x_{i}\right)\right) \\\\
b_{i}^{y}=\operatorname{sign}\left(H^{y}\left(y_{i}\right)\right)
\end{array}
\end{aligned}
\end{equation}

\begin{algorithm}
  \caption{Learning algorithm for DCGH.} 
  \renewcommand{\algorithmicrequire}{\textbf{Input:}}
  \renewcommand{\algorithmicensure}{\textbf{Output:}}
  \begin{algorithmic}[1]
    \REQUIRE 
    Training dataset $\bm{D}$; Binary codes length $K$; Hyper-parameters: $\alpha$, $\beta$.
    \ENSURE 
    Network Parameters: $\theta_{x}$, $\theta_{y}$.
    
    \STATE Initialize network parameters $\theta_{x}$ and $\theta_{y}$, maximum iteration number $epoch$, mini-batch size 128, proxies $P=\left\{p_{1}, p_{2}, \ldots \ldots p_{C}\right\}$.
    \STATE Construct a similarity matrix $\bm{S}$ from multi-label set $\left \{\bm{l}_i \right \}^{N}_{i=1}$;
    \WHILE{$iter < epoch$}
    \STATE Capture feature vector $F^{x}$ and $F^{y}$ by forward propagation.
    \STATE Compute proxy loss $\mathcal{L}_{\text {proxy }}$ by Eq.( 5).
    \STATE Compute pairwise loss $\mathcal{L}_{\text {pair }}$ by Eq.( 11).
    \STATE Compute Variance Constraint $\mathcal{L}_{\text {var }}$ by Eq.( 15).
    \STATE Update proxies $P=\left\{p_{1}, p_{2}, \ldots \ldots p_{C}\right\}$ by backpropagations.
    \STATE Update the network parameters $\theta_{x}$ and $\theta_{y}$ by backpropagations.
    \ENDWHILE
    \RETURN The trained DCGH model.
  \end{algorithmic} \label{Alg:optimization}
\end{algorithm}

\begin{algorithm}
  \caption{Learning hash codes for DCGH.} 
  \renewcommand{\algorithmicrequire}{\textbf{Input:}}
  \renewcommand{\algorithmicensure}{\textbf{Output:}}
  \begin{algorithmic}[2]
    \REQUIRE 
    Query samples $q_{i}$, Parameters for DCGH.
    \ENSURE 
    Binary hash code for $q_{i}$.
    \STATE Calculate binary-like hash codes by feeding the query data $q_{i}$ into the trained DCGH model.
    \STATE Calculating hash codes by using the $\operatorname{sign}$ function.
  \end{algorithmic} \label{Alg:optimization}
\end{algorithm}

\section{Experiments} \label{sec:experiments}
In this section, we will present and analyze the experimental results of the proposed method alongside those of several state-of-the-art competitors. Firstly, the details of experimental setting are introduced in~\ref{subsec:experimental-setting}. We then proceed to discuss performance comparison, ablation studies, sensitivity to parameters, training and encoding time and visualization in~\ref{subsec:performance-conparison},~\ref{subsec:ablation-stuides},~\ref{subsec:sensitivity-of-parameters},~\ref{subsec:Training and Encoding time} and~\ref{subsec:visualization}, respectively.

\subsection{Experimental Setting} \label{subsec:experimental-setting}

\textbf{Datasets.} Our experiments were conducted on three common used benchmark datasets, i.e., MIRFLICKR-25K~\cite{MIRFLICKR}, NUS-WIDE~\cite{NUSWIDE} and MS COCO~\cite{MSCOCO}. The brief introduction of them are listed here:
\begin{itemize}
  \item \textbf{MIRFLICKR-25K} is a small-scale cross-modal multi-label dataset collected from the Flickr website. It includes 24581 image-text pairs corresponding to 24 classes, in which each sample pair belongs to at least one category.
  \item \textbf{NUS-WIDE}  dataset contains 269,648 image-text pairs, where each of them belongs to at least one of the 81 categories. To enhance the dataset's practicality and compatibility with other research methods, we conducted a selection process, removing categories with fewer samples and choosing 21 common categories. This subset contains 195,834 image-text pairs, with each pair belonging to at least one of the categories.
  \item \textbf{MS COCO} dataset is a highly popular large-scale dataset in computer vision research. Comprising of 82,785 training images and 40,504 validation images, each image is accompanied by corresponding textual descriptions and labels. It covers 80 different categories. In our study, the training and validation sets are combined, with each sample containing both image and text modalities, and each sample belonging to at least one category.
\end{itemize}

\begin{figure*}
  \centering
  \includegraphics[width=1.0\linewidth]{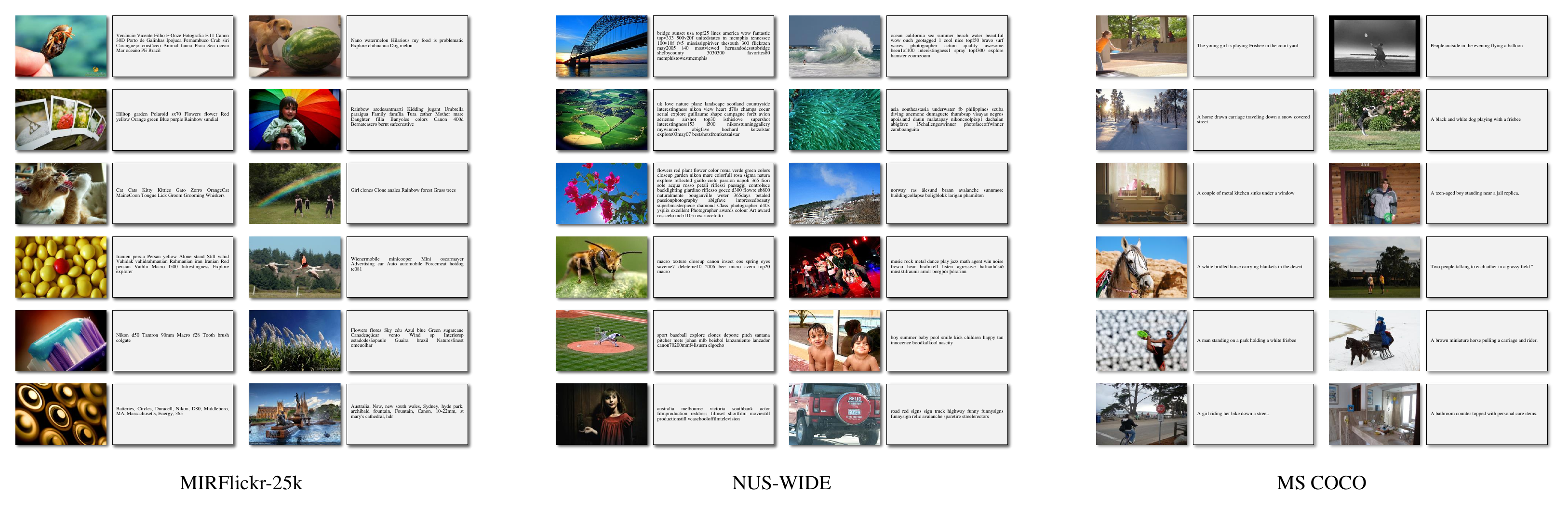}
  \vspace{-2mm}
  \caption{\small  Some examples of image-text paris in MIRFLICKR-25K, NUS-WIDE and MS COCO.}
  \label{fig:samples-of-datasets}
\end{figure*}

The statistics of these three datasets are reported in Table~\ref{tab:statistics-of-datasets}, and Fig.~\ref{fig:samples-of-datasets} displays some examples of image-text pairs of them.

\textbf{Implementation Details.}  In our experiments, we implemented a unified sampling strategy across the three datasets. Initially, we randomly selected 5,000 image-text pairs from dataset as the query set, with the remainder serving as the database set. During the model training phase, we randomly chose 10,000 image-text pairs as the training set. To ensure consistency and fairness in the experiments, we performed the same preprocessing operations on the images and text for all datasets: the image sizes were adjusted to 224×224, and the text was represented through BPE~\cite{BPE} encoding.

To ease reading, we reports the the detailed configuration of each components in the proposed DCGH framework by Table~\ref{tab:network_configuration}.

\begin{table}[]
	\caption{Configuration of the proposed architecture.}\label{tab:network_configuration}
	\begin{tabular}{p{1.65cm}lp{0.65cm}c}
		\toprule 
		\textbf{Stage}        & \textbf{Module}         & \textbf{Layer}          & \textbf{Configuration }      \\ 
		\midrule
		
		\multirow{1}{*}{Stage 1}  & Feature Extraction     & CLIP & 512             \\ 
        \midrule
		\multirow{7}{*}{Stage 2} & \multirow{3}{*}{Hash Function (Img)}     & ${fc}$   & $K$  \\
	    &                                    & $Droup$          & $0.2$                \\ 
		&                                    & $Tanh$           &                    \\ \cmidrule(lr){2-4} 
		& \multirow{3}{*}{Hash Function (Txt)} & $fc$   & $K$               \\ 
		&                                    & $Droup$          & $0.2$                \\ 
        &                                    & $Tanh$           &                    \\ 
		\bottomrule
	\end{tabular}
\end{table}

\textbf{Experimental Environment.} 
We implemented our DCGH via Pytorch, with GPU leveraging NVIDIA RTX 3090. For network configuration, DCGH's two Transformer encoders, ViT~\cite{VIT} and GPT-2~\cite{GPT}, are initialized with pre-trained CLIP (ViT-B/32)~\cite{CLIP}. Our model employs an adaptive moment estimation (Adam) optimizer to update the network parameters until it converges~\cite{ADAM}. The learning rate of two backbone transformer encoders ViT and GPT-2 was set as 0.00001, while the learning rate of the hash learning module in DCGH is set to 0.001. The two hyper-parameters, $\alpha$ and $\beta$, are set to 0.05 and 0.8, respectively. The batch size is 128.

\begin{table*}
  \caption{Statistics of MIRFLICKR-25K, NUS-WIDE and MS COCO.}\label{tbl1}
  \begin{tabular*}{\tblwidth}{@{} LLLLLLL@{} }
  \toprule
  Dataset & Total & Training & Query & Retrieval & Categories & CLIP Feature\\
  \midrule
  MIRFLICKR-25K & 24,581 & 10,000 & 5,000 & 19,581 & 24 & 512\\
  NUS-WIDE & 195,834 & 10,000 & 5,000 & 190,834 & 21 & 512\\
  MS COCO & 123,289 & 10,000 & 5,000 & 118,289 & 80 & 512\\
  \bottomrule
  \end{tabular*} \label{tab:statistics-of-datasets}
  \end{table*}

\textbf{Baseline Methods.} In our experiments, we selected 14 state-of-the-art deep
cross-modal hashing methods for comparison, which contain Deep Cross-Modal Hashing~\cite{DCMH}, Self-Supervised Adversarial Hashing~\cite{SSAH}, Cross-Modal Hamming
Hash~\cite{CMHH}, Adversary Guided Asymmetric Hashing~\cite{AGAH}, Deep Adversarial Discrete Hashing~\cite{DADH}, Self-Constraining Attention Hashing Network~\cite{SCAHN}, Multi-label Enhancement Self-supervised Deep Cross-modal Hashing~\cite{MESDCH}, Differentiable Cross-modal Hashing via Multi-modal Transformers~\cite{DCHMT}, Modality-Invariant Asymmetric Networks~\cite{MIAN}, Data-Aware Proxy Hashing~\cite{DAPH}, Deep Semantic-aware Proxy Hashing~\cite{DSPH}, Deep Neighborhood-aware Proxy Hashing~\cite{DNPH}, Deep Hierarchy-aware Proxy Hashing~\cite{DHaPH}. Semantic Channel Hash~\cite{SCH}. Due to some methods are not open source, we directly cite the results from the published papers. Here is a brief introduction to each baseline:
\begin{itemize}
  \item \textbf{DCMH} approach integrates feature learning and hash code learning into an
end-to-end framework, which uses a negative log-likelihood loss function to maintain
the similarity of the original data.
  \item \textbf{SSAH} combines self-supervised and adversarial learning to bridge
the heterogeneity gap better.
  \item \textbf{CMHH} is to penalize significantly on similar cross-modal pairs with Hamming distance larger than the Hamming radius threshold, by designing a pairwise focal loss based on the exponential distribution.
  \item \textbf{AGAH} employs an adversarial learning guided multi-label attention module to enhance the feature learning part and adopt a triplet-margin constraint and a cosine quantization technique for Hamming space similarity preservation.
  \item \textbf{DADH} uses adversarial training to learn features across modalities and utilizes a discrete hashing strategy to directly learn the discrete binary codes without any relaxation.
  \item \textbf{SCAHN} integrates the label constraints from early and late-stages as well as their fused features into the hash representation and hash function learning.
  \item \textbf{MESDCH}  first propose a multi-label semantic affinity preserving module, which uses ReLU transformation to unify the similarities of learned hash representations and the corresponding multi-label semantic affinity of original instances and defines a positive-constraint Kullback–Leibler loss function to preserve their similarity.
  \item \textbf{DCHMT} propose a differentiable cross-modal hashing method that utilizes the multi-modal transformer as the backbone to capture the location information in an image when encoding the visual content.
  \item \textbf{MIAN} propose a novel Modality-Invariant Asymmetric Networks architecture, which explores the asymmetric intra- and inter-modal similarity preservation under a probabilistic modality alignment framework.
  \item \textbf{DAPH} train a data-aware proxy network to generate class-based data-aware proxy hash codes, and then train the hash network using the proxy hash codes.
  \item \textbf{DSPH} further explore the semantic relationships of unrelated samples based on the proxy loss.
  \item \textbf{DNPH} based on the proxy loss, to enhance the quality of the obtained binary codes, a uniform distribution constraint is developed to ensure that each hash bit independently follows a discrete uniform distribution.
  \item \textbf{DHaPH} further considers the potential semantic hierarchical relationships between multi-modal samples, introduces learnable shared hierarchical proxies in hyperbolic space, and proposes a hierarchical perception proxy loss algorithm.
  \item \textbf{SCH} takes into account the utilization rate of the Hamming space and sets upper and lower bounds on the similarity between samples to prevent the loss function from oscillating.
\end{itemize}

\begin{table*}
  \scriptsize
  \centering
  \caption{Comparison with baselines in terms of mAP w.r.t. 16bits, 32bits, 64bits on MIRFLICKR-25K, NUS-WIDE, and MS COCO. The best results are in bold font.}\label{tab:hamming-ranking}
  \begin{tabular}{*{12}{c}}
    \toprule
    \multirow{2}*{Task} & \multirow{2}*{Methods} & \multicolumn{3}{c}{MIRFLICKR-25K} & \multicolumn{3}{c}{NUS-WIDE} & \multicolumn{3}{c}{MS COCO} &\\
    \cmidrule(lr){3-5}\cmidrule(lr){6-8}\cmidrule(lr){9-12}
    & & 16 bits & 32 bits & 64 bits & 16 bits & 32 bits & 64 bits & 16 bits & 32 bits & 64 bits \\
    \midrule
    \multirow{16}*{Img2Txt} 
    &DCMH (CVPR17)      & 0.7687 & 0.7736 & 0.7797 & 0.5379 & 0.5513 & 0.5617 & 0.5399 & 0.5444 & 0.5627\\
    &SSAH (CVPR18)  & 0.8079 & 0.8129 & 0.8220 & 0.6032 & 0.6058 & 0.6095 & 0.5411 & 0.4855 & 0.5395\\
    &CMHH (ECCV18)      & 0.6932 & 0.6979 & 0.6984 & 0.5439 & 0.5546 & 0.5520 & 0.5145 & 0.4509 & 0.5209\\
    &AGAH (ICMR19)     & 0.7248 & 0.7217 & 0.7195 & 0.3945 & 0.4107 & 0.4258 & 0.5501 & 0.5515 & 0.5518\\
    &DADH (ICMR20)     & 0.8098 & 0.8162 & 0.8193 & 0.6350 & 0.6568 & 0.6546 & 0.5952 & 0.6118 & 0.6237\\
    &SCAHN (Neurocomputing20) & 0.7955 & 0.8248 & 0.8297 & 0.6463 & 0.6616 & 0.6645 & 0.6376 & 0.6475 & 0.6519\\
    &MESDCH (Neurocomputing22)  & 0.7258 & 0.7438 & 0.7421 & 0.5638 & 0.5801 & 0.5875 & 0.5759 & 0.5670 & 0.5636\\
    &DCHMT (ICM22)  & 0.8201 & 0.8253 & 0.8222 & 0.6596 & 0.6706 & 0.6863 & 0.6309 & 0.6216 & 0.6553\\
    &MIAN (TKDE22)  & 0.8262 & 0.8444& 0.8501 & 0.6431 & 0.6548& 0.6625 & 0.5952 & 0.5987 & 0.6121\\
    &DAPH (SIGIR23)    & 0.7939 & 0.8071 & 0.8168 & 0.6642 & 0.6835 & 0.6925 & 0.6344 & 0.7030 & 0.7300\\
    &DSPH (TCSVT23) & 0.8129 & 0.8435 & 0.8541& 0.6830 & 0.6934 & 0.7087 & 0.6985 & 0.7466 & 0.7635\\
    &DNPH (TOMM24)    & 0.8108 & 0.8269 & 0.8289 & 0.6689 & 0.6811 & 0.6939 & 0.6438 & 0.6910 & 0.7294\\
    &DHaPH (TKDE24)   & 0.8299 & 0.8437 & 0.8531 & 0.6678 & 0.6799 & 0.6901 & \textbf{0.7284} & 0.7415 & 0.7475\\
    &SCH (AAAI24)   & \textbf{0.8363} & \textbf{0.8507} & 0.8602 & 0.6436 & 0.6677 & 0.6823 & 0.5792 & 0.6642 & 0.7217\\
    &DCGH (Ours)    & 0.8245 & 0.8482 & \textbf{0.8633} & \textbf{0.6869}
    & \textbf{0.7038} & \textbf{0.7214} & 0.7120 & \textbf{0.7492} & \textbf{0.7737}\\
    \midrule
    \multirow{16}*{Txt2Img}  
    &DCMH (CVPR17)      & 0.7857 & 0.7998 & 0.8029 & 0.5747 & 0.5810 & 0.5853 & 0.5271 & 0.5424 & 0.5450\\
    &SSAH (CVPR18)  & 0.8089 & 0.8127 & 0.8017 & 0.6011 & 0.6058 & 0.6167 & 0.4901 & 0.4798 & 0.5053\\
    &CMHH (ECCV18)     & 0.7181 & 0.7104& 0.7294 & 0.4956& 0.4831& 0.4820 & 0.4910 & 0.4930 & 0.4889\\
    &AGAH (ICMR19)     & 0.7082 & 0.7182 & 0.7344 & 0.4344 & 0.3980 & 0.4382 & 0.5012 & 0.5146 & 0.5191\\
    &DADH (ICMR20)     & 0.8019 & 0.8101 & 0.8137 & 0.6111 & 0.6182 & 0.6218 & 0.5649 & 0.5790 & 0.5870\\
    &SCAHN (Neurocomputing20)  & 0.7826 & 0.8066 & 0.8064 & 0.6587 & 0.6626 & 0.6648 & 0.6377 & 0.6512 & 0.6493\\
    &MESDCH (Neurocomputing22)  & 0.7385 & 0.7387 & 0.7437 & 0.5562 & 0.5720 & 0.5803 & 0.5044 & 0.5083 & 0.5085\\
    &DCHMT (ICM22)  & 0.7983 & 0.8048 & 0.8031 & 0.6761 & 0.6837 & 0.6943 & 0.6241 & 0.6212 & 0.6486\\
    &MIAN (TKDE22)  & 0.8139 & 0.8180 & 0.8216 & 0.6733 & 0.6922 & 0.6936 & 0.5947 & 0.6000 & 0.6067\\
    &DAPH (SIGIR23)    & 0.7949 & 0.8134 & 0.8212 & 0.6611 & 0.6813 & 0.7016 & 0.6307 & 0.6870 & 0.7180\\
    &DSPH (TCSVT23) & 0.8000 & 0.8216 & 0.8294 & 0.6997 & 0.7116 & 0.7326 & 0.7002 & 0.7483 & 0.7696\\
    &DNPH (TOMM24)   & 0.8015 & 0.8176 & 0.8166 & 0.6871 & 0.6994 & 0.7182 & 0.6468 & 0.7012 & 0.7388\\
    &DHaPH (TKDE24)   & 0.8148 & 0.8165 & 0.8229 & 0.6725 & 0.6951 & 0.7055 & 0.6935 & 0.7069 & 0.7154\\
    &SCH (AAAI24)   & \textbf{0.8251} &\textbf{0.8376} & \textbf{0.8463} & 0.6722 & 0.6880 & 0.7087 & 0.5758 & 0.6601 & 0.7265\\
    &DCGH (Ours)    & 0.8108 & 0.8297 & 0.8413 & \textbf{0.7014}
    & \textbf{0.7153} & \textbf{0.7329} & \textbf{0.7072} & \textbf{0.7536} & \textbf{0.7786}\\
    \bottomrule
  \end{tabular}
\end{table*}

\begin{figure*}
    \centering
    \begin{subfigure}[b]{0.24\textwidth}
        \includegraphics[width=\textwidth]{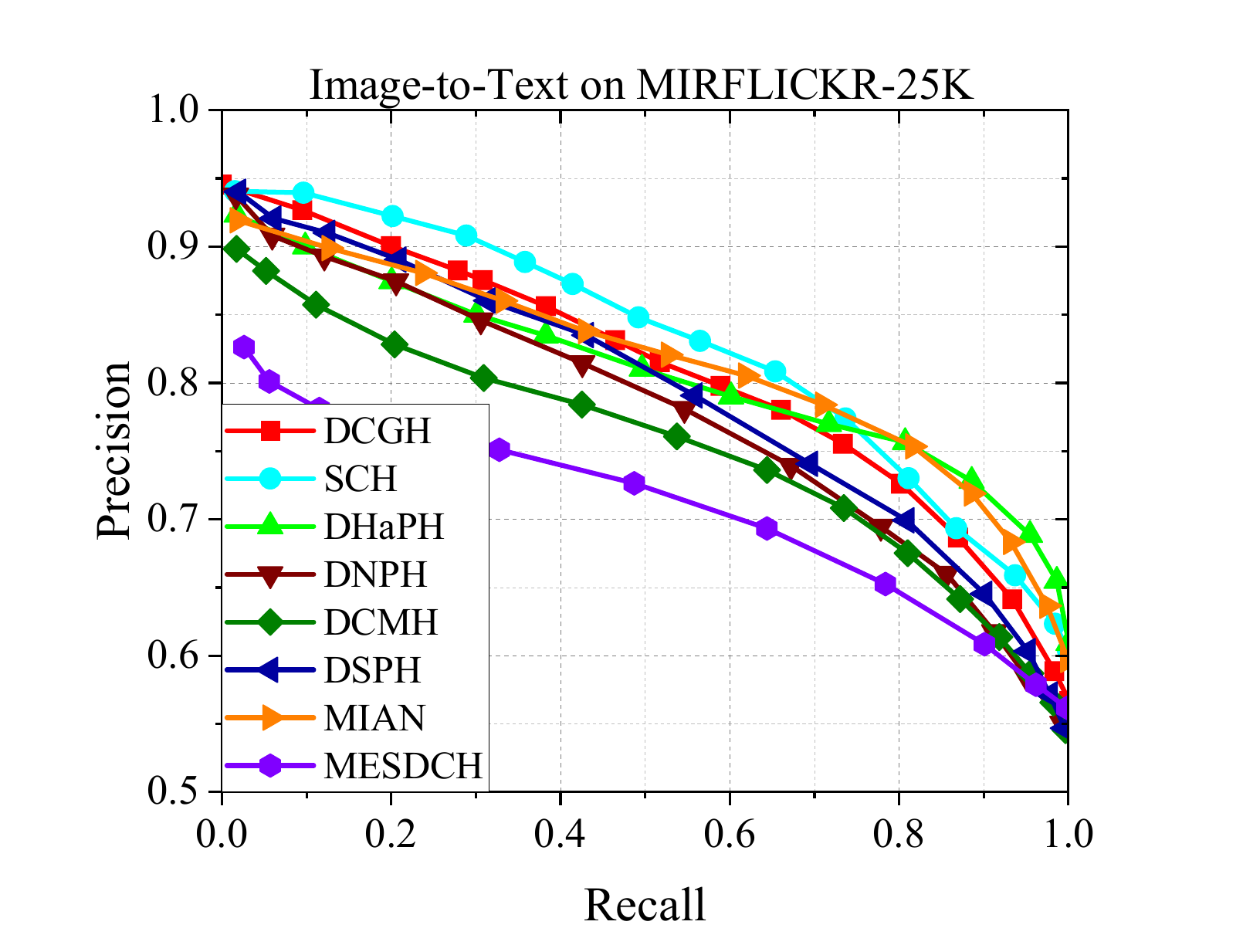}
        \caption{PR@16bits}
        \label{fig:16bitsPR-MIRI-T}
    \end{subfigure}
    \hfill
    \begin{subfigure}[b]{0.24\textwidth}
        \includegraphics[width=\textwidth]{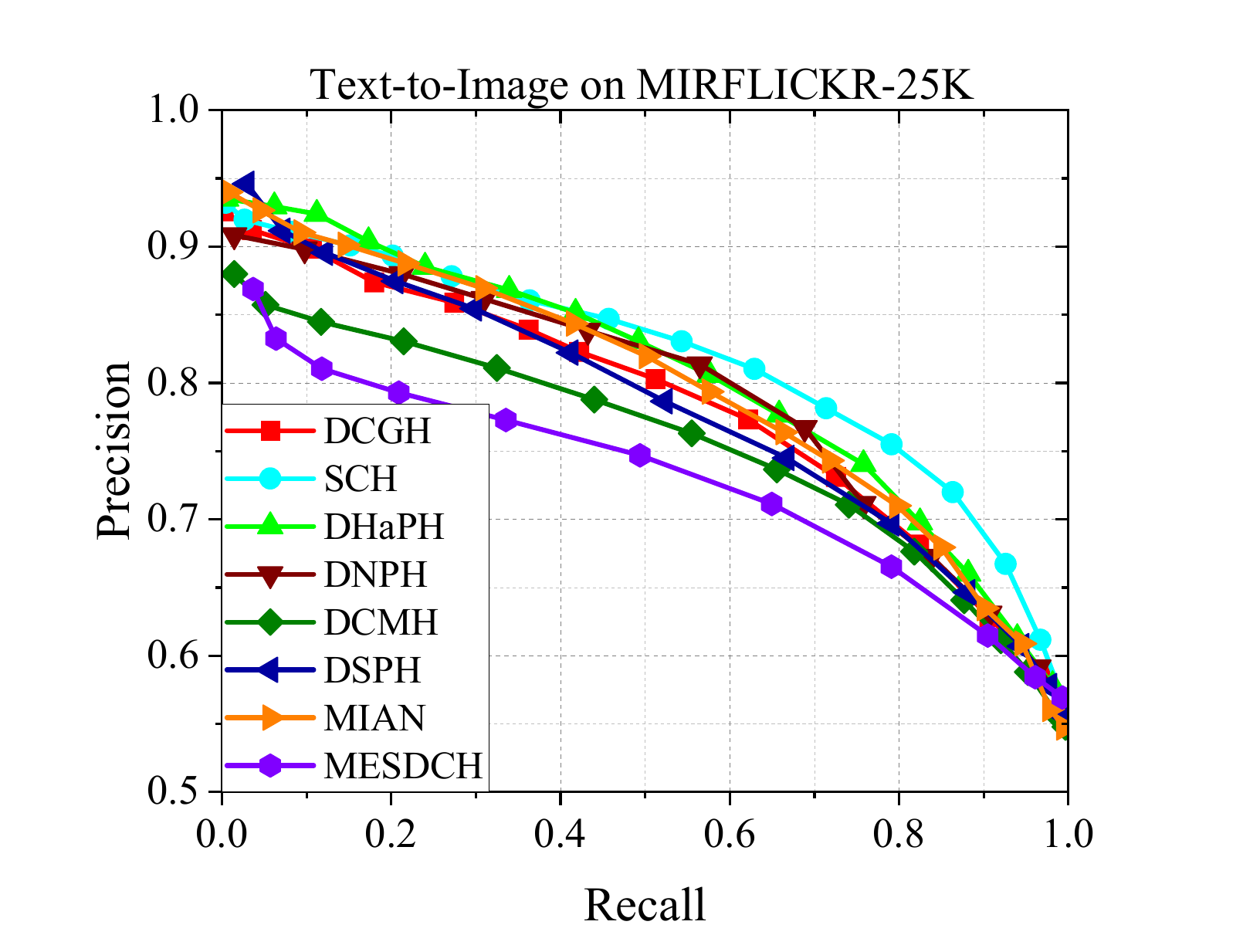}
        \caption{PR@16bits}
        \label{fig:16bitsPR-MIRT-I}
    \end{subfigure}
    \hfill
    \begin{subfigure}[b]{0.24\textwidth}
        \includegraphics[width=\textwidth]{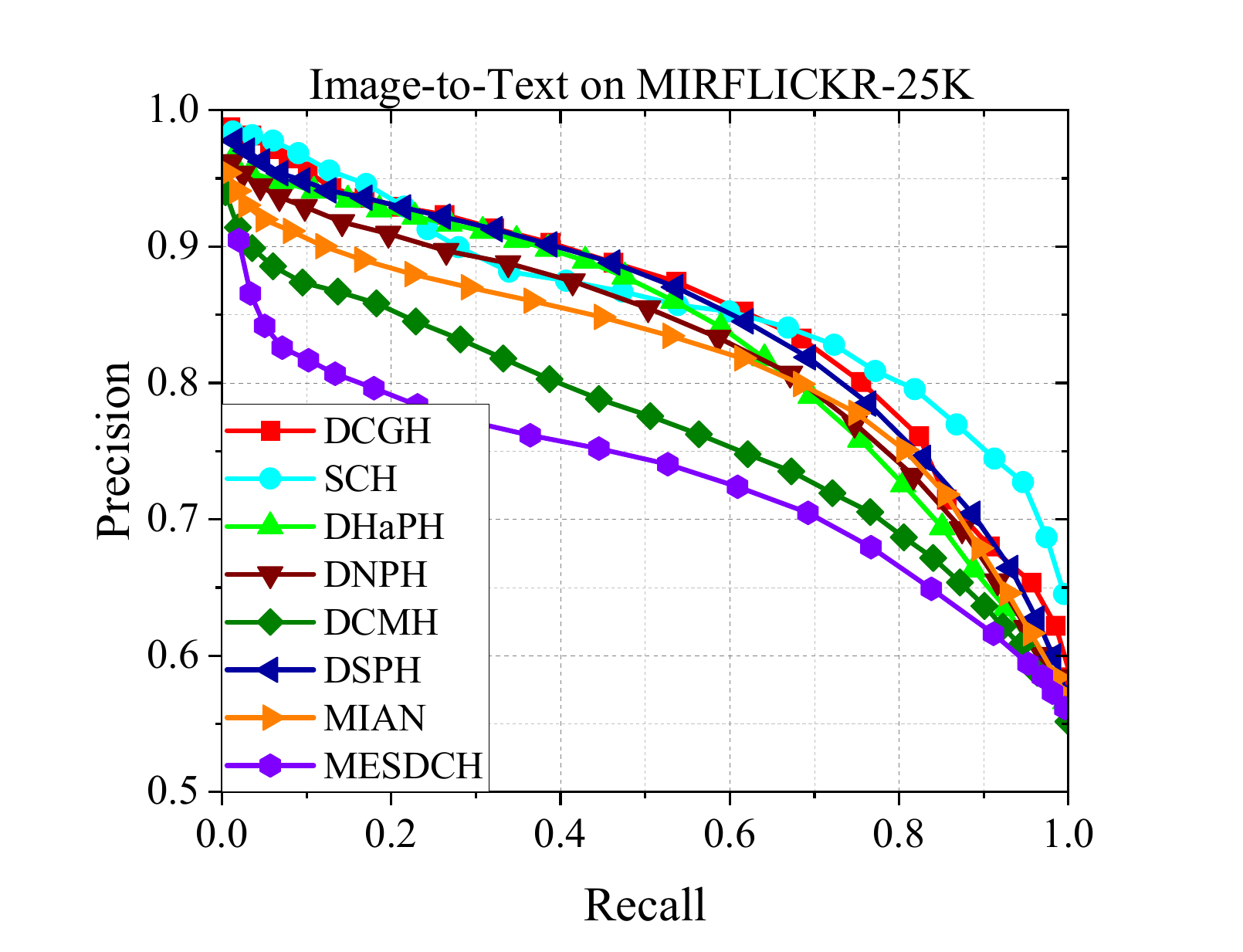}
        \caption{PR@32bits}
        \label{fig:PRMIR32BitsI→T}
    \end{subfigure}
    \hfill
    \begin{subfigure}[b]{0.24\textwidth}
        \includegraphics[width=\textwidth]{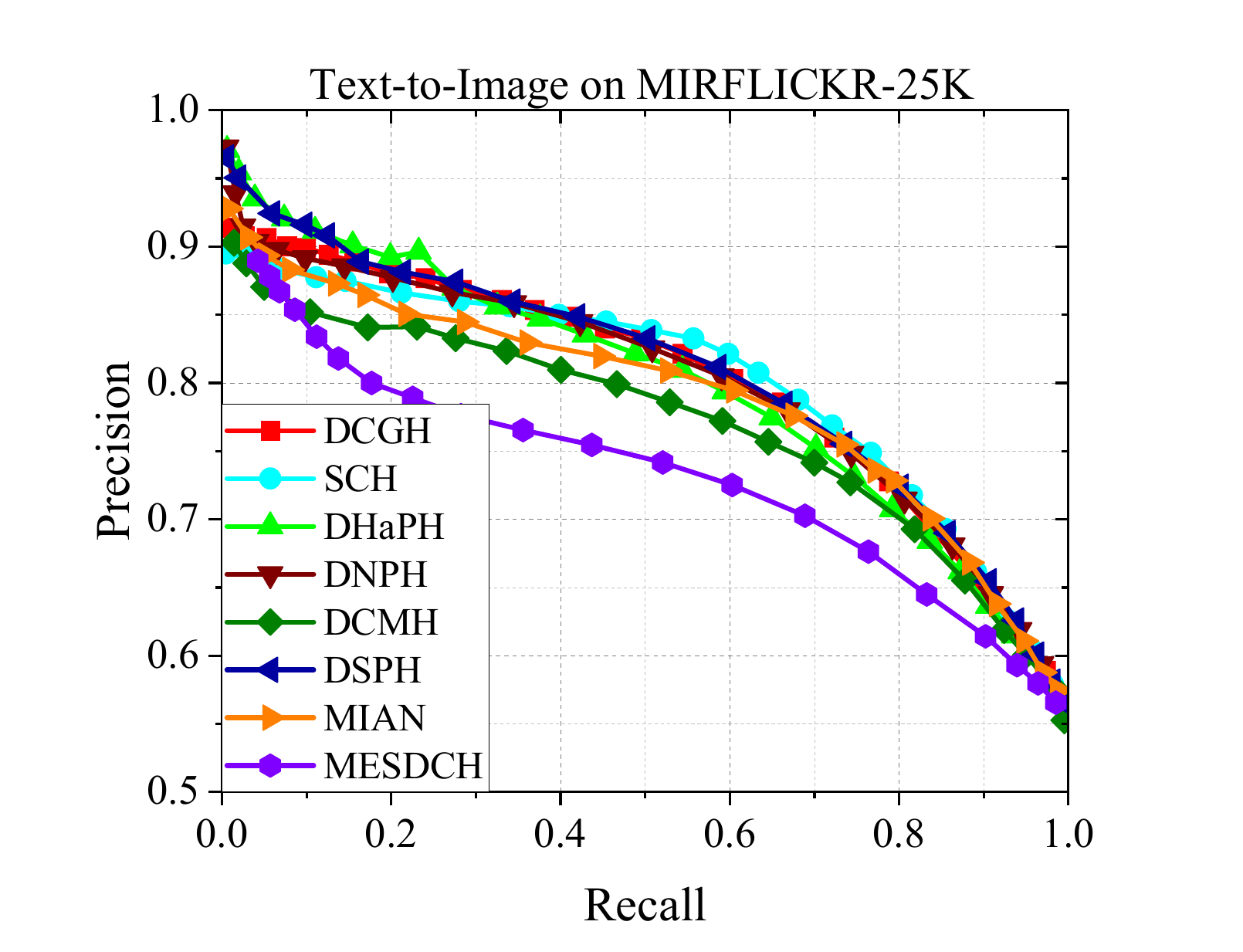}
        \caption{PR@32bits}
        \label{fig:PRMIR32BitsT→I}
    \end{subfigure}
    \caption{\small Results of Precision-Recall curves on MIRFLICKR-25K w.r.t.16bits and 32bits.}
    \label{fig:mir-flickr25k}
\end{figure*}

\begin{figure*}
    \centering
    \begin{subfigure}[b]{0.24\textwidth}
        \includegraphics[width=\textwidth]{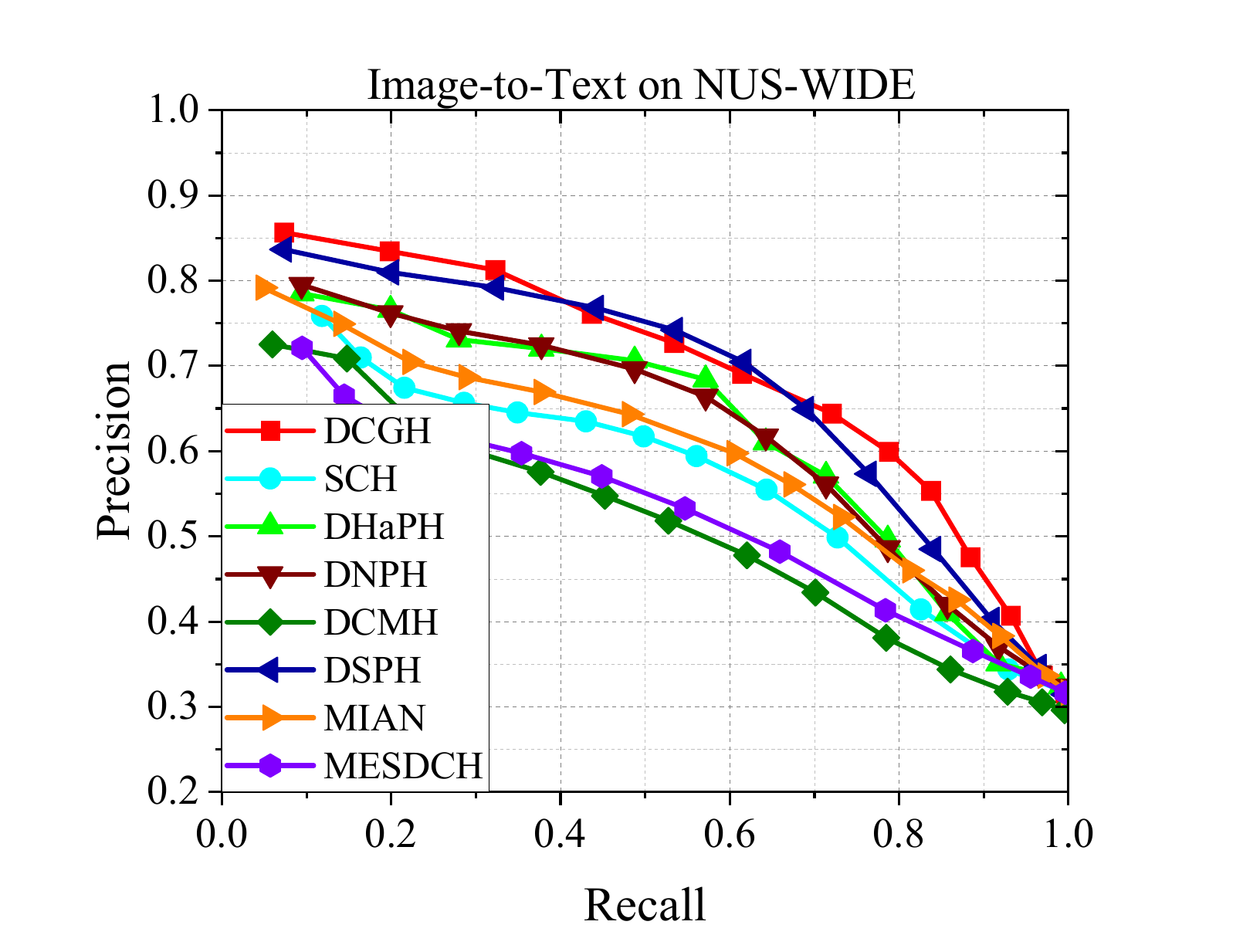}
        \caption{PR@16bits}
        \label{fig:16bitsPR-NUSI-T}
    \end{subfigure}
    \hfill
    \begin{subfigure}[b]{0.24\textwidth}
        \includegraphics[width=\textwidth]{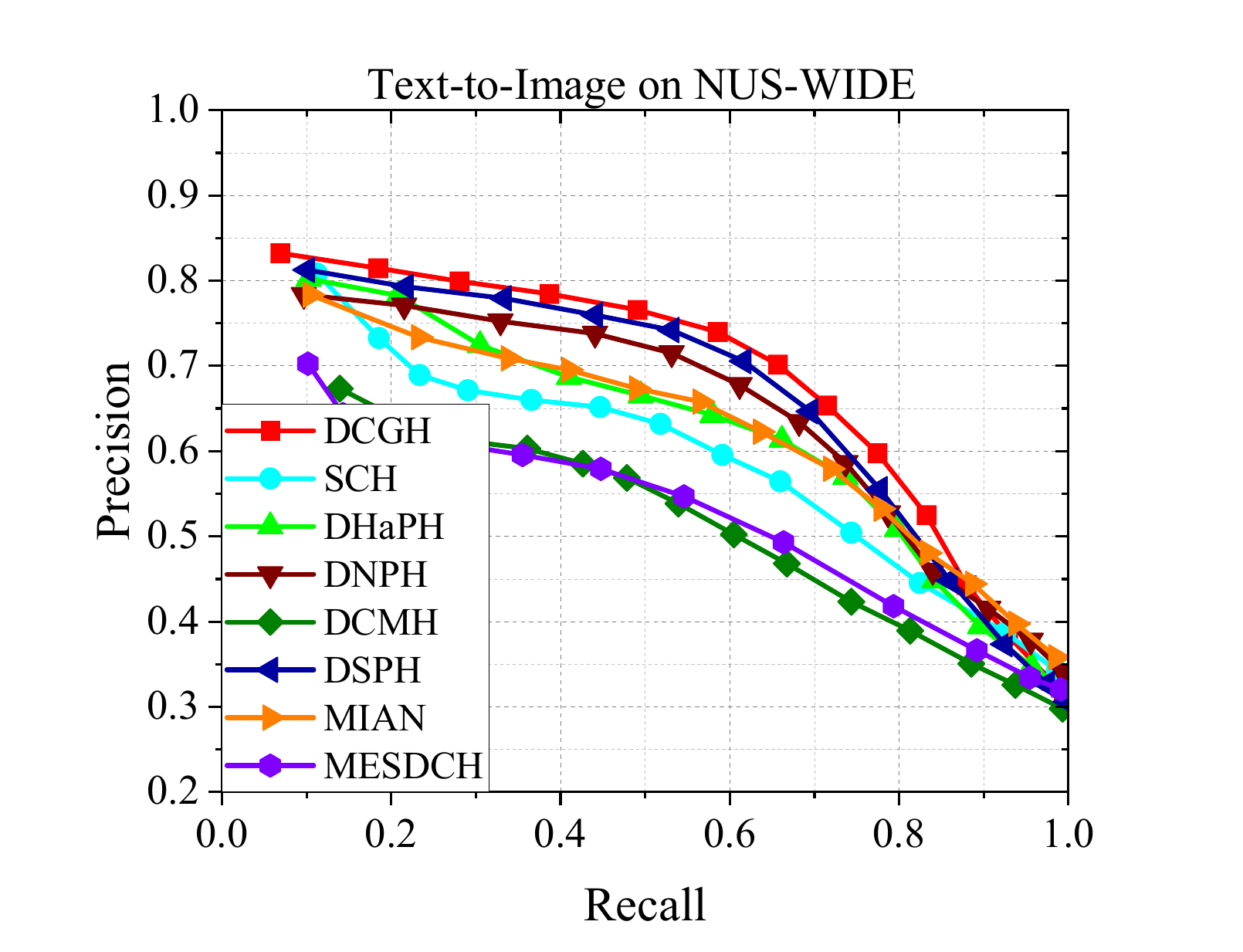}
        \caption{PR@16bits}
        \label{fig:16bitsPR-NUST-I}
    \end{subfigure}
    \hfill
    \begin{subfigure}[b]{0.24\textwidth}
        \includegraphics[width=\textwidth]{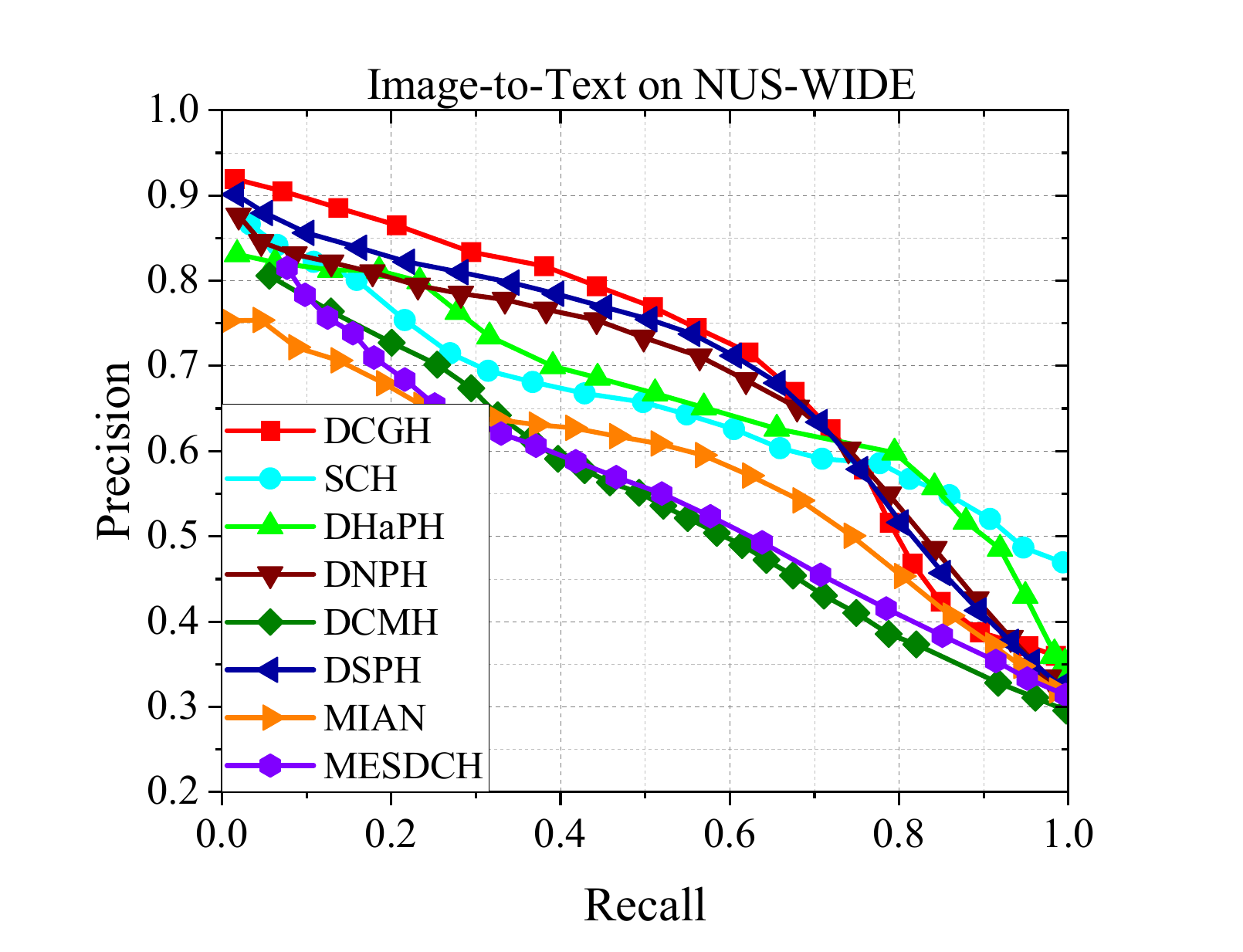}
        \caption{PR@32bits}
        \label{fig:PRNUS32BitsI→T}
    \end{subfigure}
    \hfill
    \begin{subfigure}[b]{0.24\textwidth}
        \includegraphics[width=\textwidth]{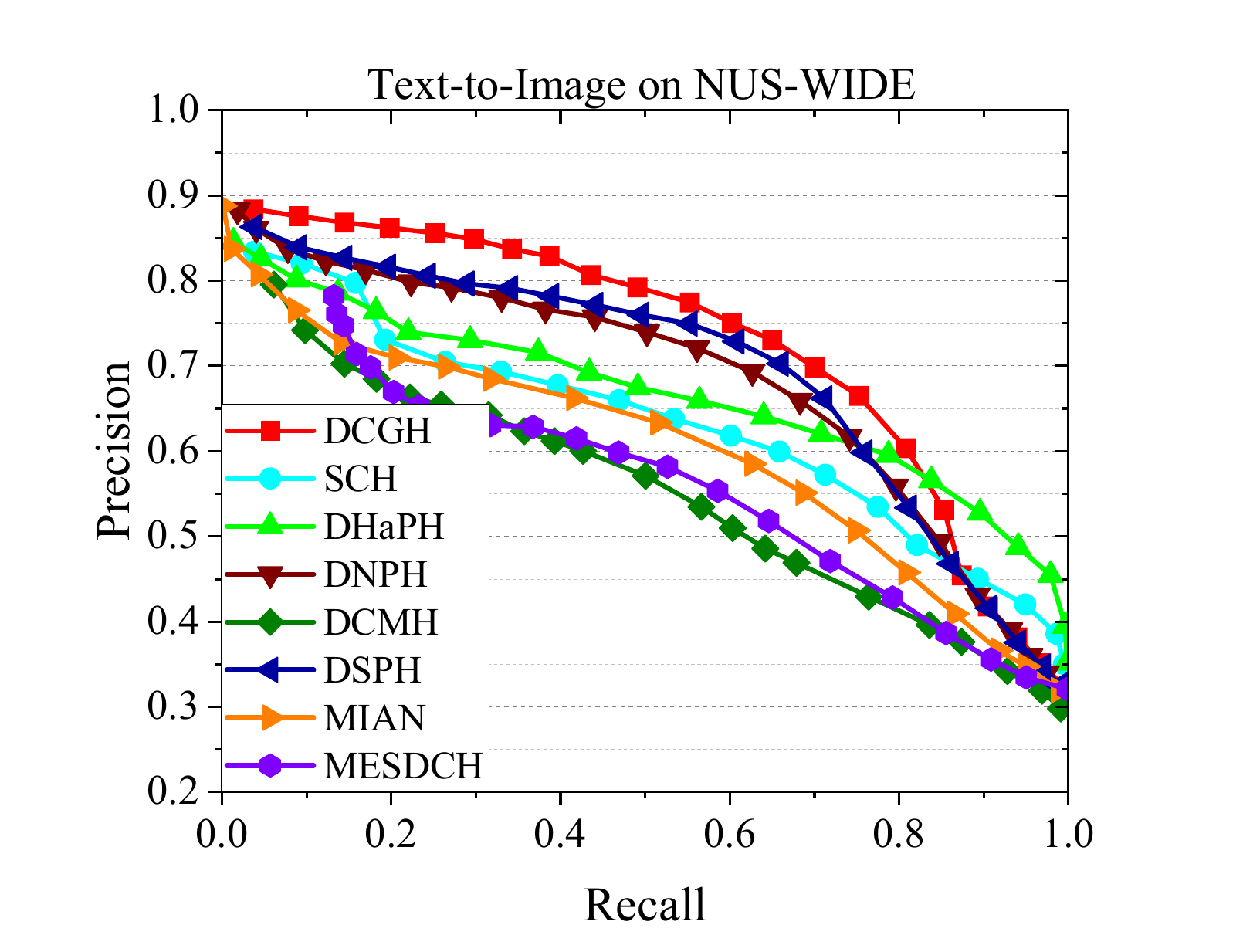}
        \caption{PR@32bits}
        \label{fig:PRNUS32BitsT→I}
    \end{subfigure}
    \caption{\small Results of Precision-Recall curves on NUS-WIDE w.r.t.16bits and 32bits.}
    \label{fig:nus-wide}
\end{figure*}

\begin{figure*}
    \centering
    \begin{subfigure}[b]{0.24\textwidth}
        \includegraphics[width=\textwidth]{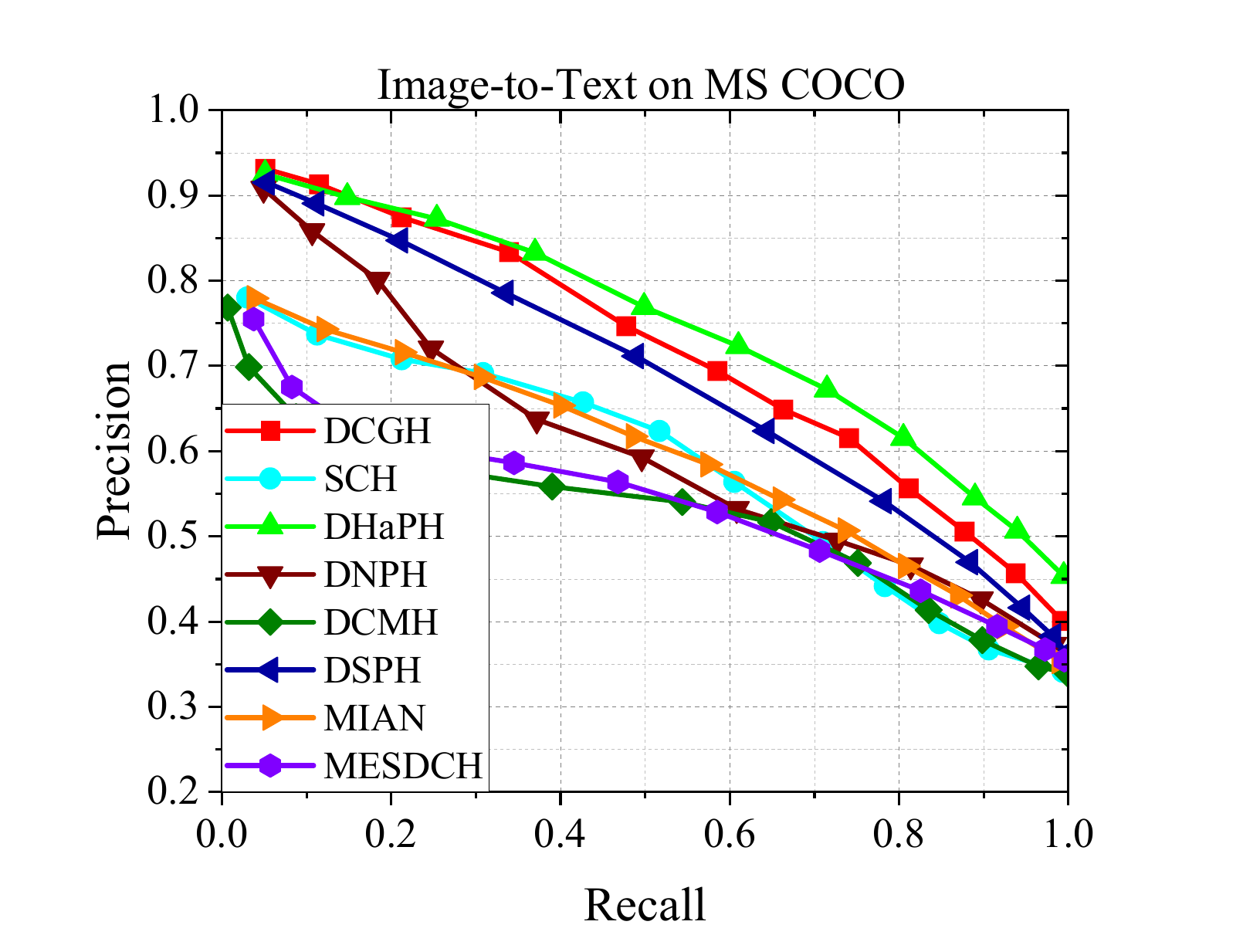}
        \caption{PR@16bits}
        \label{fig:16bitsPR-COCOI-T}
    \end{subfigure}
    \hfill
    \begin{subfigure}[b]{0.24\textwidth}
        \includegraphics[width=\textwidth]{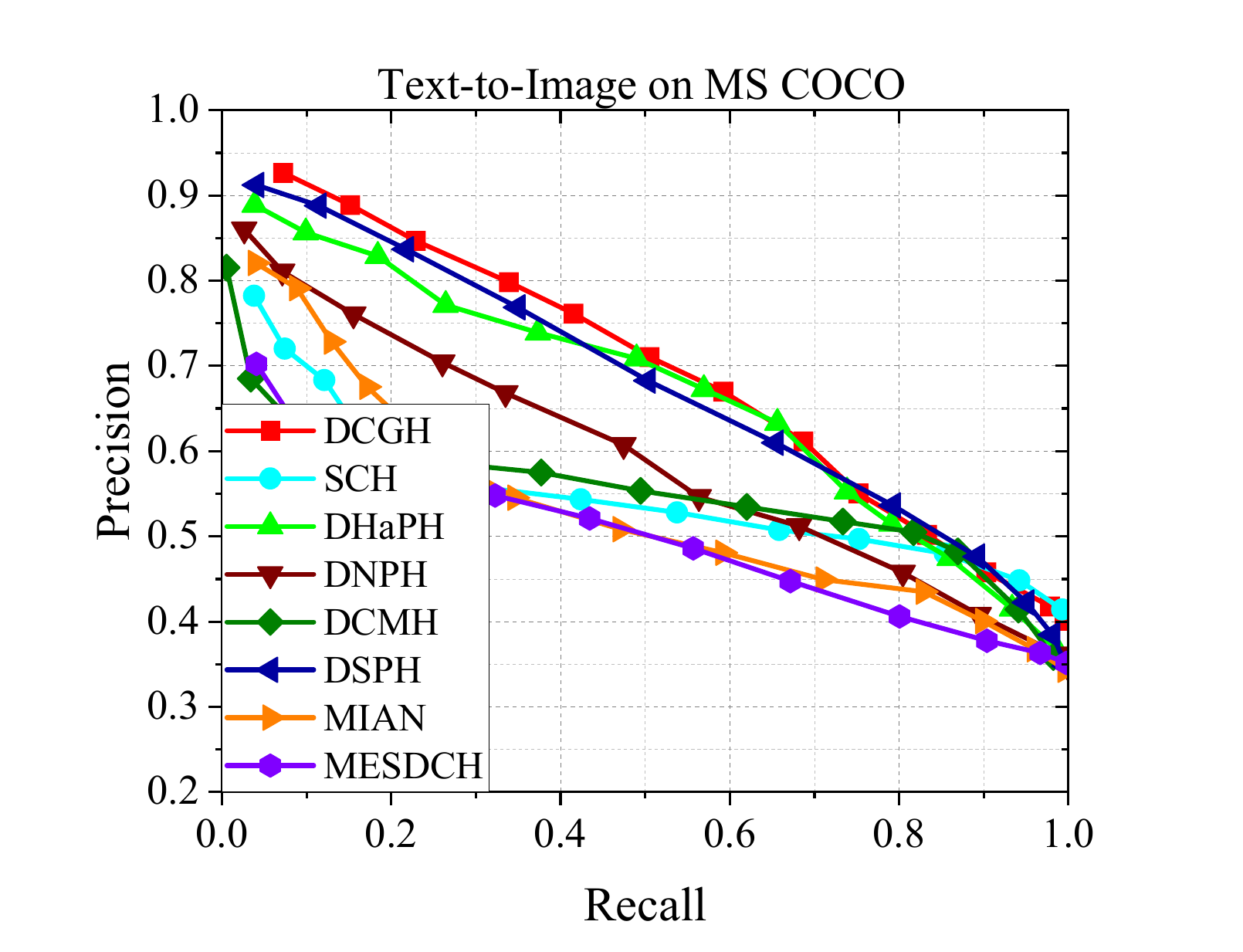}
        \caption{PR@16bits}
        \label{fig:16bitsPR-COCOT-I}
    \end{subfigure}
    \hfill
    \begin{subfigure}[b]{0.24\textwidth}
        \includegraphics[width=\textwidth]{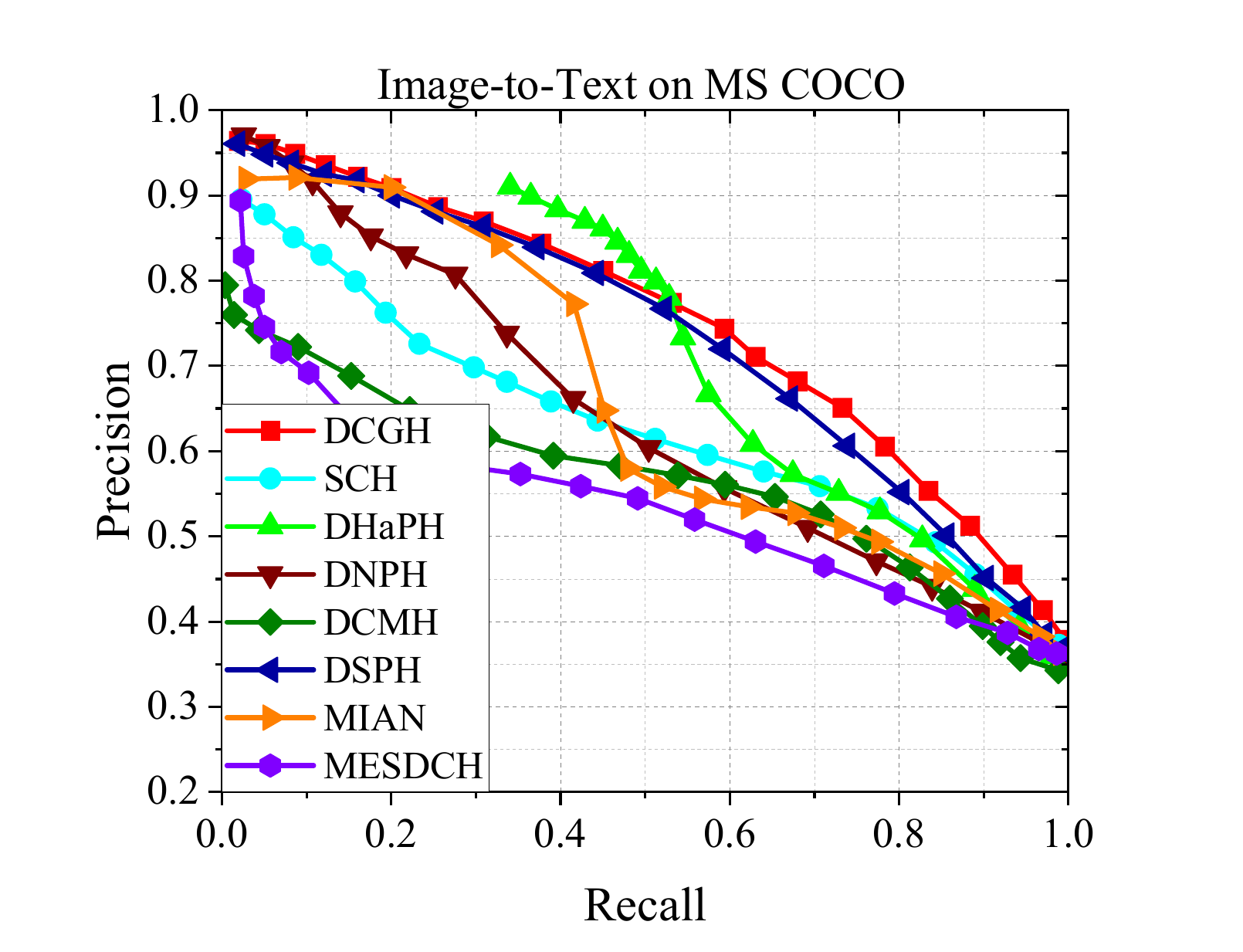}
        \caption{PR@32bits}
        \label{fig:PRCOCO32BitsI→T}
    \end{subfigure}
    \hfill
    \begin{subfigure}[b]{0.24\textwidth}
        \includegraphics[width=\textwidth]{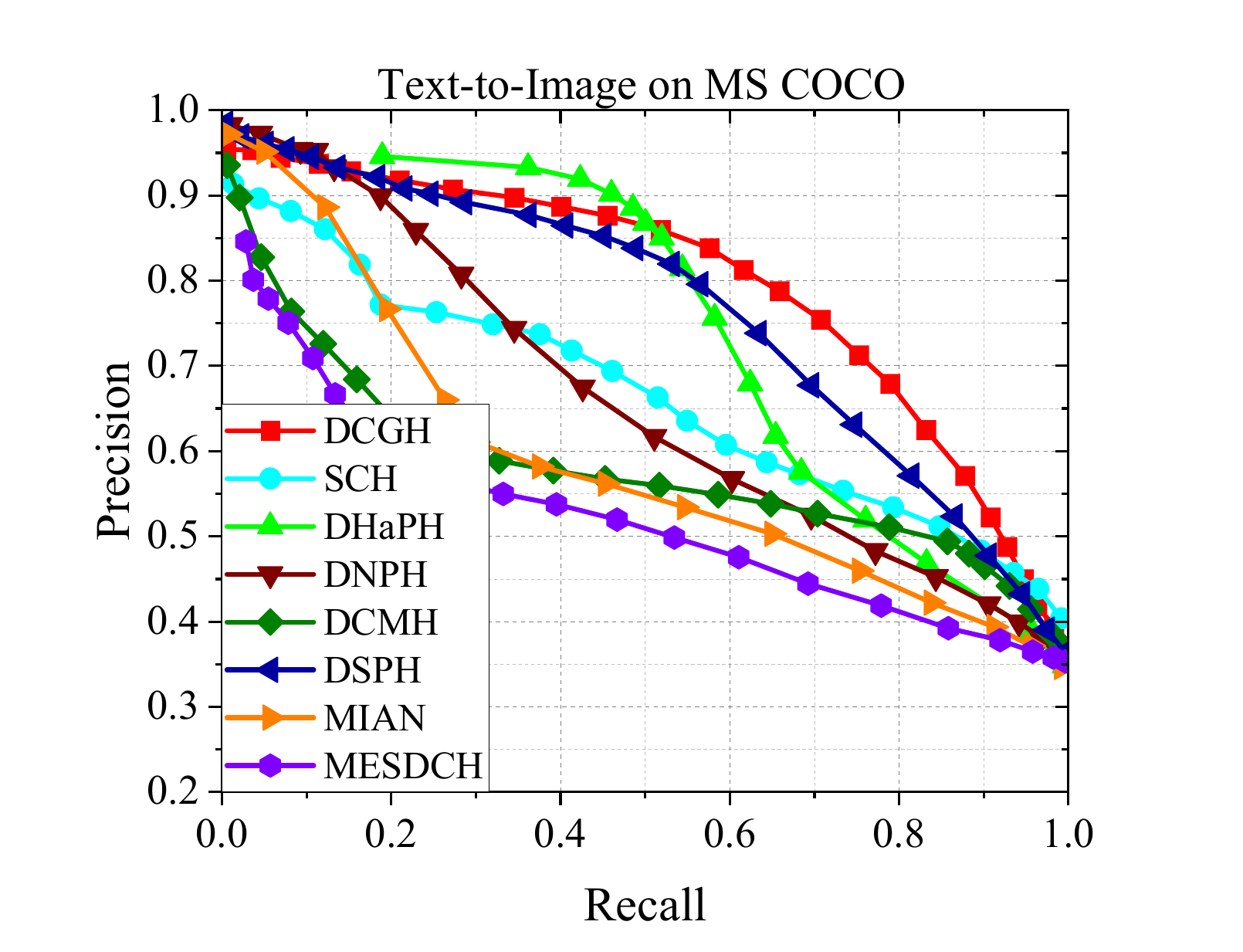}
        \caption{PR@32bits}
        \label{fig:PRCOCO32BitsT→I}
    \end{subfigure}
    \caption{\small Results of Precision-Recall curves on MS COCO w.r.t.16bits and 32bits.}
    \label{fig:ms coco}
\end{figure*}

\begin{figure*}
    \centering
    \begin{subfigure}[b]{0.3\textwidth}
        \includegraphics[width=\textwidth]{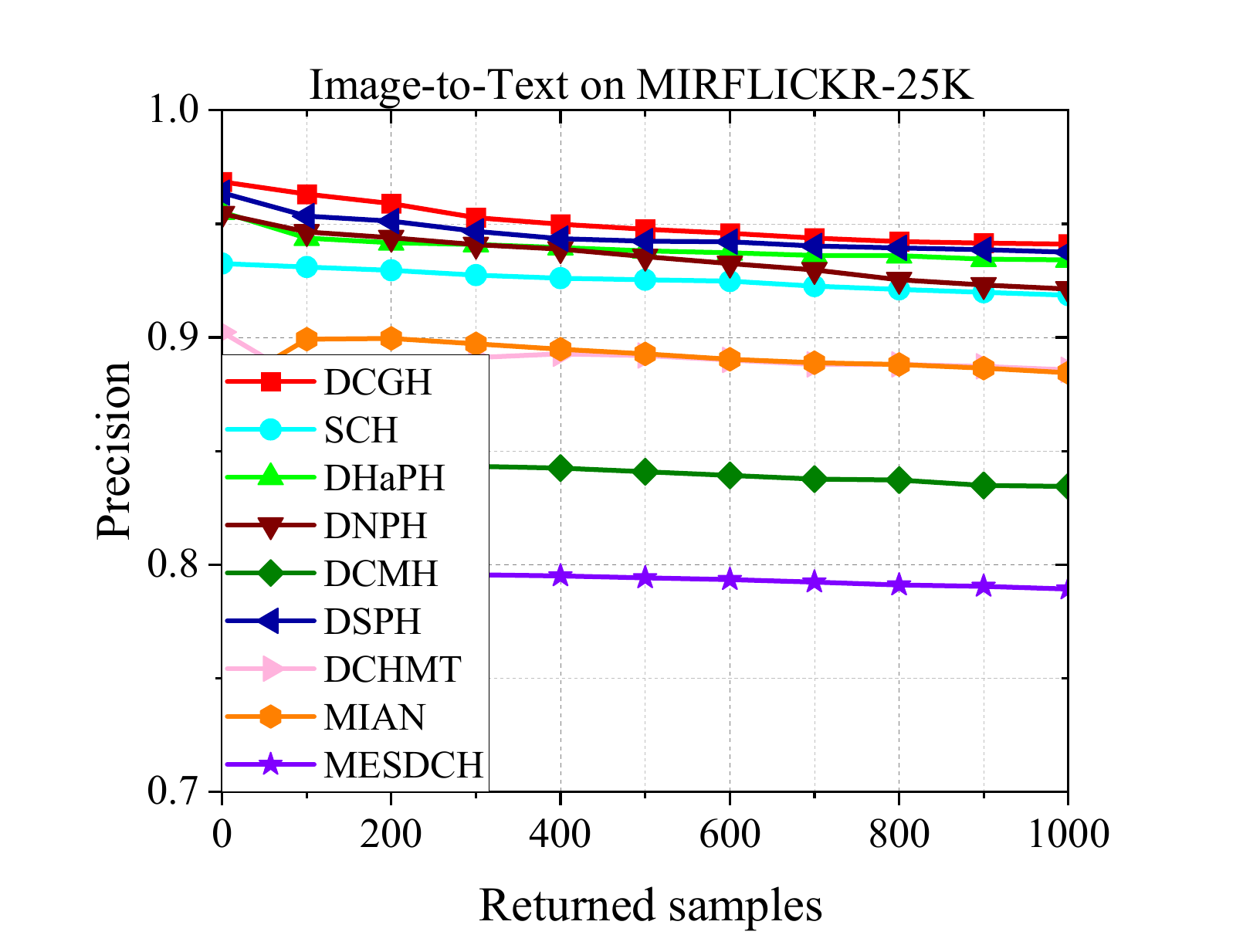}
        \caption{}
        \label{fig:P-TOPMIR32BitsI→T}
    \end{subfigure}
    \hfill
    \begin{subfigure}[b]{0.3\textwidth}
        \includegraphics[width=\textwidth]{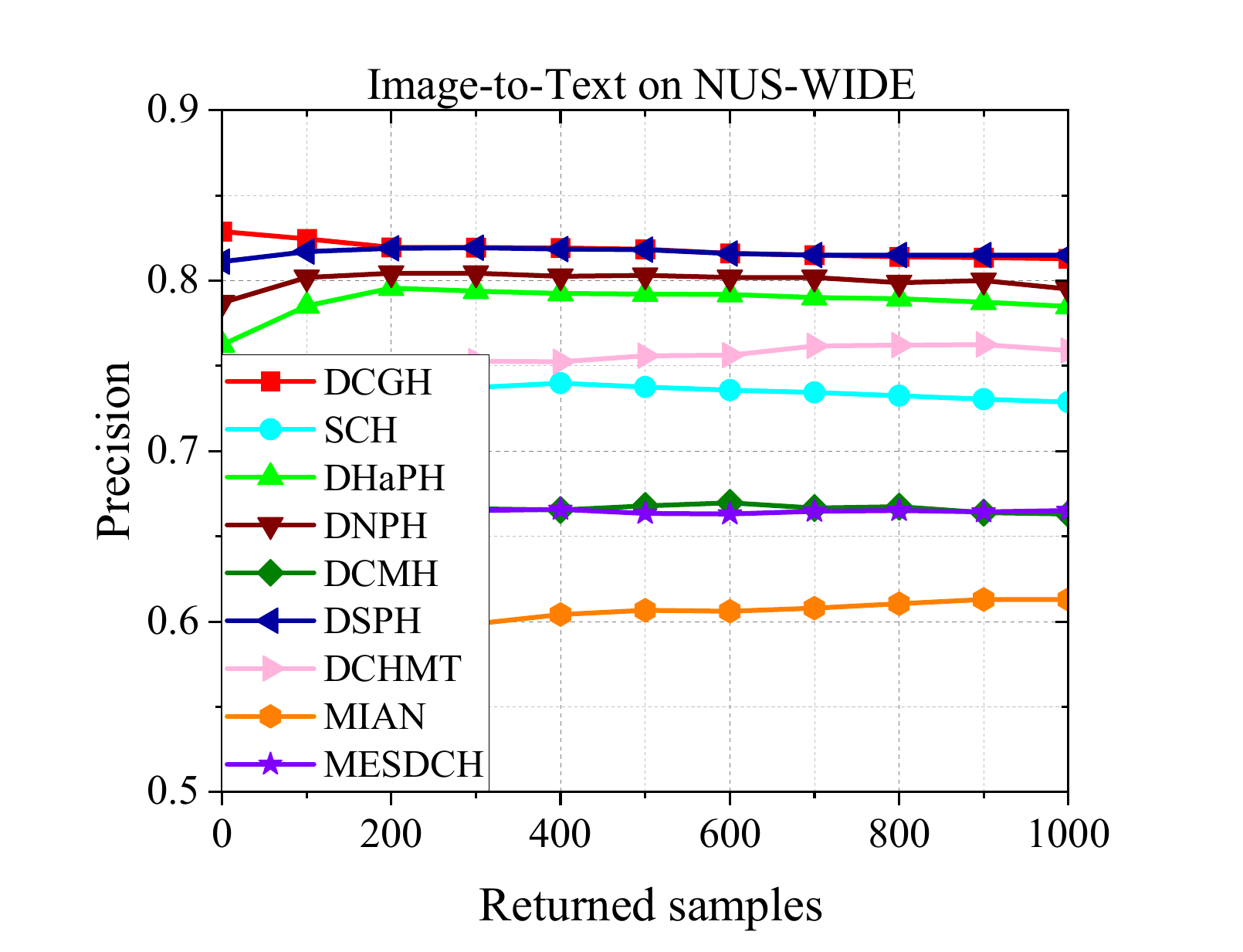}
        \caption{}
        \label{fig:P-TOPNUS32BitsI→T}
    \end{subfigure}
    \hfill
    \begin{subfigure}[b]{0.3\textwidth}
        \includegraphics[width=\textwidth]{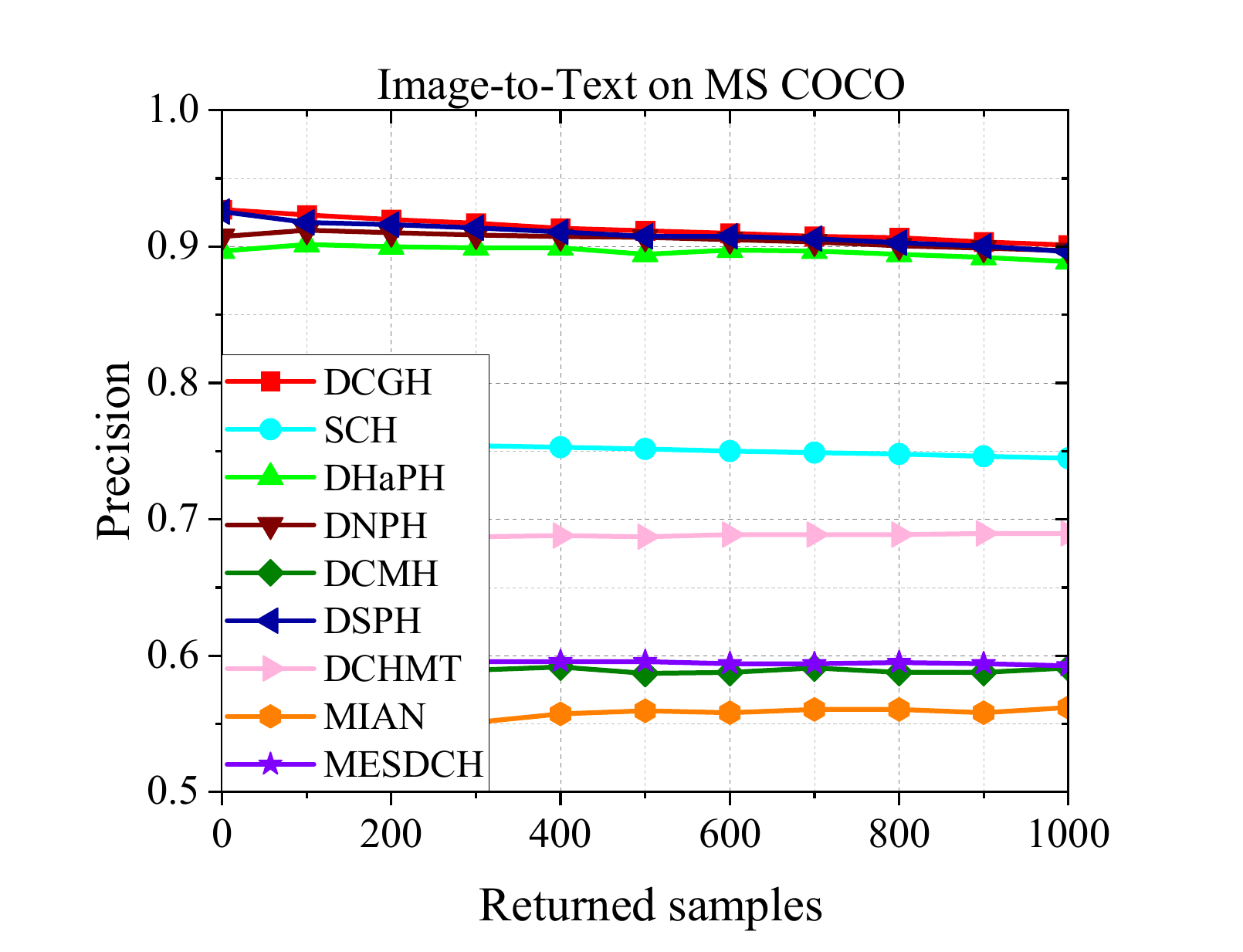}
        \caption{}
        \label{fig:P-TOPCOCO32BitsI→T}
    \end{subfigure}
    
    \begin{subfigure}[b]{0.3\textwidth}
        \includegraphics[width=\textwidth]{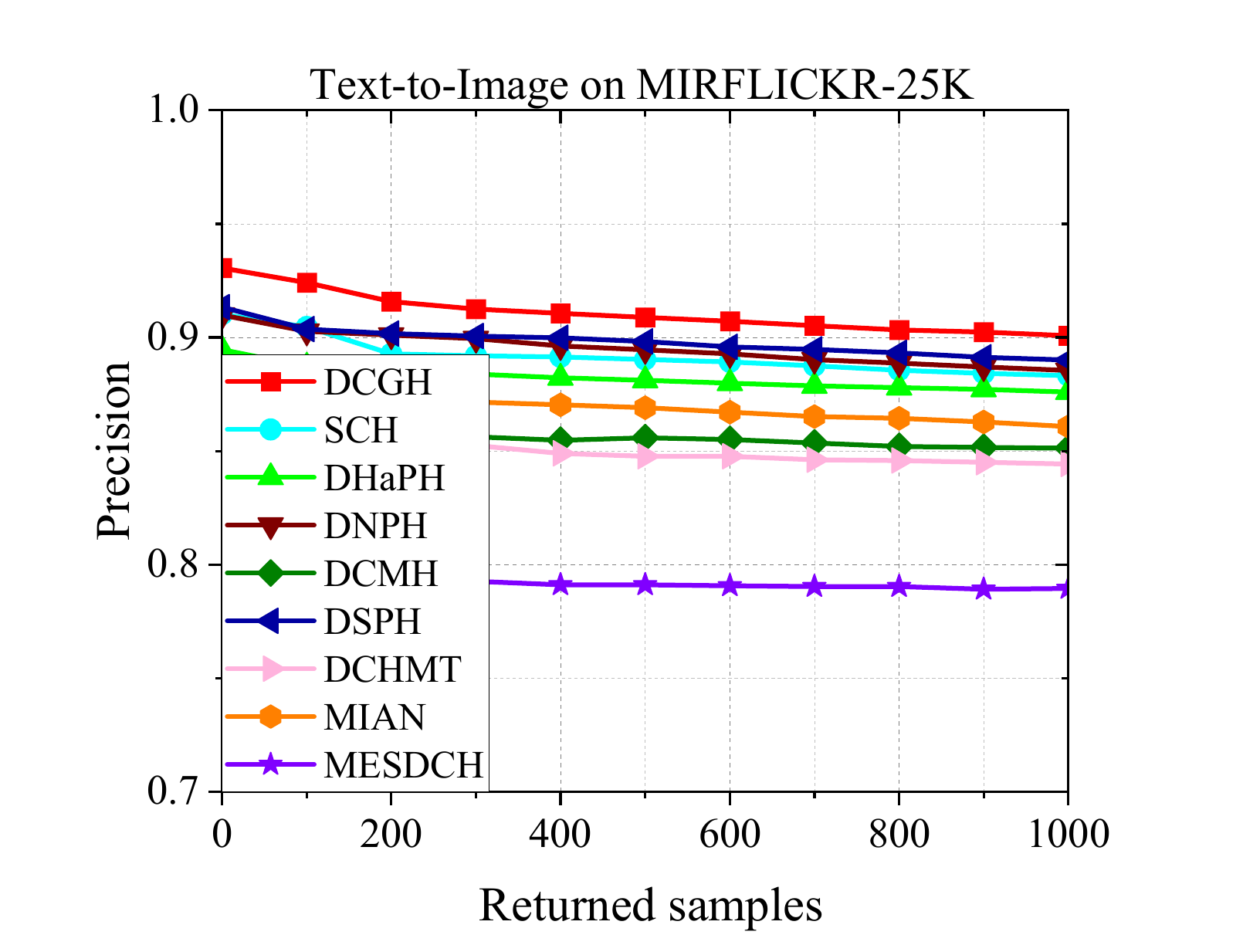}
        \caption{}
        \label{fig:P-TOPMIR32BitsT→I}
    \end{subfigure}
    \hfill
    \begin{subfigure}[b]{0.3\textwidth}
        \includegraphics[width=\textwidth]{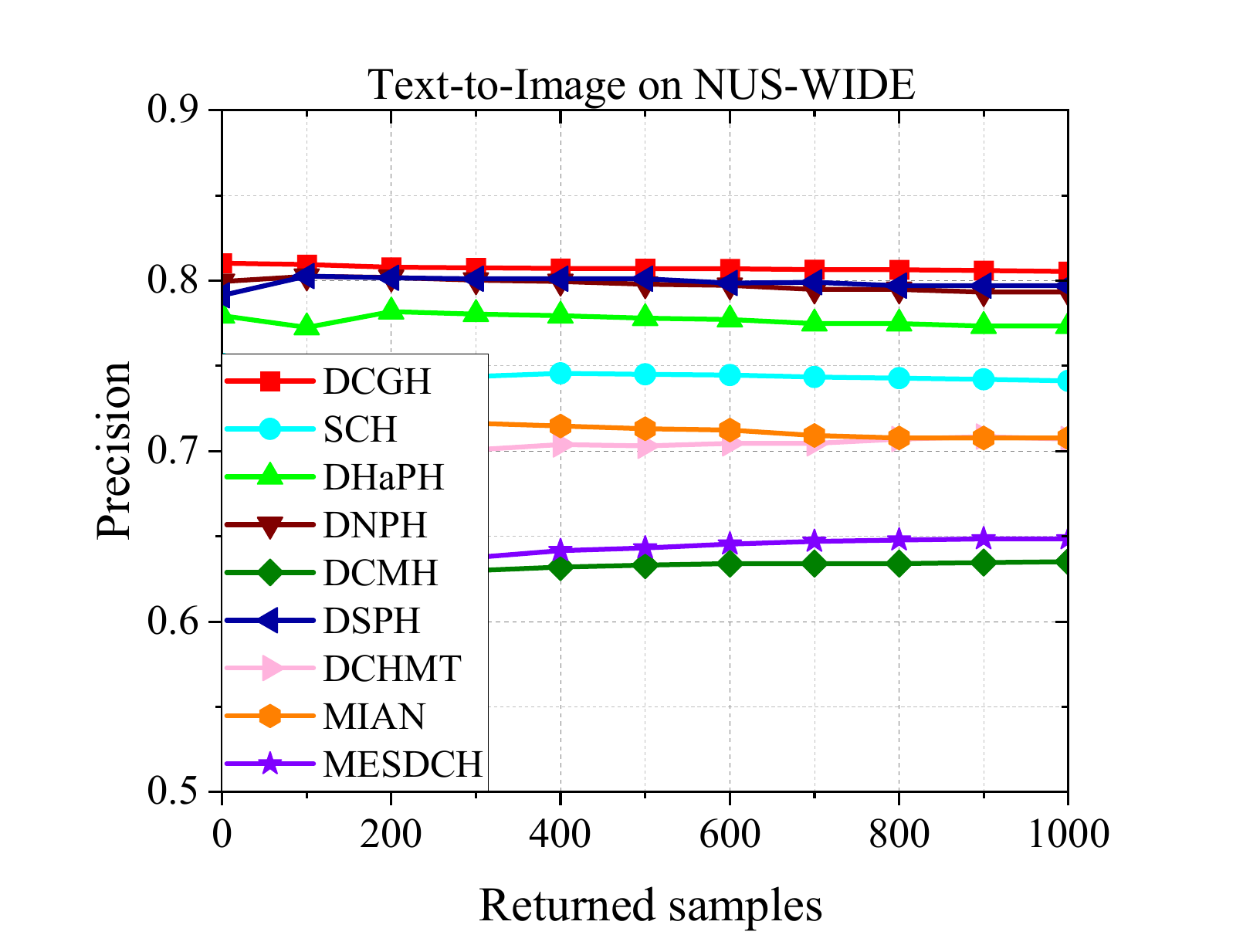}
        \caption{}
        \label{fig:P-TOPNUS32BitsT→I}
    \end{subfigure}
    \hfill
    \begin{subfigure}[b]{0.3\textwidth}
        \includegraphics[width=\textwidth]{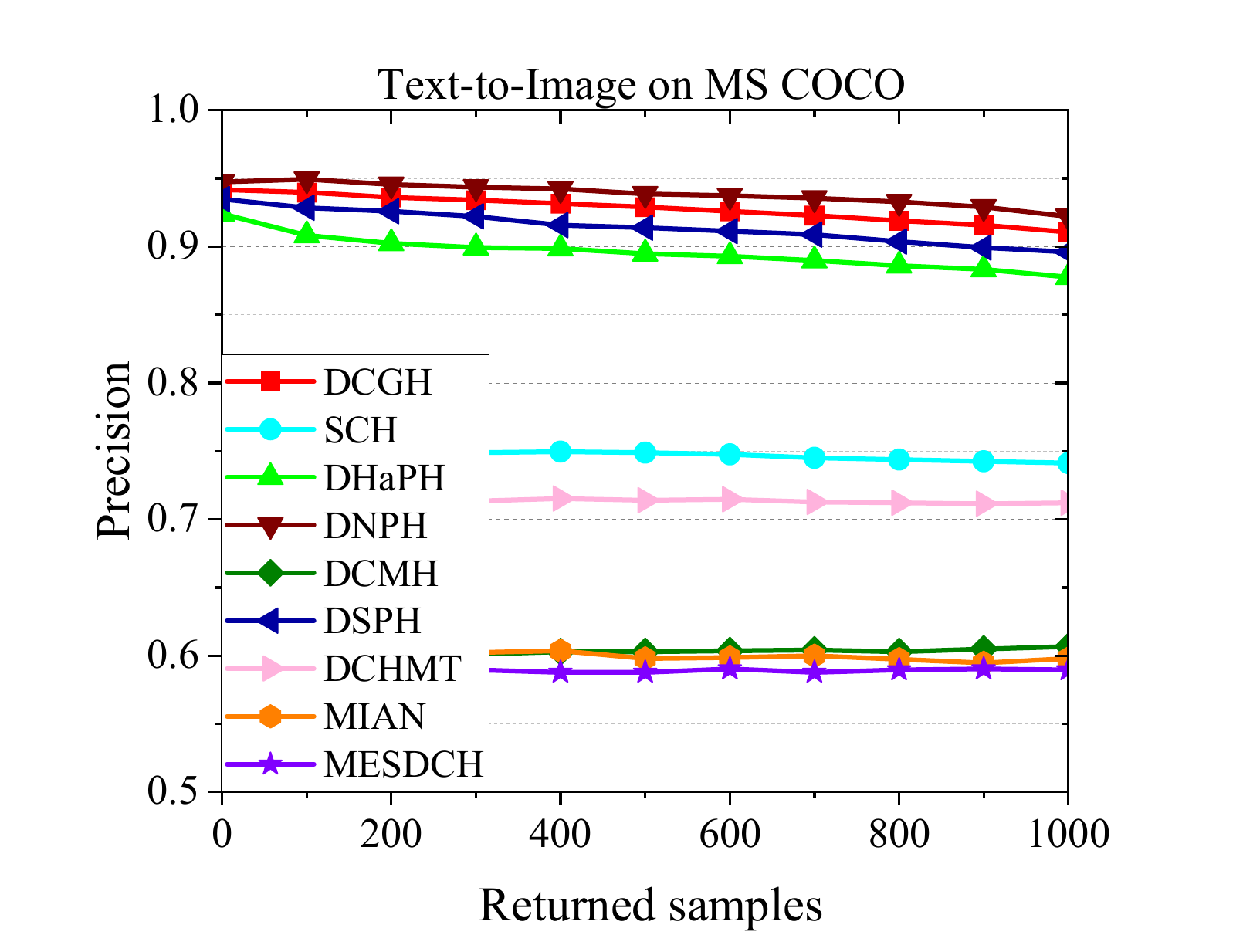}
        \caption{}
        \label{fig:P-TOPCOCO32BitsT→I}
    \end{subfigure}
    
    \caption{\small Results of TopN-precision curves on  MIRFLICKR-25K, NUS-WIDE, and MS COCO datasets w.r.t.32bits.}
    \label{fig:P-TOP}
\end{figure*}

\begin{figure*}
    \centering
    \begin{subfigure}[b]{0.3\textwidth}
        \includegraphics[width=\textwidth]{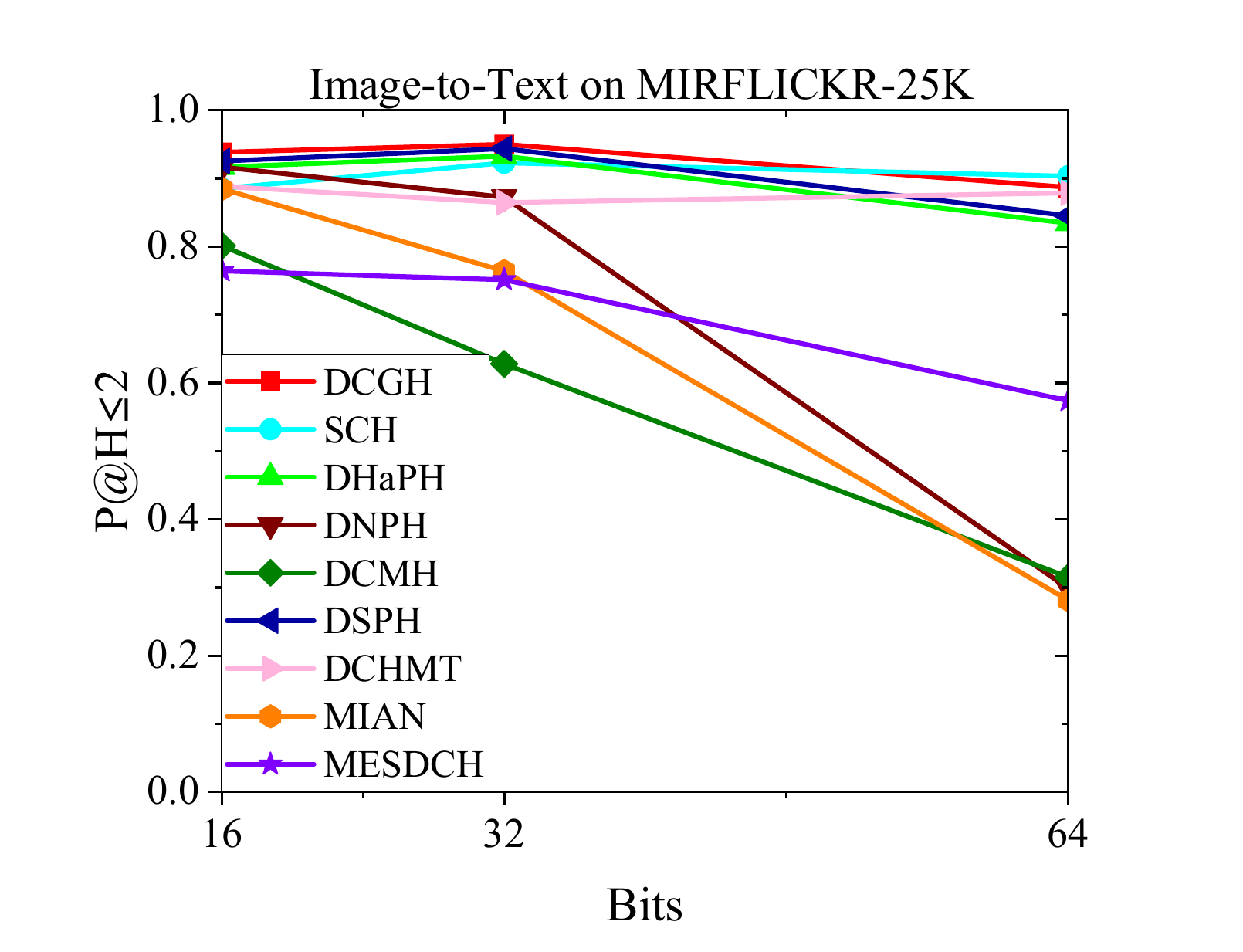}
        \caption{}
        \label{fig:R2mirI-T}
    \end{subfigure}
    \hfill
    \begin{subfigure}[b]{0.3\textwidth}
        \includegraphics[width=\textwidth]{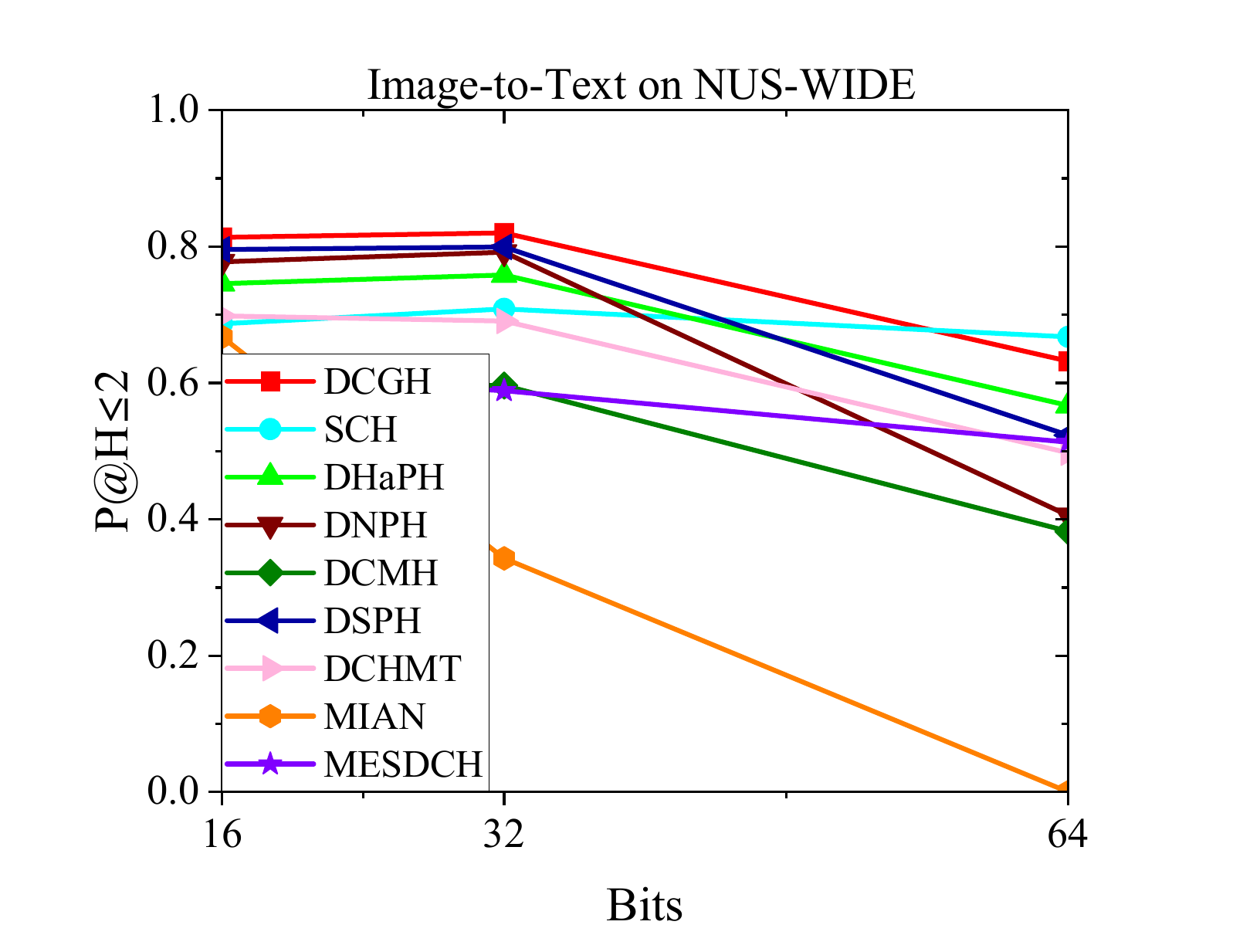}
        \caption{}
        \label{fig:R2NUSI-T}
    \end{subfigure}
    \hfill
    \begin{subfigure}[b]{0.3\textwidth}
        \includegraphics[width=\textwidth]{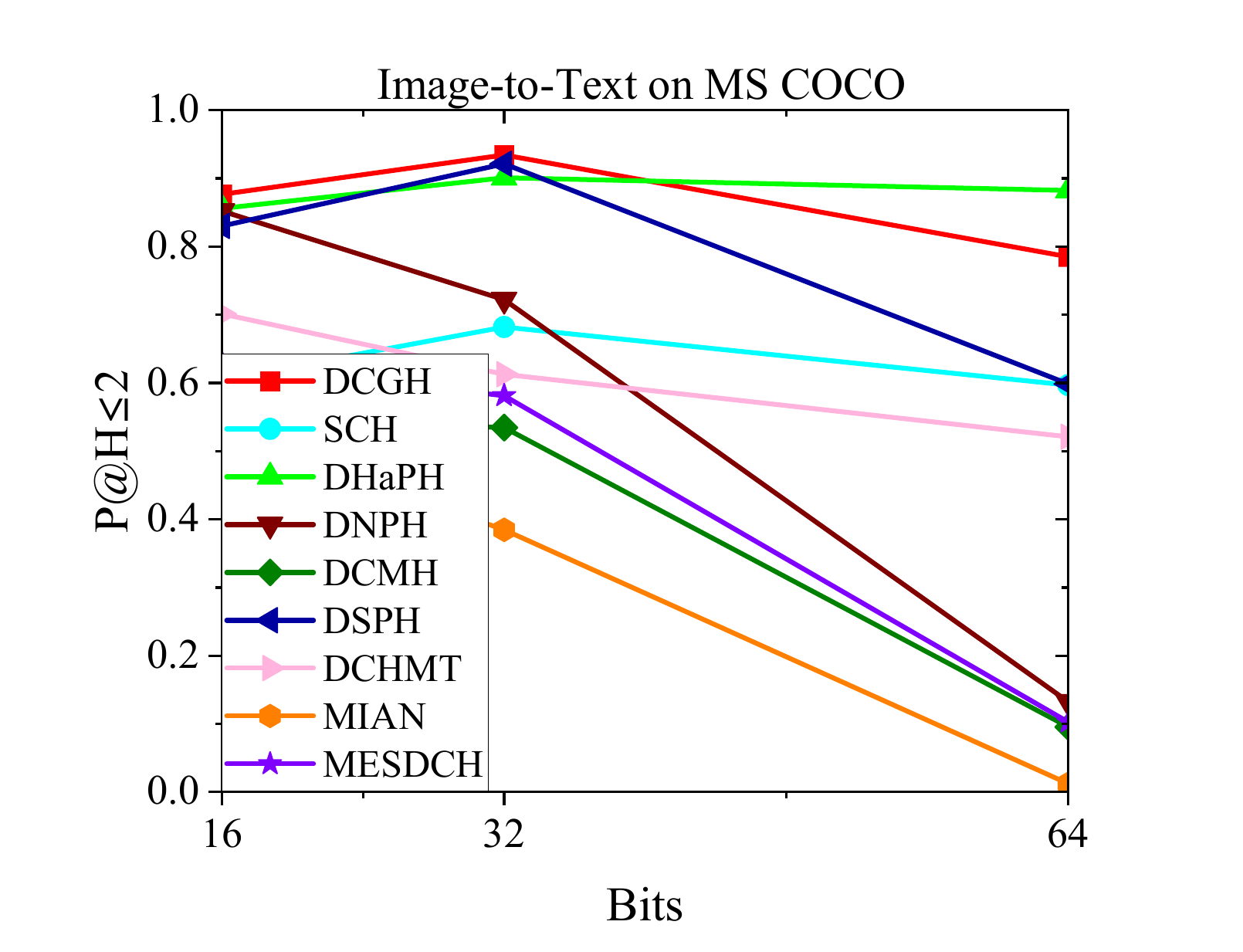}
        \caption{}
        \label{fig:R2COCOI→T}
    \end{subfigure}
    
    \begin{subfigure}[b]{0.3\textwidth}
        \includegraphics[width=\textwidth]{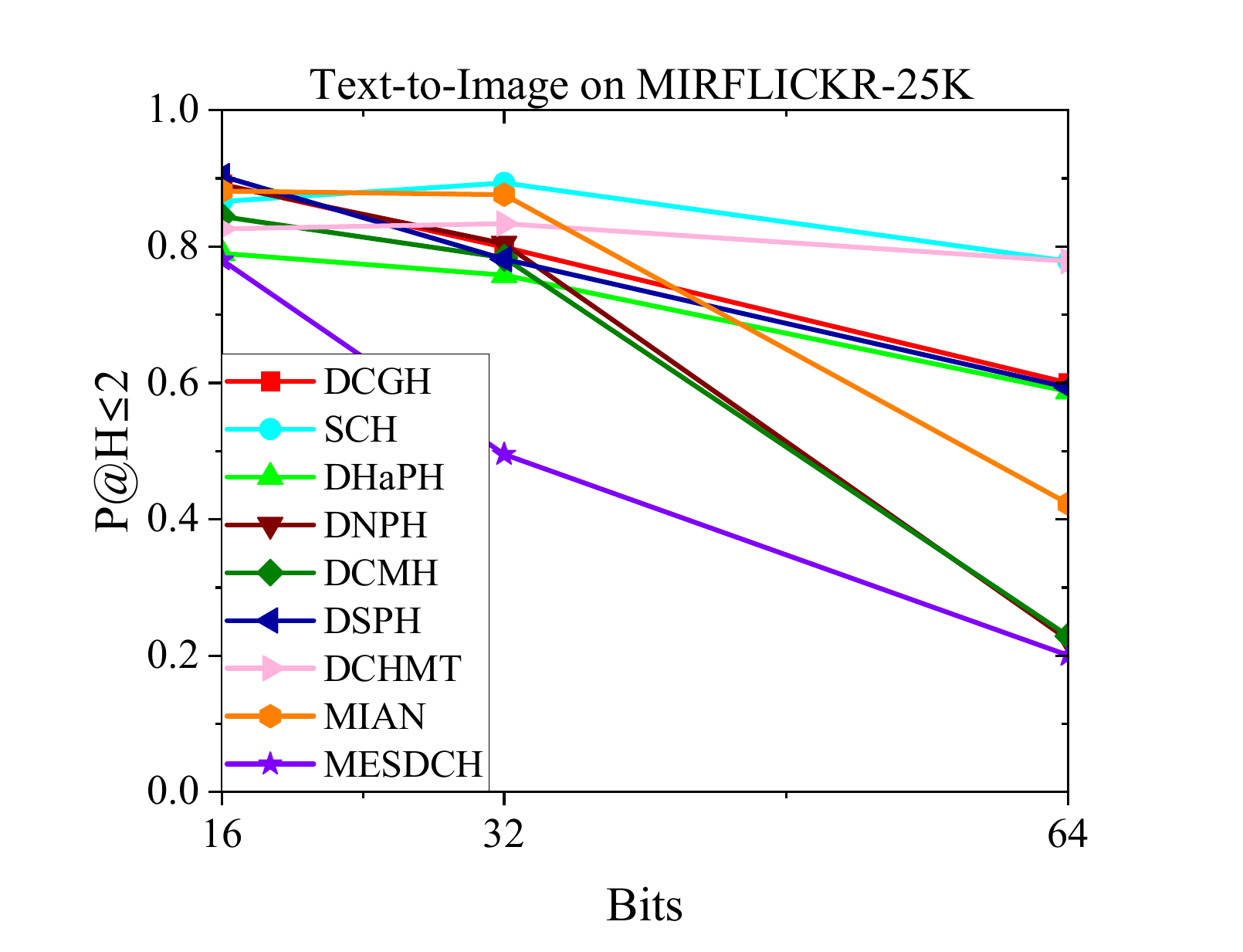}
        \caption{}
        \label{fig:R2MIRT→I}
    \end{subfigure}
    \hfill
    \begin{subfigure}[b]{0.3\textwidth}
        \includegraphics[width=\textwidth]{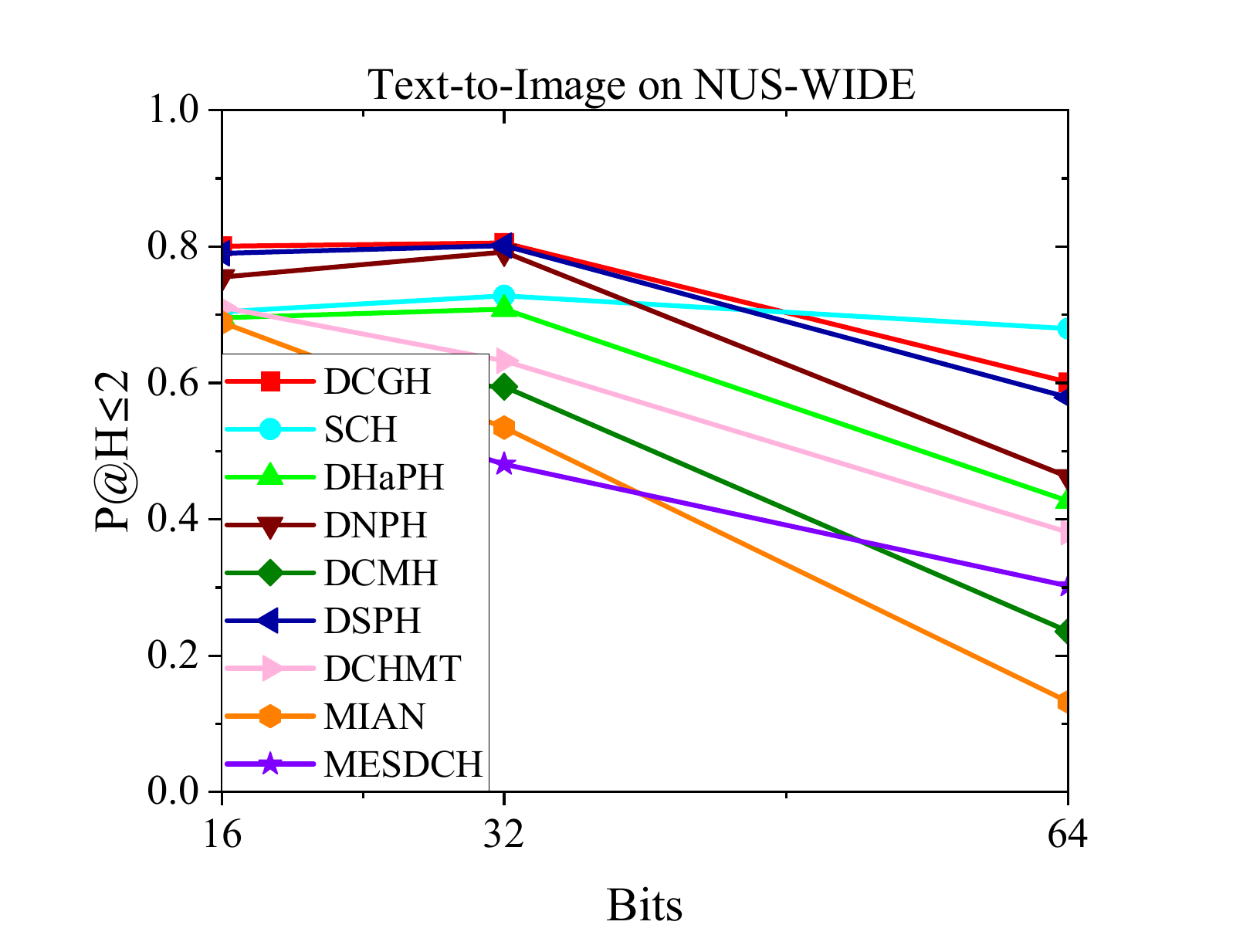}
        \caption{}
        \label{fig:R2NUST-I}
    \end{subfigure}
    \hfill
    \begin{subfigure}[b]{0.3\textwidth}
        \includegraphics[width=\textwidth]{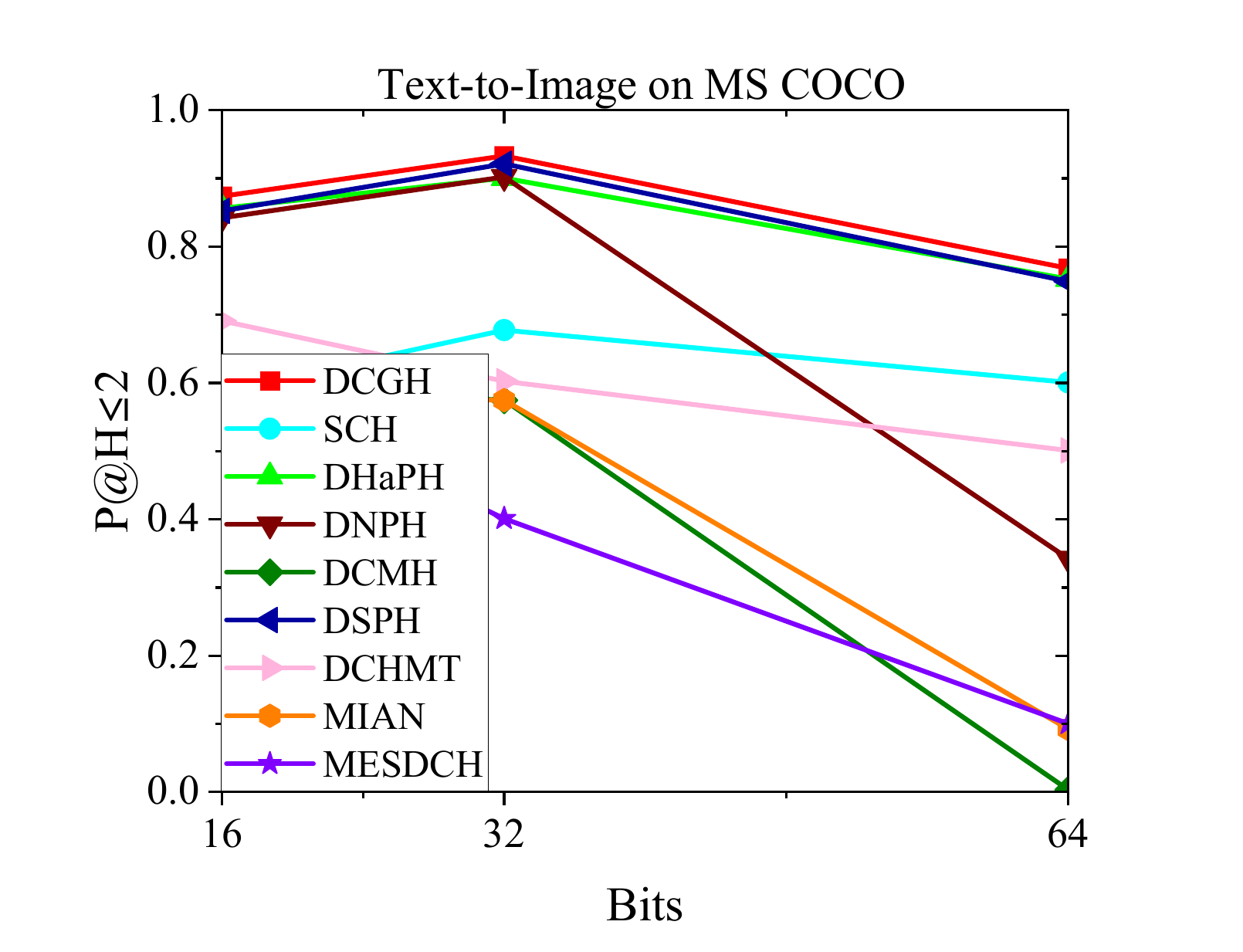}
        \caption{}
        \label{fig:R2COCOT→I}
    \end{subfigure}
    \caption{\small The mAP@H$\leq$2 results w.r.t. different code lengths on MIRFLICKR-25K, NUS-WIDE, and MS COCO datasets.}
    \label{fig:R2}
\end{figure*}

\begin{table*}
  \scriptsize
  \centering
  \caption{Comparison with baselines in terms of NDCG@1000 w.r.t. 16bits, 32bits, 64bits on MIRFLICKR-25K, NUS-WIDE, and MS COCO and the best results are shown in boldface.}\label{tab:NDCG@1000}
  \begin{tabular}{*{12}{c}}
    \toprule
    \multirow{2}*{Task} & \multirow{2}*{Methods} & \multicolumn{3}{c}{MIRFLICKR-25K} & \multicolumn{3}{c}{NUS-WIDE} & \multicolumn{3}{c}{MS COCO} &\\
    \cmidrule(lr){3-5}\cmidrule(lr){6-8}\cmidrule(lr){9-12}
    & & 16 bits & 32 bits & 64 bits & 16 bits & 32 bits & 64 bits & 16 bits & 32 bits & 64 bits \\
    \midrule
    \multirow{9}*{Img2Txt} 
    &DCMH (CVPR17)      & 0.4002 & 0.4156 & 0.4300 & 0.4396 & 0.4794 & 0.4794 & 0.1646 & 0.1783 & 0.1705\\
    &MESDCH (Neurocomputing22)  & 0.3550 & 0.3650 & 0.3529 & 0.3934 & 0.4202 & 0.4279 & 0.2311 & 0.2324 & 0.2322\\
    &DCHMT (ICM22)  & 0.4310 & 0.3876 & 0.4137 & 0.4534 & 0.4708 & 0.4729 & 0.2557 & 0.2328 & 0.3059\\
    &MIAN (TKDE22)  & 0.4517 & 0.4886& 0.5055 & 0.4444 & 0.3769& 0.4101 & 0.1652 & 0.1703 & 0.1554\\
    &DSPH (TCSVT23) & 0.5214 & 0.5668 & 0.5846& 0.5245 & 0.5424 & 0.5742 & 0.4701 & 0.5386 & 0.5932\\
    &DNPH (TOMM24)   &  \textbf{0.5396} & 0.5703 & 0.5811 & \textbf{0.5258} & 0.5506 & 0.5533 & \textbf{0.5116} & \textbf{0.5764} & 0.5986\\
    &DHaPH (TKDE24)   & 0.4785 & 0.4906 & 0.5070 & 0.4566 & 0.4732 & 0.4735 & 0.3919 & 0.4137 & 0.4104\\
    &SCH (AAAI24)   & 0.4921 & 0.5168 & 0.5288 & 0.4461 & 0.4663 & 0.5090 & 0.1815 & 0.3097 & 0.4190\\
    &DCGH (Ours)    & 0.5344 & \textbf{0.5743} & \textbf{0.5859} &0.5266
    & \textbf{0.5547} & \textbf{0.5857} & 0.4714 & 0.5497 & \textbf{0.5998}\\
    \midrule
    \multirow{9}*{Txt2Img}  
    &DCMH (CVPR17)      & 0.4111 & 0.4200 & 0.4345 & 0.4316 & 0.4414 & 0.4383 & 0.1853 & 0.1962 & 0.1986\\
    &MESDCH (Neurocomputing22)  & 0.3731 & 0.3645 & 0.3728 & 0.3820 & 0.3990 & 0.4060 & 0.2236 & 0.2234 & 0.2191\\
    &DCHMT (ICM22)  & 0.3772 & 0.3869 & 0.3663 & 0.4442 & 0.4431 & 0.4618 & 0.2554 & 0.2550 & 0.3079\\
    &MIAN (TKDE22)  & 0.4282 & 0.4462 & 0.4448 & 0.4533 & 0.4669 & 0.4762 & 0.1854 & 0.1891 & 0.1790\\
    &DSPH (TCSVT23) & 0.4495 & 0.4827 & 0.4822 & 0.5122 & 0.5364 & 0.5465 & 0.4700 & 0.5480 & 0.6000\\
    &DNPH (TOMM24)   & \textbf{0.4709} & \textbf{0.4977} & 0.5082 & 0.5102 & 0.5482 & 0.5527 & \textbf{0.5103} & \textbf{0.6045} & \textbf{0.6115}\\
    &DHaPH (TKDE24)   & 0.4543 & 0.4582 & 0.4602 & 0.4007 & 0.4015 & 0.4157 & 0.4049 & 0.4189 & 0.4362\\
    &SCH (AAAI24)   & 0.4317 & 0.4446 & 0.4530 & 0.4557 & 0.4760 & 0.5070 & 0.1886 & 0.3058 & 0.4106\\
    &DCGH (Ours)    & 0.4540 & 0.4755 & \textbf{0.5092} & \textbf{0.5219}
    & \textbf{0.5497} & \textbf{0.5530} & 0.4715 & 0.5590 & 0.6031\\
    \bottomrule
  \end{tabular}
\end{table*}

\textbf{Evaluation Protocols.} In our experiments, we employed five commonly used evaluation metrics to assess the performance of cross-modal similarity search, which include mean Average Precision (mAP), Normalized Discounted Cumulative Gain with top 1000 returned samples (NDCG$@$1000), Precision with Hamming radius 2 (P$@$H$\leq $2) curve, Precision-Recall (PR) curve and Precision$@$Top N curve. The mAP is calculated as the aver-
age of the AP across all query samples. The PR curve represents the variation curve between recall and precision. Top-N Precision curve illustrates the proportion of truly relevant data among the top N results returned by the system. NDCG$@$1,000 is a comprehensive metric that assesses ranking performance of the top 1,000 retrieval results. Precision within Hamming radius 2 describes the precision of the samples retrieved within the specified Hamming radius. The experimental results from the aforementioned evaluation metrics demonstrate that the DCGH method achieves excellent performance in cross-modal similarity search.

\subsection{Performance Comparison} \label{subsec:performance-conparison}

 We validate the performance of DCGH by comparing it with state-of-the-art deep cross-modal hashing methods on image-text retrieval tasks across three public datasets. The mAP results are shown in Table~\ref{tab:hamming-ranking}, where "Img2Txt" indicates image-to-text retrieval, and "Txt2Img" indicates text-to-image retrieval. In most cases, DCGH outperforms other baseline methods, achieving satisfactory performance. On the MIRFLICKR-25K dataset, compared to the baseline methods, the SCH method performs the best. In the image-to-text retrieval task, the SCH method is on average 0.37\% higher than ours, and in the text-to-image retrieval task, the SCH method is on average 0.9\% higher than ours. However, on the NUS-WIDE and MS COCO datasets, as the amount of data increases, the issue of intra-class dispersion leads to unsatisfactory performance for the SCH method. In contrast, our DCGH method considers both intra-class aggregation and inter-class structural relationships, and it essentially achieves the best performance compared to other baseline methods. This confirms the effectiveness of our method. The performance of this method in the multi-label retrieval scenario was evaluated using the NDCG@1000 evaluation metric. 
 
 The results of NDCG@1000 can be found in Table~\ref{tab:NDCG@1000}. From Table~\ref{tab:NDCG@1000}, it can be observed that our method and the DNPH method have comparable scores with each having their own advantages and disadvantages. The DNPH method, which introduces a uniform distribution constraint on the basis of proxy loss, achieves the optimal NDCG@1000 scores for the image-text retrieval tasks at 16bits and 32bits on three public datasets in most cases. However, at 64bits, as the length of the hash code increases, the discrete space becomes sparser, and DCGH can obtain higher-quality ranking results compared to DNPH.

 To further evaluate the performance of our method, Figures~\ref{fig:mir-flickr25k}, Figures~\ref{fig:nus-wide}, and Figures~\ref{fig:ms coco} show the PR curves on the MIRFLICKR-25K, NUS-WIDE, and MS COCO datasets at 16 bits and 32 bits, and Figures~\ref{fig:P-TOP} show the
 results of TopN-precision curves on MIRFLICKR-25K, NUS-WIDE, and MS COCO datasets w.r.t.32bits. 
 The mAP@H$\leq$2 results for different hash code lengths on the MIRFLICKR-25K, NUS-WIDE, and MS COCO datasets are shown in Figure~\ref{fig:R2}, where (a), (b), and (c) display the mAP@H$\leq$2 results for the image retrieval text task. (d), (e), and (f) show the mAP@H$\leq$2 results for the text retrieval image task.  Compared with the state-of-the-art methods in the baseline, our method achieves comparable or even better results.

\subsection{Ablation Studies} \label{subsec:ablation-stuides}
To validate the effectiveness of the DCGH method, we implemented three variations to calculate the mAP values for the tasks of image retrieval of text and text retrieval of images. Specifically: (1) DCGH-$P$-$V$: using only proxy loss $\mathcal{L}_{proxy}$ to train the model. (2) DCGH-$X$-$V$: using only pairwise loss $\mathcal{L}_{pair}$ to train the model. Since the proxy loss was not considered, $\alpha$ and $\beta$ are set to 1. (3) DCGH-$V$: Variance constraint $\mathcal{L}_{var}$ was not used.

The ablation experiment results are shown in Table~\ref{tab:Ablation Studies}. Comparing the results of DCGH-$P$-$V$ and DCGH-$X$-$V$ on three benchmark datasets reveals that DCGH-$P$-$V$, which only uses proxy loss, generally performs better on large datasets like NUS-WIDE and MS COCO than DCGH-$X$-$V$, which only uses pairwise loss. As the amount of data increases, the issue of intra-class dispersion caused by pairwise loss can significantly affect the effectiveness of the hash codes. Comparing the performance of DCGH-$V$ with DCGH-$P$-$V$ and DCGH-$X$-$V$ across the three datasets shows that DCGH-$V$, which combines proxy loss and pairwise loss to consider both intra-class aggregation and inter-class structural relationship preservation, significantly outperforms DCGH-$P$-$V$ and DCGH-$X$-$V$, which only use one type of loss each. This confirms the rationality of our combination approach. By comparing the results of DCGH and DCGH-$V$ on the three benchmark datasets, it is evident that introducing a variance constraint to prevent semantic bias leads to better results, confirming the effectiveness of the variance constraint.By comparing three variables, the effectiveness of each component of the DCGH method was verified. By combining proxy loss, pairwise loss and variance constraints, DCGH is able to learn excellent hash codes.

\begin{table*}
  \scriptsize
  \centering
  \caption{mAP Result of DCGH and Its Variants on MIRFLICKR-25K, NUS-WIDE, and MS COCO w.r.t. 16 bits, 32 bits, and 64 bits.}\label{tab:Ablation Studies}
  \begin{tabular}{*{12}{c}}
    \toprule
    \multirow{2}*{Task} & \multirow{2}*{Methods} & \multicolumn{3}{c}{MIRFLICKR-25K} & \multicolumn{3}{c}{NUS-WIDE} & \multicolumn{3}{c}{MS COCO} &\\
    \cmidrule(lr){3-5}\cmidrule(lr){6-8}\cmidrule(lr){9-12}
    & & 16 bits & 32 bits & 64 bits & 16 bits & 32 bits & 64 bits & 16 bits & 32 bits & 64 bits \\
    \midrule
    \multirow{4}*{Img2Txt} 
    &DCGH-$P$-$V$      & 0.8154 & 0.8258 & 0.8412 & 0.6678 & 0.6910 & 0.7122 & 0.6562 & 0.7116 & 0.7447\\
    &DCGH-$X$-$V$  & 0.7902 & 0.8288 & 0.8457 & 0.6723 & 0.6893 & 0.6989 & 0.6273 & 0.6718 & 0.7096\\
    &DCGH-$V$      & 0.8164 & 0.8430 & 0.8604 & 0.6867 & 0.7021 & 0.7158 & 0.7077 & 0.7428 & 0.7722\\
    &DCGH (Ours)    & \textbf{0.8245} & \textbf{0.8482} & \textbf{0.8633} & \textbf{0.6869}
    & \textbf{0.7038} & \textbf{0.7214} & \textbf{0.7120} & \textbf{0.7492} & \textbf{0.7737}\\
    \midrule
    \multirow{4}*{Txt2Img}  
    &DCGH-$P$-$V$      & 0.8087 & 0.8192 & 0.8313 & 0.6890 & 0.7127 & 0.7304 & 0.6614 & 0.7259 & 0.7592\\
    &DCGH-$X$-$V$  & 0.7853 & 0.8184 & 0.8329 & 0.6848 & 0.6985 & 0.7079 & 0.6256 & 0.6784 & 0.7183\\
    &DCGH-$V$     & 0.8022 & 0.8229& 0.8351 & 0.7001& 0.7147& 0.7321 & 0.7065 & 0.7456 & 0.7774\\
    &DCGH (Ours)    & \textbf{0.8108} & \textbf{0.8297} & \textbf{0.8413} & \textbf{0.7014}
    & \textbf{0.7153} & \textbf{0.7329} & \textbf{0.7072} & \textbf{0.7536} & \textbf{0.7786}\\
    \bottomrule
  \end{tabular}
\end{table*}

\subsection{Sensitivity to Hyperparameters} \label{subsec:sensitivity-of-parameters}
\begin{figure*}
    \centering
    \begin{subfigure}[b]{0.3\textwidth}
        \includegraphics[width=\textwidth]{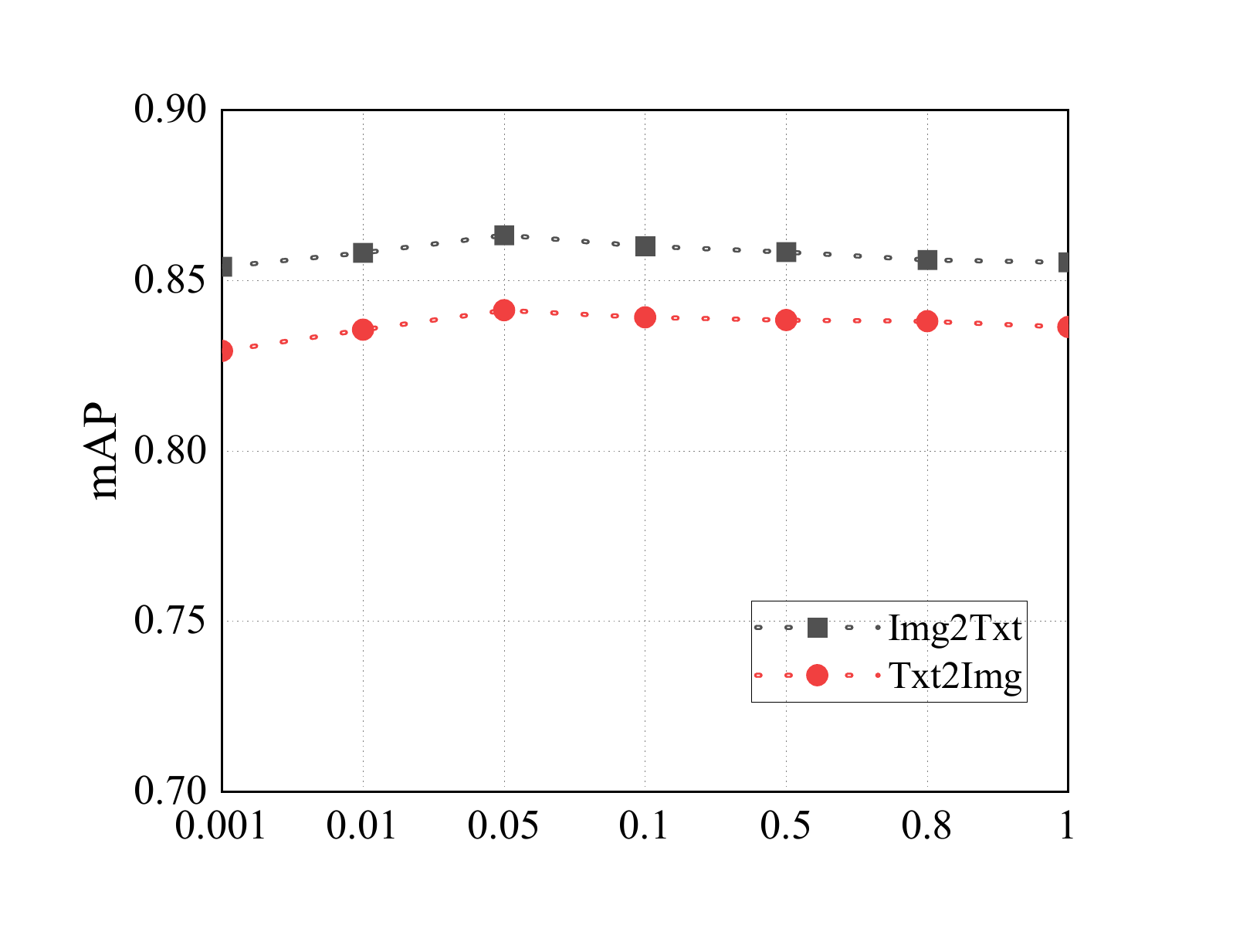}
        \caption{$\alpha$ on MIR@64 bits}
        \label{fig:a-mirflickr64bit}
    \end{subfigure}
    \hfill
    \begin{subfigure}[b]{0.3\textwidth}
        \includegraphics[width=\textwidth]{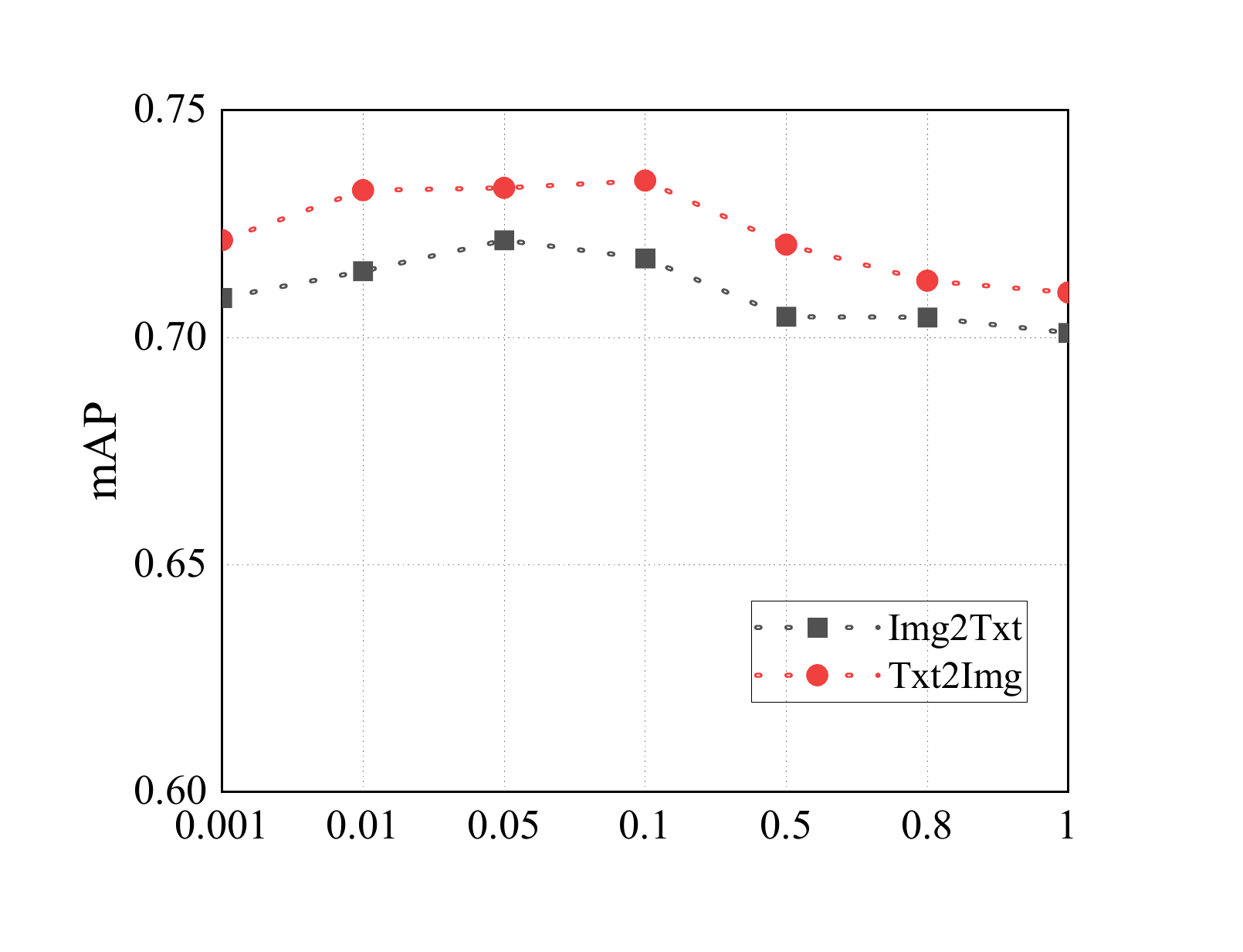}
        \caption{$\alpha$ on NUS@64 bits}
        \label{fig:a-nuswide64bit}
    \end{subfigure}
    \hfill
    \begin{subfigure}[b]{0.3\textwidth}
        \includegraphics[width=\textwidth]{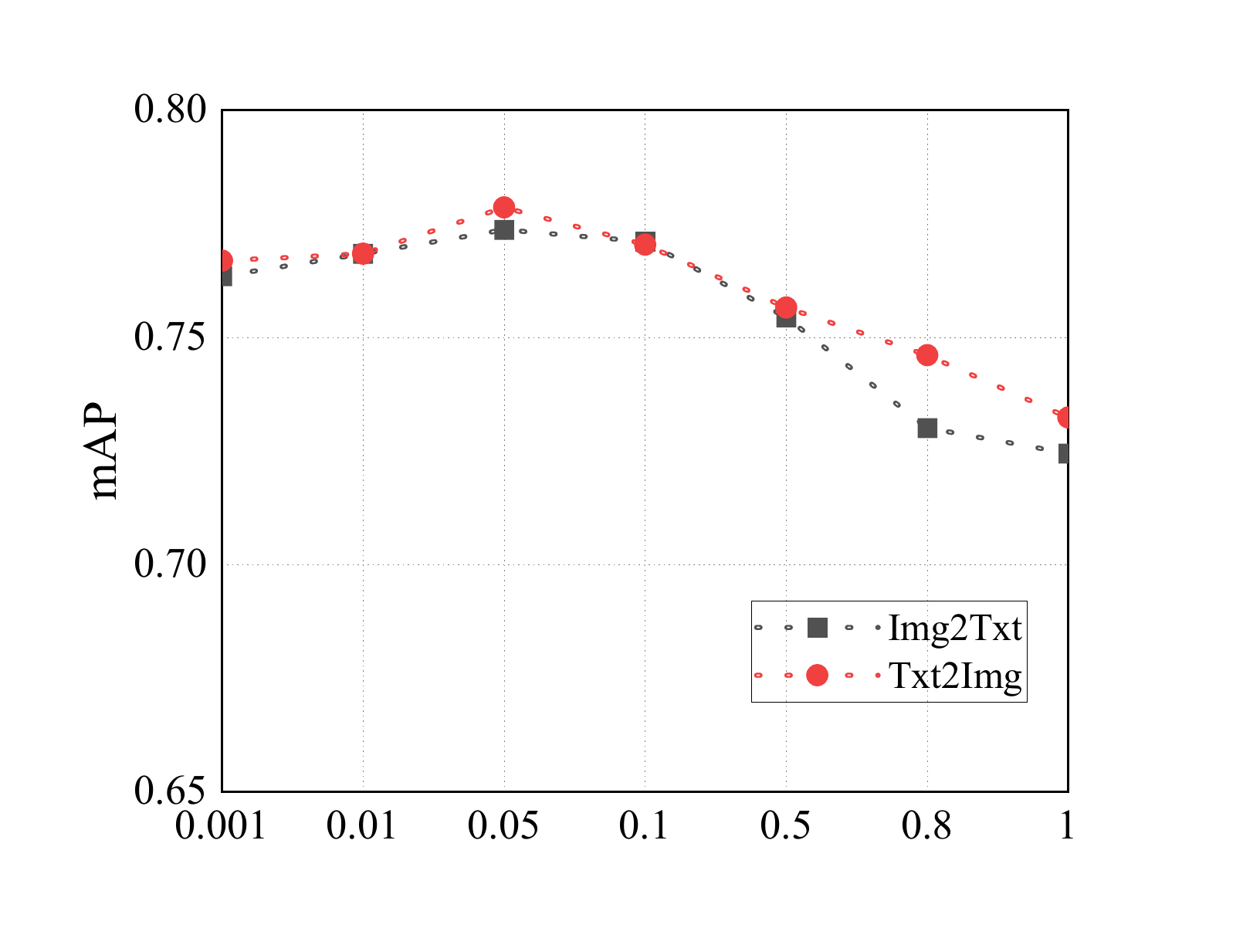}
        \caption{$\alpha$ on COCO@64 bits}
        \label{fig:a-coco64bit}
    \end{subfigure}
    
    \begin{subfigure}[b]{0.3\textwidth}
        \includegraphics[width=\textwidth]{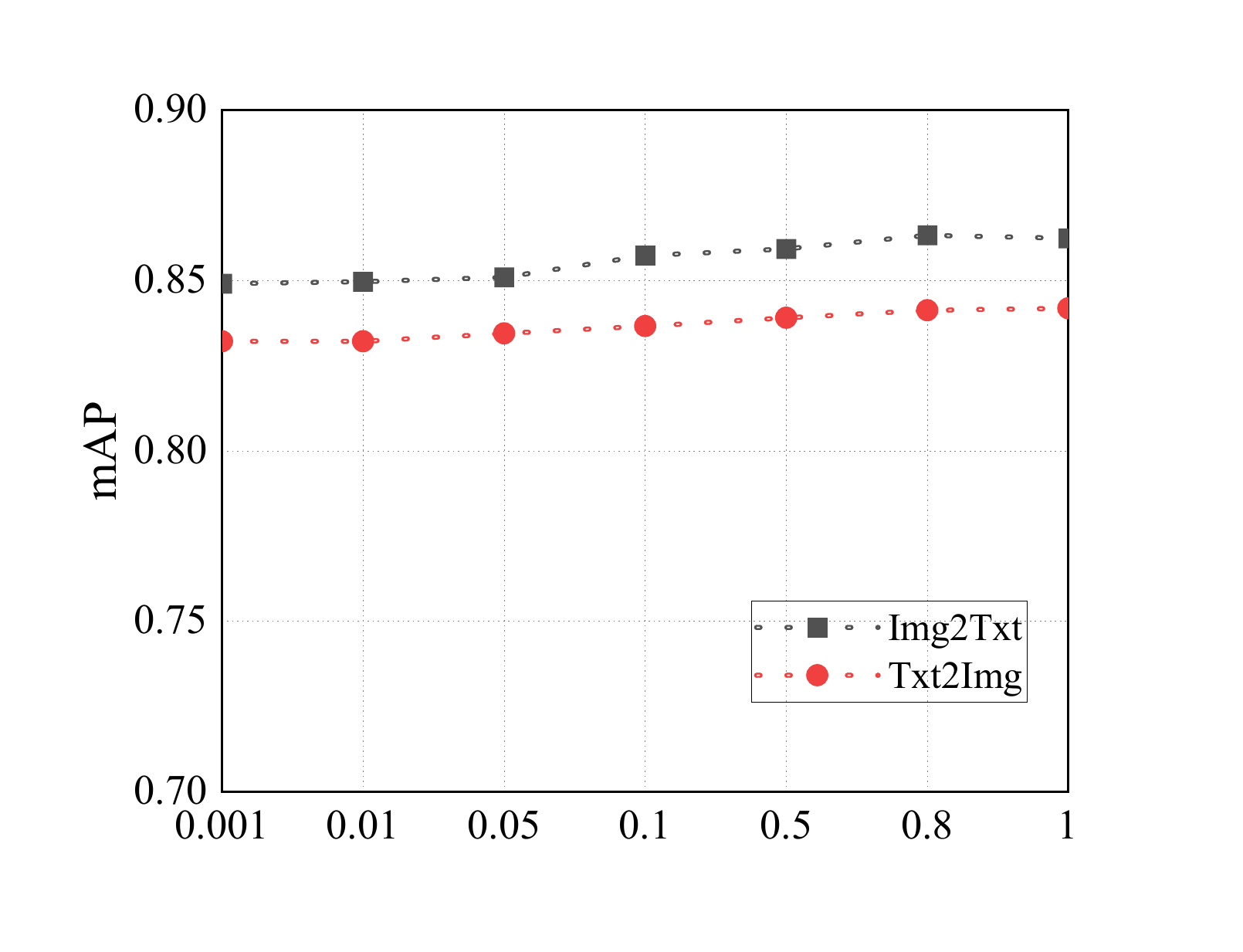}
        \caption{$\beta$ on MIR@64 bits}
        \label{fig:b-mirflickr64bit}
    \end{subfigure}
    \hfill
    \begin{subfigure}[b]{0.3\textwidth}
        \includegraphics[width=\textwidth]{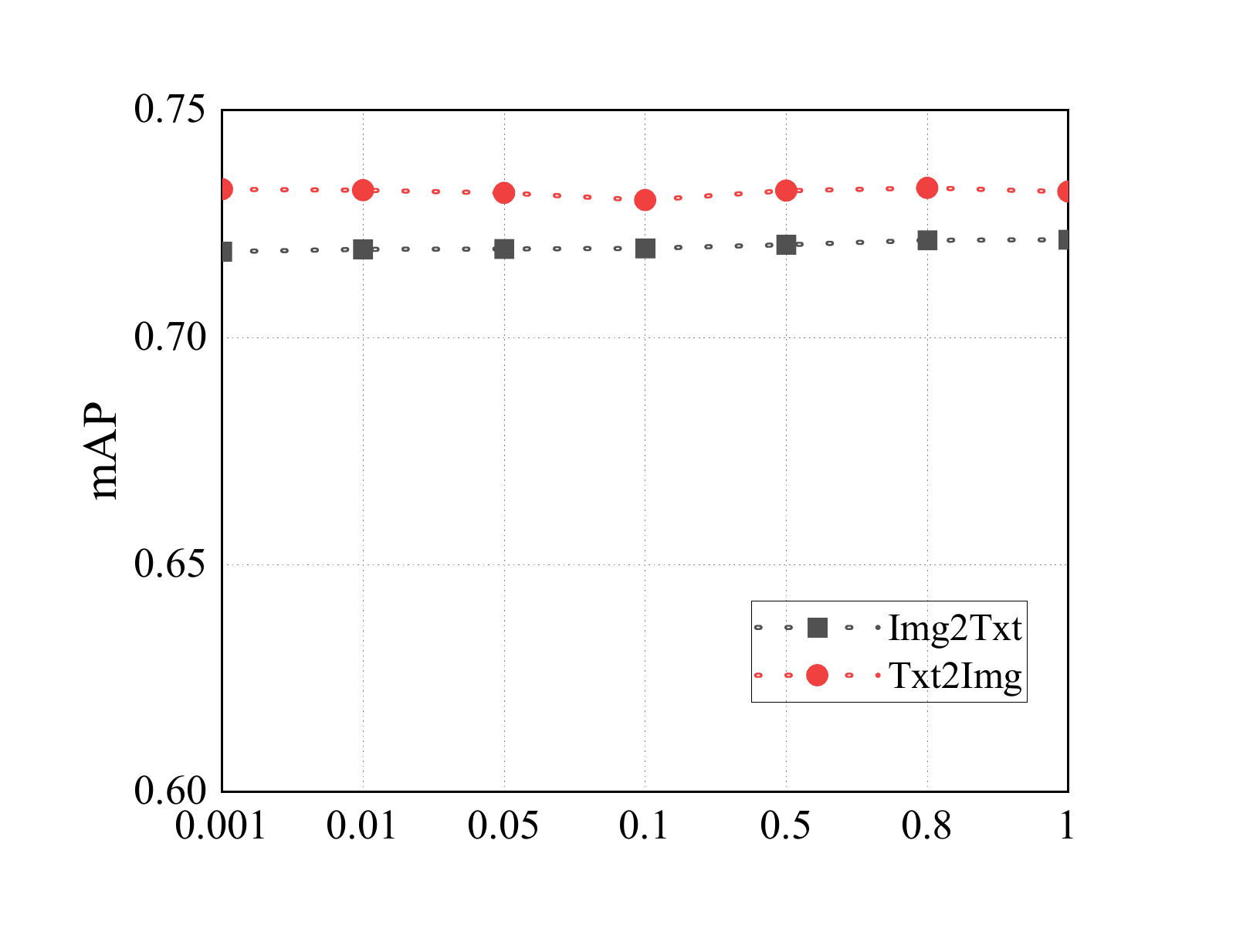}
        \caption{$\beta$ on NUS@64 bits}
        \label{fig:b-nuswide64bit}
    \end{subfigure}
    \hfill
    \begin{subfigure}[b]{0.3\textwidth}
        \includegraphics[width=\textwidth]{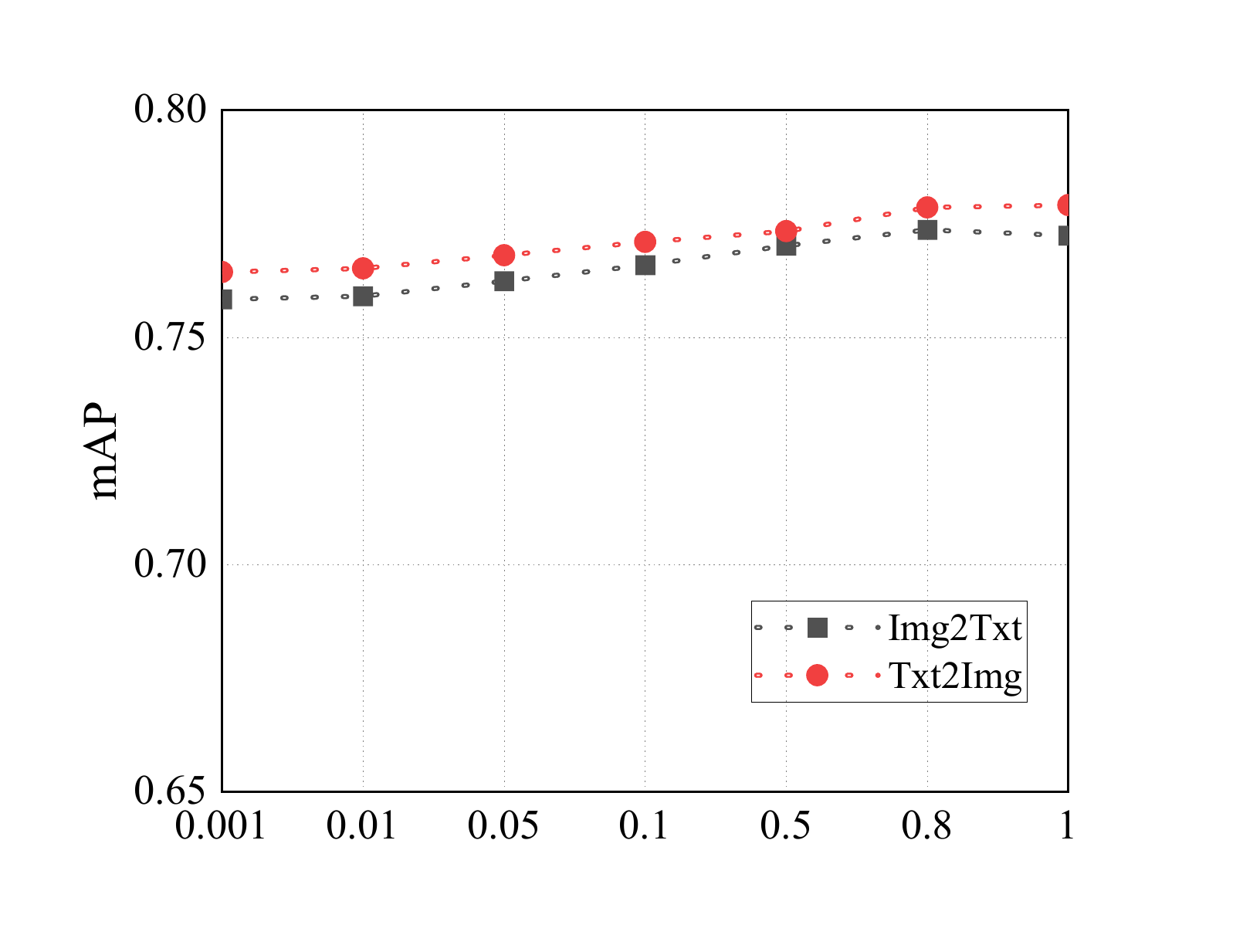}
        \caption{$\beta$ on COCO@64 bits}
        \label{fig:b-coco64bit}
    \end{subfigure}
    
    \caption{\small Parameter sensitive of $\alpha$ and $\beta$ on MIRFLICKR-25K, NUS-WIDE, and MS COCO datasets. The code length is 64.}
    \label{fig:SH}
\end{figure*}

We also investigate the sensitivity of the parameters $\alpha$ and $\beta$.
We set their ranges to \{0.001,0.01,0.05,0.1,0.5,0.8,1\} and report
the results in Fig~\ref{fig:SH}. From Figures~\ref{fig:SH} (a) and (d), it can be observed that the trend of alpha and beta on the MIRFLICKR-25K dataset is quite intuitive, with the best performance occurring at alpha and beta values of 0.05 and 0.8, respectively. Figures~\ref{fig:SH} (b) and (c) show that on large-scale datasets, as the value of alpha increases, meaning the constraint between positive samples becomes stronger, the proxy loss can no longer dominate the training of the hashing network, leading to intra-class dispersion and a significant drop in mAP scores. Figure~\ref{fig:SH} (f) indicates that due to the more relaxed constraint on negative samples, the increase of beta has a smaller impact on the hashing network. From Figure~\ref{fig:SH} (e), we also find an interesting phenomenon that the value of beta has almost no effect on the mAP scores on the NUS-WIDE dataset, which we believe may be due to the overwhelming size of the NUS-WIDE dataset with only 21 categories, and the proportion of unrelated samples is too small. This also demonstrates the ingenuity of our approach to set separate weight parameters for positive and negative sample constraints in the pairwise loss.

\subsection{Training and Encoding time} \label{subsec:Training and Encoding time}

\begin{figure}
    \centering
    \begin{subfigure}[b]{0.23\textwidth}
        \includegraphics[width=\textwidth]{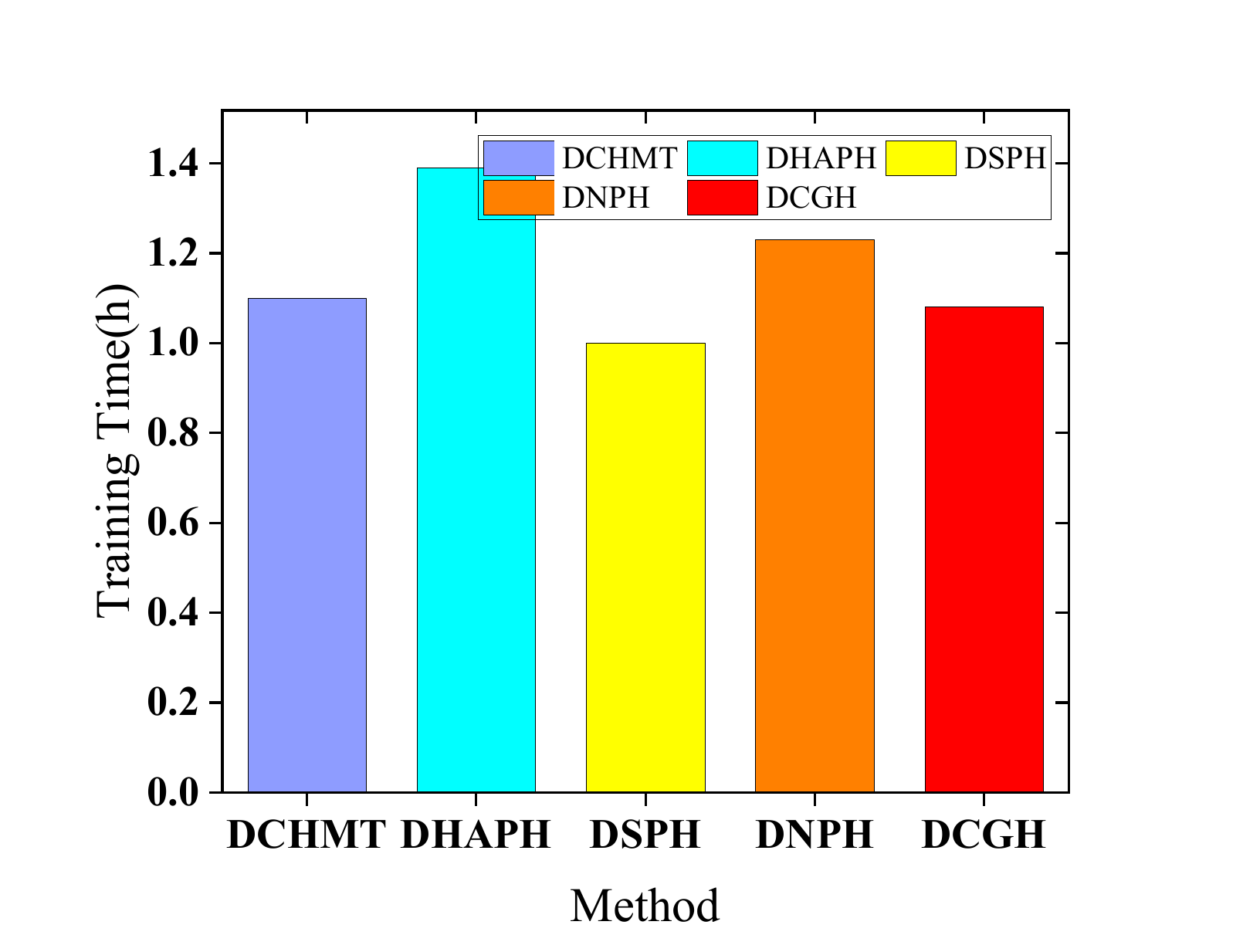}
        \caption{Training time}
        \label{fig:Training Time}
    \end{subfigure}
    \begin{subfigure}[b]{0.24\textwidth}
        \includegraphics[width=\textwidth]{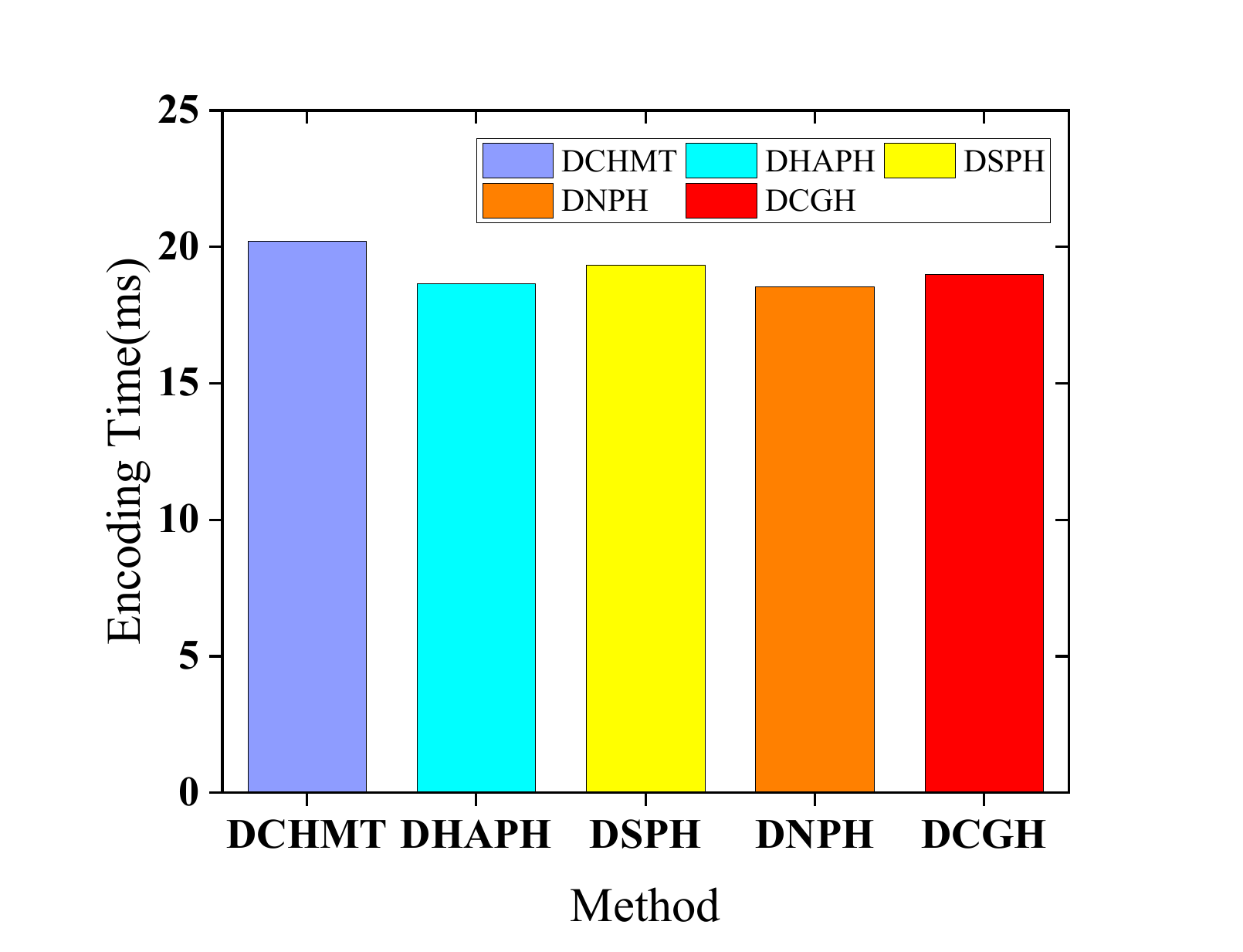}
        \caption{Encoding time}
        \label{fig:Encoding Time}
    \end{subfigure}
    \caption{\small Training time and Encoding time of DCGH on
MIRFLICKR-25K.}
    \label{fig:time}
\end{figure}

To investigate the efficiency of the model, we compared the training time and encoding time of the DCGH algorithm with other transformer-based cross-modal hashing algorithms on the MIRFLICKR-25K dataset using 32-bit binary codes, with the results shown in Figure~\ref{fig:time}. During the optimization of the DCGH model, a proxy loss algorithm was proposed, which has a time complexity of $O(NC)$, and $C$ is the number
of categories. Additionally, the time complexities for the pairwise loss and variance constraint are $O(N^2)$ and $O(\lambda NC)$, respectively. Here, $\lambda$ represents the ratio of the number of related data-proxy pairs to the total number of data-proxy pairs, with $\lambda$ being less than 1. Therefore, the total time complexity of our algorithm is $O\left(NC+\lambda NC+N^{2}\right)$, which is approximately equal to $O(N^2)$. As shown in Figure~\ref{fig:time} (a), compared with several other state-of-the-art methods, our training time is average, and since the training process of these methods is offline, the training time does not affect the performance of the method. As shown in the comparison of encoding time in Figure~\ref{fig:time} (b), the encoding time of all methods is within the millisecond range, indicating that our method achieves comfortable encoding time.

\subsection{Visualization} \label{subsec:visualization}

\begin{figure*}
    \centering
    \begin{subfigure}[b]{0.3\textwidth}
        \includegraphics[width=\textwidth]{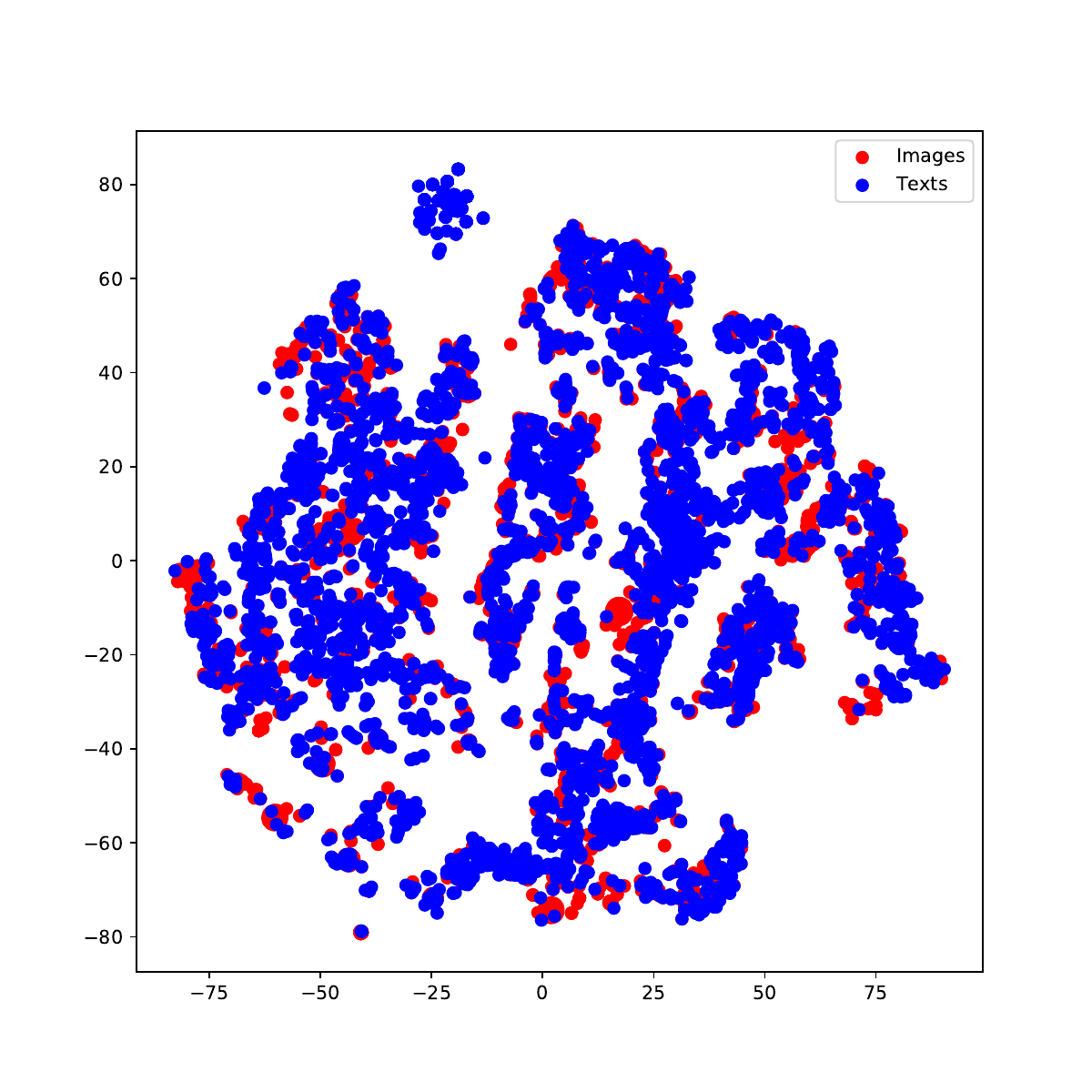}
        \caption{MIR@64 bits}
        \label{fig:tsne-flickr25k}
    \end{subfigure}
    \hfill
    \begin{subfigure}[b]{0.3\textwidth}
        \includegraphics[width=\textwidth]{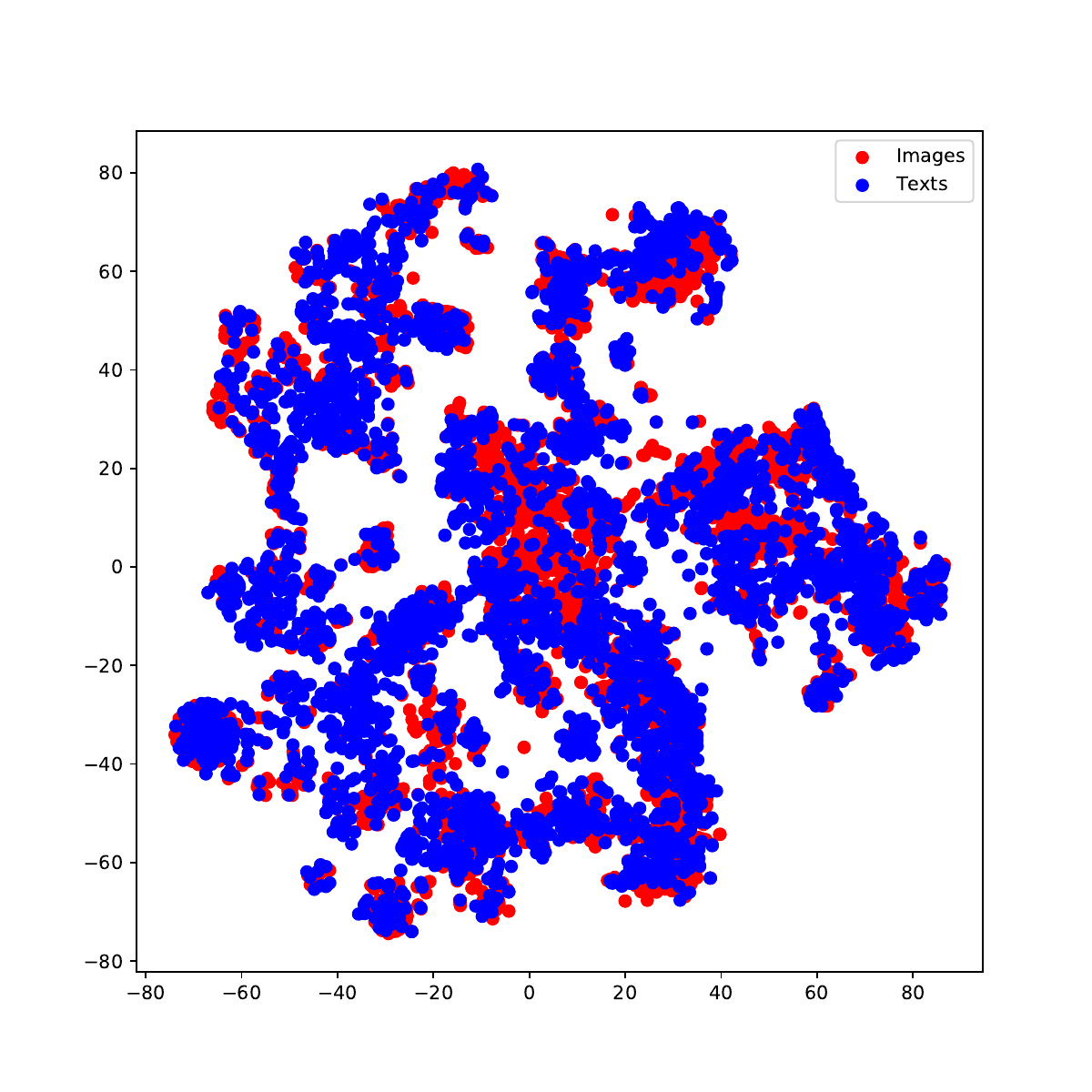}
        \caption{NUS@64 bits}
        \label{fig:tsne-nuswide}
    \end{subfigure}
    \hfill
    \begin{subfigure}[b]{0.3\textwidth}
        \includegraphics[width=\textwidth]{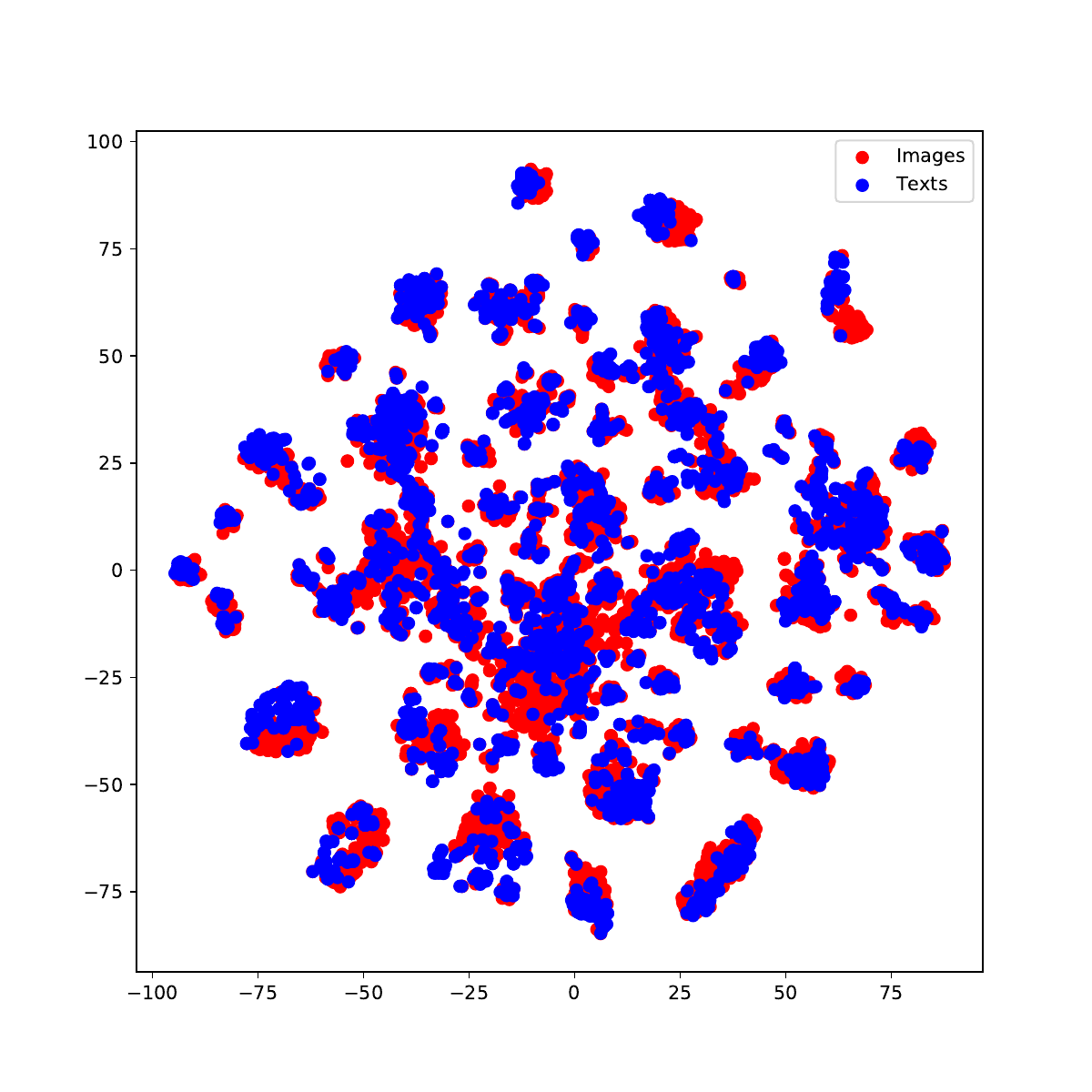}
        \caption{COCO@64 bits}
        \label{fig:tsne-coco}
    \end{subfigure}
    \caption{\small T-sne visualization of hash codes of image and text modalities
on MIRFLICKR-25K, NUS-WIDE, and MS COCO datasets. The code length is 64.}
    \label{fig:TSNE-IT}
\end{figure*}

To explore the ability of our model to bridge the semantic gap between different modalities, we used the T-sne~\cite{T-SNE} technique to project discrete hash codes into a two-dimensional space on three public datasets, as shown in Figure~\ref{fig:TSNE-IT}. We can see that the alignment between text and images is quite good. 

\begin{figure*}
    \centering
    \begin{subfigure}[b]{0.3\textwidth}
        \includegraphics[width=\textwidth]{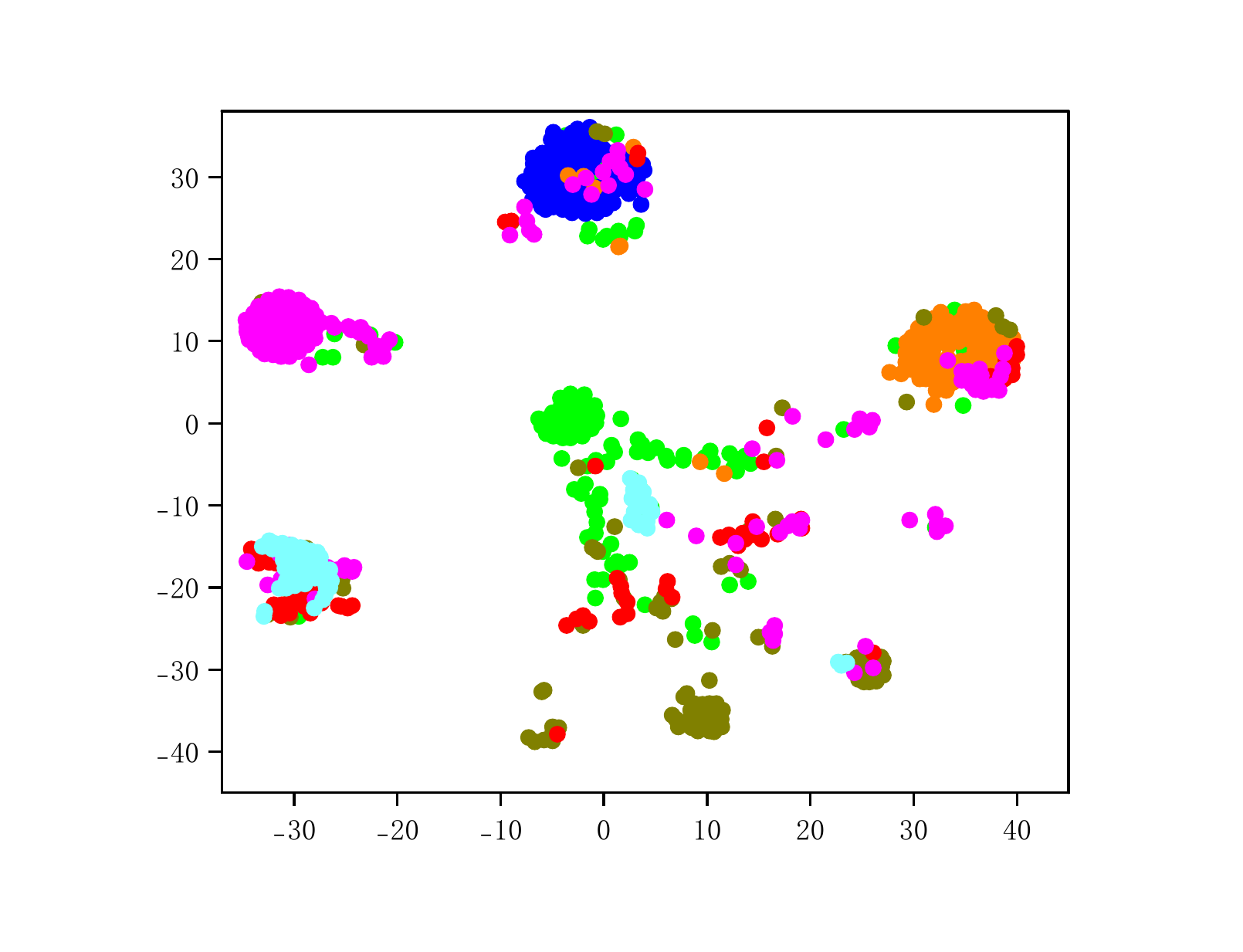}
        \caption{DCHMT}
        \label{fig:TSNE-DCHMT}
    \end{subfigure}
    \hfill
    \begin{subfigure}[b]{0.3\textwidth}
        \includegraphics[width=\textwidth]{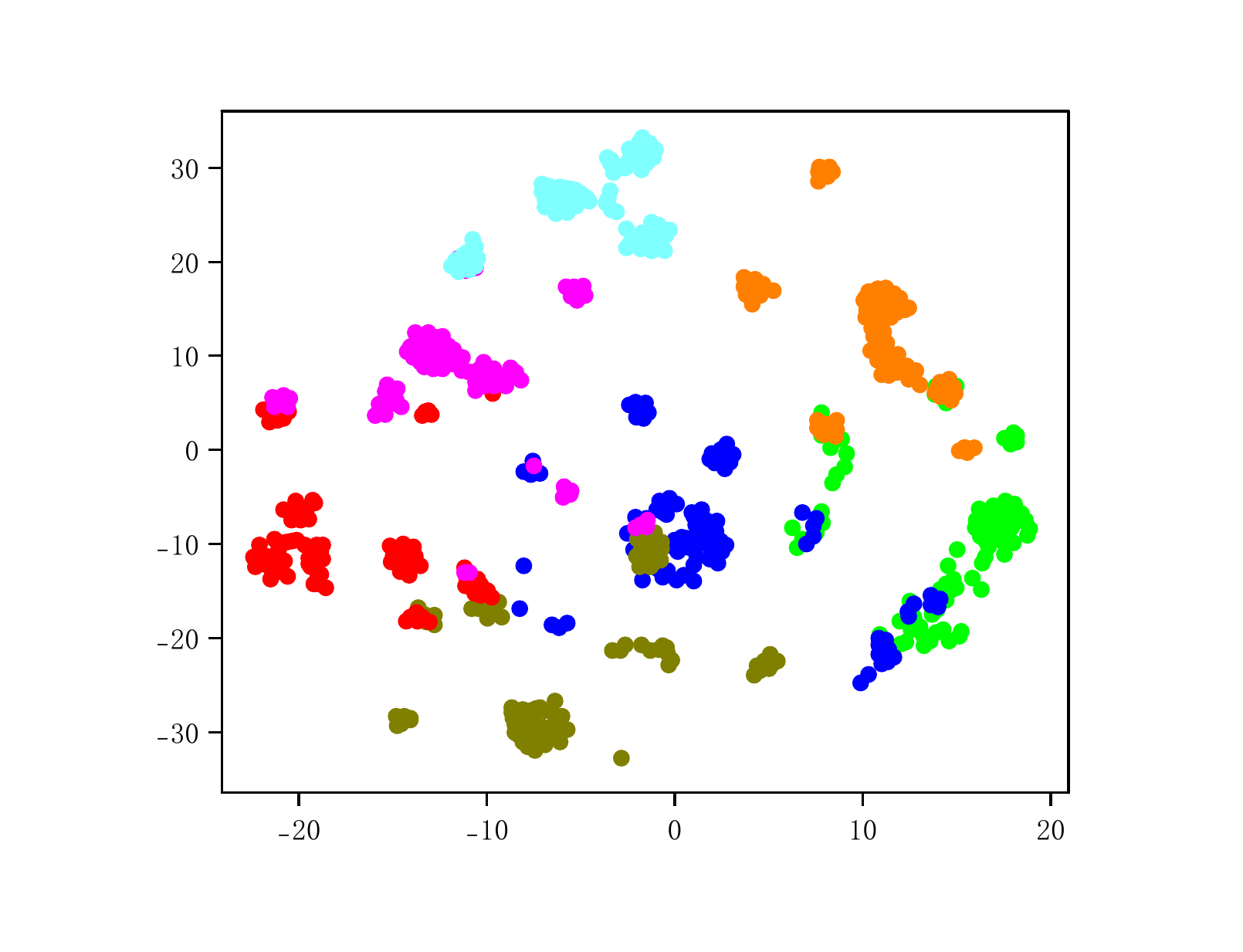}
        \caption{SCH}
        \label{fig:T-SNE-SCH}
    \end{subfigure}
    \hfill
    \begin{subfigure}[b]{0.3\textwidth}
        \includegraphics[width=\textwidth]{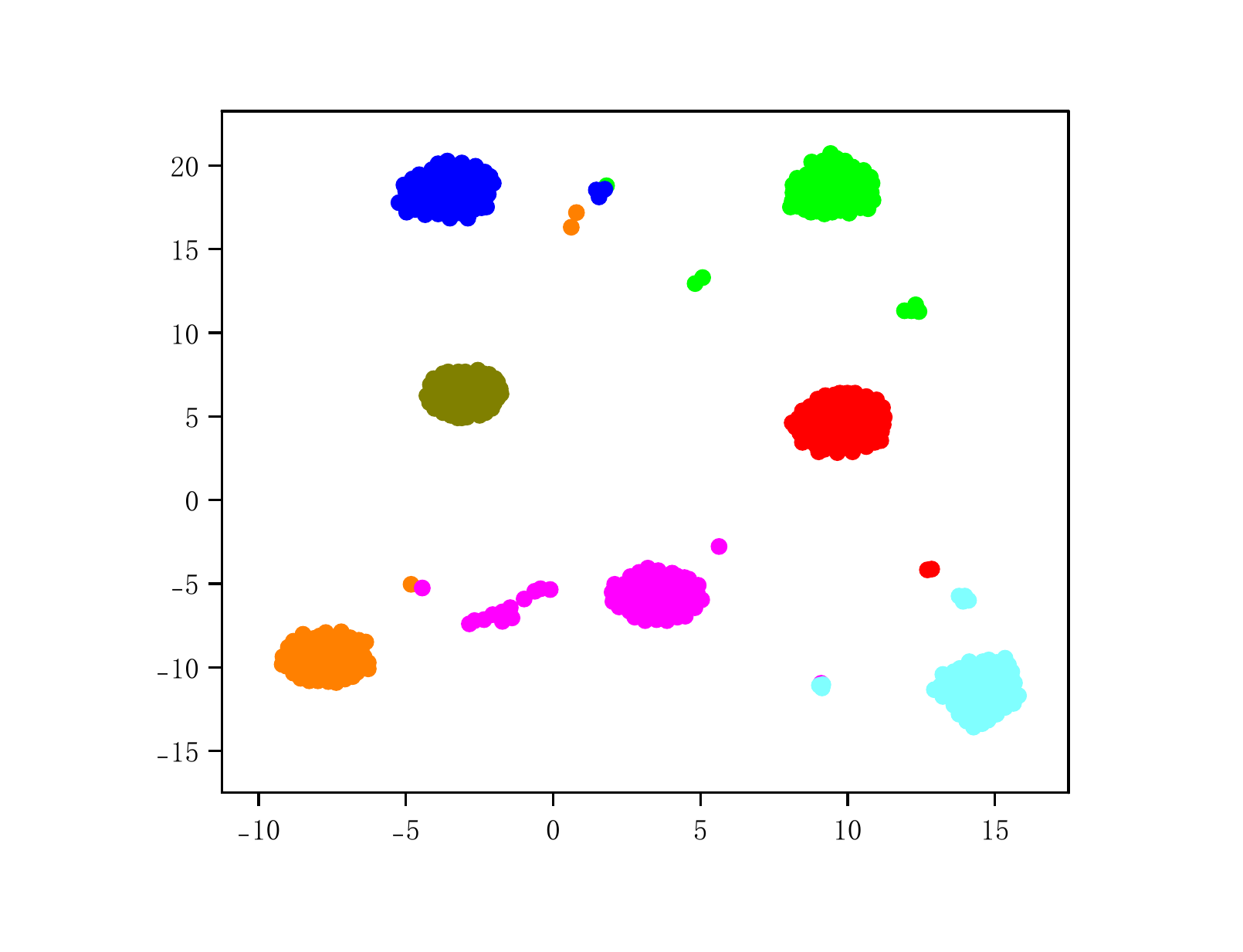}
        \caption{DCGH}
        \label{fig:TSNE-DCGH}
    \end{subfigure}
    \caption{\small T-sne result of the DCHMT, the SCH, and the DCGH on NUS-WIDE w.r.t. 16 bits.}
    \label{fig:TSNE-DB}
\end{figure*}

To explore the quality of the hash codes and the intra-class aggregation, we selected samples from seven different single-label categories in the NUS-WIDE dataset and performed T-sne visualization on 16-bit hash codes using the DCHMT, SCH, and DCGH methods. The results are shown in Figure~\ref{fig:TSNE-DB}. Through comparison and analysis, our method can obtain higher quality and more intra-class aggregated binary hash codes due to the joint training of the model with proxy loss, pairwise loss, and variance constraint terms.

\section{Conclusion} \label{sec:conclusion}
This paper proposes a cross-modal retrieval method based on DCGH, which takes into account both the intra-class aggregation and inter-class structural relationship preservation of hash codes, generating superior hash codes. In the feature learning module, the framework proposed in this paper adopts an image semantic encoder and a text semantic encoder based on the Transformer encoder structure, respectively capturing the latent semantic information of images and text. For hash learning, the method explores the relationships between points and classes as well as points and points through the fusion of proxy loss and pairwise loss, and introduces a variance constraint to further address the semantic bias issue that may exist after fusion. Extensive experiments on the MIRFLICKR-25K, NUS-WIDE, and MS COCO datasets have proven that the DCGH method has good retrieval performance. We introduce the DCGH method to the image-text cross-modal retrieval task. However, multi-modal data sources involve more than just images and text. In the future, we will extend the DCGH framework to achieve image-text-video multi-modal retrieval.



\section*{Declaration of competing interest}
The authors declare that they have no known competing financial interests or personal relationships that could have appeared to influence the work reported in this paper. 

\section*{Data availability}
Data will be made available on request.

\bibliographystyle{cas-model2-names}

\bibliography{cas-refs}


\end{document}